\documentclass[aps,prd,floatfix,nofootinbib,superscriptaddress,noeprint,11pt]{revtex4-2}
\pdfoutput=1 
\usepackage[T1]{fontenc}
\usepackage[utf8]{inputenc}
\usepackage[greek,english]{babel}
\usepackage{CJKutf8}
\usepackage{amsfonts}
\usepackage{tikz}
\usetikzlibrary{decorations}
\usepackage{amsmath}
\usepackage{amssymb}
\usepackage{verbatim,graphicx}
\usepackage{mathtools}
\usepackage{color}
\usepackage[all]{xy}
\usepackage{simplewick}
\usepackage{setspace}
\usepackage{array}
\usepackage{tkz-euclide}
\usepackage{amsthm}
\usepackage{enumitem}
\usepackage{bbm}
\usepackage{bbold}
\usepackage{subfigure}
\usepackage[%
bookmarks=true,        
bookmarksopen=true,    
bookmarksnumbered=true]{hyperref}

\newcounter{amg}

\newcounter{ls}

\newcounter{jvcc}

\newcounter{zjp}

\newcommand{\bi} {\langle i|}
\newcommand{\bj} {\langle j}
\newcommand{\bk} {\langle k|}
\newcommand{\bl} {\langle l|}
\renewcommand{\bm} {\langle m|}
\newcommand{\ki} {|i \rangle}
\newcommand{\kj} {|j \rangle}
\newcommand{\kk} {|k \rangle}
\newcommand{\kl} {|l \rangle }
\newcommand{\km} {|m \rangle } 
\newcommand{\edoc}{\end{document}}

\newcommand{\Tr}{{\rm Tr}}

\newcommand{\eref}[1]{(\ref{#1})}
\newcommand{\nn}{\nonumber}
\newcommand{\sign}{{\rm sign}}
\newcommand{\be}{\begin{eqnarray}}
\newcommand{\ee}{\end{eqnarray}}

\newcommand{\ket}[1]{| #1\rangle}

\newcommand{\sC}{\mathcal{C}}
\newcommand{\rmK}{\mathrm{K}}
\newcommand{\rmR}{\mathrm{R}}
\newcommand{\rmA}{\mathrm{A}}
\newcommand{\rmH}{\mathrm{H}}
\newcommand{\rmT}{\mathrm{T}}
\renewcommand{\(}{\left(}
\renewcommand{\)}{\right)}

\newcommand{\heav}[1]{\,\Theta\!\left(#1\right)}
\newcommand{\bmat}{\left ( \begin{array}{cc} }
	\newcommand{\emat}{\end{array} \right ) }

\newcommand{\beq}{\begin{equation}}
\newcommand{\beqs}{\begin{equation*}}
\newcommand{\eeq}{\end{equation}}
\newcommand{\eeqs}{\end{equation*}}

\begin{document}

\title{Keldysh Wormholes and Anomalous Relaxation in the Dissipative Sachdev-Ye-Kitaev Model}
\author{Antonio M. Garc\'\i a-Garc\'\i a}
\email{amgg@sjtu.edu.cn}
\affiliation{Shanghai Center for Complex Physics,
	School of Physics and Astronomy, Shanghai Jiao Tong
	University, Shanghai 200240, China}
\author{Lucas S\'a}
\email{lucas.seara.sa@tecnico.ulisboa.pt}
\affiliation{CeFEMA, Instituto Superior T\'ecnico, Universidade de Lisboa, Av.\ Rovisco Pais, 1049-001 Lisboa, Portugal}
\author{Jacobus J. M. Verbaarschot}
\email{jacobus.verbaarschot@stonybrook.edu}
\affiliation{Center for Nuclear Theory and Department of Physics and Astronomy, Stony Brook University, Stony Brook, New York 11794, USA}
\author{Jie Ping Zheng\begin{CJK*}{UTF8}{gbsn}
		(郑杰平)
\end{CJK*}}
\email{jpzheng@sjtu.edu.cn}
\affiliation{Shanghai Center for Complex Physics,
	School of Physics and Astronomy, Shanghai Jiao Tong
	University, Shanghai 200240, China}

\begin{abstract}
\vspace{0.5cm}
We study the out-of-equilibrium dynamics of a Sachdev-Ye-Kitaev (SYK) model, $N$ fermions with a $q$-body interaction of infinite range, coupled to a Markovian environment.
Close to the infinite-temperature steady state, the real-time Lindbladian dynamics of this system is identical to the near-zero-temperature dynamics in Euclidean time of a two-site non-Hermitian SYK with intersite coupling whose gravity dual has been recently related to wormhole configurations. We show that the saddle-point equations in the real-time formulation are identical to those in Euclidean time. Indeed, an explicit calculation of Green's
functions at low temperature, numerical for $q = 4$ and analytical for $q = 2$ and large $q$, illustrates this equivalence. 
Only for very strong coupling does the decay rate approach the linear dependence on the coupling characteristic of a dissipation-driven approach to the steady state. 
For $q > 2$, we identify a potential gravity dual of the real-time dissipative SYK model: a double-trumpet configuration in a near-de Sitter space in two dimensions with matter. This configuration, which we term a Keldysh wormhole, is responsible for a finite decay rate even in the absence of coupling to the environment.
\end{abstract}

\maketitle

\newpage

{
\setstretch{1.2}
\tableofcontents
}

\clearpage

\section{Introduction}
The existence of an environment, either thermal, quantum, or the very measurement process, makes the time evolution of a quantum system not strictly unitary, complicating the description of its time evolution. 
An elegant approach to this problem, advocated by Caldeira and Leggett~\cite{caldeira1981},
is to describe the system plus environment by an Hermitian Hamiltonian. The integration of the environmental degrees of freedom then results in nonunitary dynamics. Especially for many-body systems, it is technically hard to tackle the resulting problem, either numerically or analytically, because the tracing out of the environment leads to nonlocal interactions in time.
 
In the limit where the environment has a sufficiently short correlation time, the Liouvillian generator that governs the quantum time evolution of the density matrix of the reduced system can be written in the so-called Lindblad form~\cite{belavin1969,lindblad1976,gorini1976,breuer2002,manzano2020}. Compared to the usual von Neumann equation of motion, the Lindblad equation contains additional terms modeled by the so-called jump or Lindblad operators \cite{lindblad1976} that describe the coupling of the system to the environment. The advantage of this approach is that the quantum master equation is expressed only in terms of the system degrees of freedom, which facilitates its solution. 
In recent times, there has been a renewed interest in different aspects of the Liouvillian dynamics, especially in random systems \cite{denisov2018,can2019,can2019a,sa2019,wang2020,sommer2021,lange2021,tarnowski2021,li2022PRB,costa2022}, including the study of the spectrum \cite{denisov2018,sa2019,wang2020,tarnowski2021}, with the aim to extending the Bohigas-Giannoni-Schmit conjecture \cite{bohigas1984} to dissipative quantum systems \cite{grobe1988,fyodorov1997,sa2019csr}, the characterization of the late-time dynamics \cite{can2019,can2019a,sa2019,li2022PRB} towards a steady state~\cite{sa2019,sa2020,costa2022,tonielli2020} or the robustness of dissipative quantum chaotic features \cite{rubio2022}.

The Keldysh path integral \cite{kamenevbook,sieberer2016} is the standard procedure used to investigate the nonequilibrium time evolution of the density matrix. 
A key difference to the standard path integral, which will be very important in the following, is that, in order to set up the calculation for the density matrix, it is necessary \cite{kamenevbook,sieberer2016} to consider the time evolution of a state matrix, not a state vector. This results in two time branches, one forward and one backward in time, and therefore to the doubling of degrees of freedom of the path integral (vectorization). In this representation, the density matrix is a state in the doubled Hilbert space and the Liouvillian has two parts: the first is anti-Hermitian, corresponding to the Hermitian Hamiltonian, before any coupling to the environment but multiplied by the imaginary unit, and its conjugate such that the whole system is anti-Hermitian; the second contains Lindblad jump operators depending on fields living in both copies and includes (i) an explicit coupling between the degrees of freedom of the two copies and (ii) intrasite interactions effectively modifying the (anti-)Hermitian part.
Using standard field theory techniques, it is then possible to investigate the out-of-equilibrium time evolution of a strongly interacting system either coupled to a bath or undergoing a continuous measurement process. If we are interested in the steady state, and in sufficiently long timescales, it is plausible to assume that the initial state becomes irrelevant, leading to further simplifications, namely, a time-translational invariant relaxation.

Intriguingly, operators similar to the vectorized Liouvillian are being employed in a completely different context: as non-Hermitian two-site Sachdev-Ye-Kitaev (SYK) Hamiltonians \cite{french1970,bohigas1971,french1971,bohigas1971a,benet2001,sachdev1993,sachdev2010,kitaev2015,maldacena2016}, $N$ Majoranas with infinite range $q$-body interactions in zero dimensions, whose gravity dual is a wormhole \cite{maldacena2018,gao2016,garcia2021}. Earlier, a single-site SYK model was proposed as an example of the holographic duality
based on the fact that features such as the saturation of a bound on chaos \cite{maldacena2015}, spectral correlations described by random matrix theory \cite{garcia2016,altland2018,*altland2022}, and an exponential growth of the density of low-energy excitations \cite{cotler2016,garcia2017,stanford2017} are shared by certain near-AdS$_2$ backgrounds \cite{jackiw1985,teitelboim1983,maldacena2016a,almheiri2015} in the low-temperature limit. 
It has also been noticed \cite{maldacena2018} that the gravity dual of two identical two-site Hermitian SYKs with a weak intersite coupling is a traversable wormhole \cite{gao2016}. These traversable wormholes are the dominant
configurations only for sufficiently low temperature. At higher temperature, a transition to a black hole configuration occurs, which can also be characterized by an analysis of level statistics \cite{garcia2019}.

By contrast, a pair of non-Hermitian PT-symmetric SYKs with no explicit intersite coupling in the low-temperature limit is dual to a Euclidean wormhole \cite{garcia2021}. The presence of an explicit coupling triggers \cite{garcia2022b} a transition from Euclidean to traversable wormholes.
Both traversable and Euclidean wormholes are characterized by a gapped ground state and a first-order phase transition in the free energy separating the wormhole and black hole phases. 
Differences include a qualitatively different dependence of the gap on the parameters of the model and a different pattern of oscillations of Green's functions in real time directly related to different symmetries of the solutions: $U(1)$ for Euclidean and $SL(2,R)$ for traversable. For earlier work on wormholes, mostly in higher dimensions, see \cite{Lavrelashvili:1987jg,Hawking:1987mz,Giddings:1987cg,giddins1988,maldacena2004}. 

In this paper, we show that the near-zero-temperature dynamics of a two-site SYK Hamiltonian in Euclidean time, dual to gravitational wormholes in certain limits, and the real-time Liouvillian describing a single-site SYK coupled to a Markovian bath close to the infinite-temperature steady state coincide.
%In this paper, we show that the low-temperature dynamics of SYK Hamiltonians in %Euclidean time, dual to wormholes, and Liouvillians describing an SYK system coupled %to a Markovian bath close to the steady state coincide. 
More specifically, we show that the equations of motion in the Euclidean problem in the low-temperature limit are identical to the equations of motion in the real-time Liouvillian evolution close to the steady state. This implies that the retarded Green's function of the real-time problem is identical to the equivalent one in the Euclidean problem. As a consequence, the so-called gap in the Euclidean formulation that characterizes the wormhole phase is equal to the dominant decay rate in the real-time Liouvillian evolution. This equivalence only holds when neglecting the boundary conditions, which are generally different
for the two approaches. In practical terms, the equivalence works for sufficiently low temperatures in the Euclidean problem and for times such that the system is sufficiently close to the steady state in the real-time problem. 
 
Finally, we note that the gravitational dual of a dissipative field theory with decoherence was investigated in Ref.~\cite{delcampo2020}. We also stress that there are already several works that have studied different aspects of the non-Hermitian SYK model: from the use of the Lindblad formalism \cite{sa2022,kulkarni2022} and the investigation of the entanglement entropy growth \cite{pengfei2021,pengfei2021b} and decoherence effects \cite{pengfei2021c,xu2020,cornelius2022}, to its relation to Euclidean wormholes \cite{garcia2021,garcia2022}, replica symmetry breaking \cite{garcia2021a,garcia2022b}, and a symmetry classification \cite{garcia2022d}. However, these studies have not investigated the mentioned intriguing relation between wormholes in Euclidean time and strongly interacting dissipative systems in real time. We shall see that this finding has a strong impact on important features of the dynamics, such as an enhancement of the decay rate in the limit of very weak coupling to the environment. 

The remainder of the paper is organized as follows. In the next section, we introduce the open SYK model, which will be our main object of study, and the equations of motion in the $\Sigma G$ formulation \cite{maldacena2016} in both real and Euclidean time. 
In Sec.~\ref{sec:q4}, we calculate the long-time behavior of the Green's function by numerically solving the Schwinger-Dyson (SD) equations for $q=4$ and find an anomalously large decay rate (or gap) for small intersite coupling. 
For $q=2$, the Green's functions can be evaluated analytically and, because the model is not quantum chaotic, no anomalous relaxation is found, which is discussed in Sec.~\ref{sec:q2}. The large-$q$ limit is worked out analytically in Sec.~\ref{sec:qinfty}. 
The gravity dual of the SYK Hamiltonian coupled to an environment, dubbed Keldysh wormhole, is discussed in Sec.~\ref{sec:dual} and concluding remarks are made in Sec.~\ref{sec:conclusions}.
Additional technical details are worked out in six appendices.

\section{Lindblad equation and SYK Hamiltonians}
\label{sec:models}
We study a Hermitian SYK Hamiltonian with a $q$-body interaction of infinite range coupled to a Markovian environment. The SYK model is defined by the Hamiltonian 
\begin{align}\label{eq:sykq}
	H^{{\rm SYK}} &=
	-i^{q/2}
	\sum_{i_1<\cdots < i_q}
	J_{i_1\cdots i_q}
	\psi^{i_1} \cdots \psi^{i_q},
\end{align}
where $J_{i_1\cdots i_q}$ are random numbers extracted from a Gaussian distribution of zero average and variance $\langle J_{i_i \cdots i_q}^2 \rangle = (q-1)!J^2/N^{q-1}$, with $i_1,\dots,i_q=1,\dots,N$. We will mostly focus on the cases $q = 2$ (integrable), $q=4$, and large $q \to\infty$. The last two are quantum chaotic. The $\psi^i$ are Majorana operators defined by the commutation relation $\{\psi^i, \psi^j\}=\delta_{ij}$.

\subsection{The vectorized Lindblad equation}
The real-time evolution of the density matrix $\rho$ \cite{breuer2002} of this open SYK is simply $d\rho/dt = \mathcal{L}(\rho)$
where $\mathcal{L}$ is the Liouvillian which can be conveniently expressed in the Lindblad form: 
\begin{align} \label{eq:liou0}
	\mathcal{L}
	(\rho)
	&=
	- i [H^{{\rm SYK}}, \rho]
	+
	\sum_{\alpha}
	\left[
	L^{\ }_{\alpha}
	\rho
	L^{\dag}_{\alpha}
	-
	\frac{1}{2}
	\{
	L^{\dag}_{\alpha}L^{\ }_{\alpha},
	\rho
	\}
	\right].
\end{align}
Note the $-i$ difference in $d\rho/dt =\mathcal{L}(\rho)$ with respect to the time evolution of a state in the Schr\"odinger picture. 

The Markovian environment is described by Lindblad jump 
operators of the form 
\be
\label{jump_ope}
L_i = \sqrt{\mu}\psi^i,
\ee
with $i = 1, \dots, N$ and $\mu$ a positive real number. We believe that most of our main conclusions apply to more general jump operators but we will see that the relation to wormholes is better understood in this case.

For sufficiently long times, and taking into account that our jump operators are Hermitian, the density matrix will relax to a maximally entangled state at infinite temperature,
\be
\label{eq:steady-state}
\rho_\infty =\frac{1}{2^{N/2}} \sum_k |k\rangle \langle k|,
\ee
characterized by $\mathcal{L}(\rho_\infty)=0$. We note that for jump operators more general than those of Eq.~(\ref{jump_ope}), the system's density matrix may decay either to a maximally entangled state at finite temperature or to a nonequilibrium steady state, depending on the details of the Hamiltonian.
Our main interest is in the approach to the steady state~(\ref{eq:steady-state}). For that purpose, we study the retarded Green's function 
\be 
iG^\rmR(t) \delta_{ij} =  \Theta(t)\left\langle \Tr\left[\rho_{\infty}\{\psi^i(t),\psi^j(0)\}\right]\right\rangle,
\label{steady}
\ee
where $\Theta(t)$ is the Heaviside function.

In order to compute this quantity, it is useful to employ the Keldysh path integral that involves the doubling of the degrees of freedom and the vectorization of the Liouvillian (see Appendix~\ref{app:choi} for details).
 Here we state the final result for the partition function,
\be
\label{path_integral}
Z 
= \int \mathcal{D}\psi_L\mathcal{D}\psi_R\, e^{iS[\psi_L,\psi_R]},
\ee
where $L,R$ stand for the two copies of the original degrees of freedom, i.e., the Hilbert space has been doubled $\mathcal{H} = \mathcal{H}_L \otimes \mathcal{H}_R$. The
appropriate action is given by
\be
iS =\int_{-\infty}^{+\infty} dt \Big(-\frac{1}{2} \sum_i \psi^i_L \partial_t \psi^i_L-\frac{1}{2} \sum_i \psi^i_R \partial_t \psi^i_R +\mathcal{L} \Big),
\label{eq:II_action_real}
\ee
with the vectorized Liouvillian
\be
\label{doublelindbrad}
\mathcal{L} = -iH^{{\rm SYK}}_L + i  (-1)^\frac{q}{2} 
H^{{\rm SYK}}_R - i \mu \sum_i\psi_L^i\psi_R^i - \frac{1}{2}\mu 
N
\ee
acting on the doubled Hilbert space, where $H_L^\mathrm{SYK}$ and $H_R^\mathrm{SYK}$, given by Eq.~(\ref{eq:sykq}), have real couplings with the same probability distribution. For simplicity, we do not use the tensor product notation, and it is understood that $H^{{\rm SYK}}_L \equiv H^{{\rm SYK}} \otimes \mathbb{1}$ and so on. 
We assume that boundary conditions play no role in the dynamics, which allows us to extend the limits of integration in the action to infinity. This assumption holds if the system is sufficiently close to the steady state.
We evaluate the Green's functions of the operator $\mathcal{L}$
in two different ways: by the Keldysh approach corresponding to open real-time evolution, in the next subsection, and in Appendix~\ref{app:real_time}; and in Euclidean time, see Sec.~\ref{sect:euclid} and Appendix~\ref{app:Euclidean_time}.
In both cases, the saddle-point equations are obtained under the assumption of translational invariance. For a detailed derivation of the Keldysh approach, we refer the reader to Ref.~\cite{sa2022}.

\subsection{The Keldysh approach}
\label{sec:keldysh}

The Keldysh approach relies on the closed-contour representation of the time integration, see also Refs.~\cite{kamenevbook,sieberer2016,sa2022,tonielli2020}. We introduce fields $\psi^{i}(t^\pm)$ living on the closed-time contour $\sC=\sC^+\cup\sC^-$, with real time running from $-\infty$ to $+\infty$ (branch $\sC^+$) and then back again to $-\infty$ (branch $\sC^-$). We identify $\psi^i(t^+)=\psi^i_L(t)$, with $t^+\in\sC^+$, and $\psi^i(t^-)=i\psi^i_R(t)$, with $t^-\in\sC^-$. $\psi^i(t^+)$ ($\psi^i(t^-)$) 
corresponding to Majoranas acting on the left (right) of the density matrix. 
After disorder averaging, the action~(\ref{eq:II_action_real}) can be written in terms of the Green's functions
\begin{equation}
G_{ab}(t_1,t_2)=-\frac{i}{N}\sum_{i=1}^N \psi^i (t_1^a)\,\psi^i(t_2^b),
\end{equation}
with $a,b=+,-$ corresponding to the respective contours, and its self-energy $\Sigma_{ab}(t_1,t_2)$ (see Appendix~\ref{app:real_time} for details).
The saddle-point equations in terms of $G_{++}(t_1,t_2)$ (both times on $\sC^+$), $G_{+-}(t_1,t_2)$ (time $t_1 \in \sC^+$ and time $t_2 \in \sC^-$), and the corresponding self-energies are given by
\be
\label{eq:SD_real_I}
&&(\omega-\Sigma_{++}(\omega))G_{++}(\omega)-\Sigma_{+-}(\omega)G_{+-}(\omega)=1,\nn\\
&&(\omega-\Sigma_{++}(\omega))G_{+-}(\omega) -\Sigma_{+-}(\omega)G_{++}(\omega)=0,
\\
\label{eq:SD_real_II}
&&\Sigma_{++}(t)=-i^qJ^2 G_{++}^{q-1}(t), \qquad
 \Sigma_{+-}(t)=-i^q J^2G_{+-}^{q-1}(t)+i\mu\delta(t).
\ee
Here, we have assumed translational invariance of the solutions, so that they only depend on the relative
time $t=t_1-t_2$, which makes it possible to write the first line of equations in frequency space.
The equations can be decoupled in terms of the combinations
\be
G^\rmR = G_{++}-G_{+-}, \qquad \Sigma^\rmR = \Sigma_{++}-\Sigma_{+-} ,\\
G^\rmA = G_{++}+G_{+-} , \qquad \Sigma^\rmA = \Sigma_{++}+\Sigma_{+-} \ .
\ee
This results in
\be
G^\rmR(\omega)&=&\frac 1{\omega-\Sigma^\rmR(\omega)},\qquad
G^\rmA(\omega)=\frac 1{\omega-\Sigma^\rmA(\omega)},
\label{gra}
\ee
with self-energies given by
\be
\Sigma^\rmR(t)-\Sigma^\rmA(t) &=&-i^q 2^{2-q} {J^2}\left[G^\rmR(t) -G^\rmA(t)\right]^{q-1} -2i\mu\delta(t),\\
\label{eq:sigmaR_needs}
\Sigma^\rmR(t)+\Sigma^\rmA(t) &=&-i^q 2^{2-q} {J^2}\left[G^\rmR(t) +G^\rmA(t)\right]^{q-1}.
\ee

At the saddle point, we have that the time-ordered and lesser Green's functions are, respectively
\be
G^\rmT = \langle G_{++} \rangle
\qquad \text{\and} \qquad
G^< = \langle G_{+-} \rangle.
\ee
This results in
\begin{align}
\langle G^\rmR\rangle  &= G^\rmT - G^< =-\sign(t)\, G^< - G^< =
-2 \Theta(t) G^<,
\\
\langle G^\rmA \rangle &= G^\rmT + G^< =-\sign(t)\, G^< + G^< =
2 \Theta(-t) G^<,
\end{align}
and we recognize $\langle{G^{\rmR/\rmA}}\rangle$ as the retarded and advanced Green's functions. Moreover, at the saddle point, there is a single independent Green's function. From now on, it is understood that we are always at the saddle point and drop the brackets $\langle\cdots\rangle$ throughout.

Numerically, we solve the saddle-point equations in terms of the spectral function
\be
\rho^-(\omega)=-\frac{1}{\pi}\mathrm{Im}G^\mathrm{R}(\omega),
\ee
see Eqs.~(\ref{sdreal_rho}) and (\ref{sdreal_sigma}) in Appendix~\ref{app:real_time}. In general, we expect the system to relax exponentially to its steady state at a rate $\Gamma$ (also known as the gap). In that case, we have 
\begin{equation}
iG^\mathrm{R}(t)\propto e^{-\Gamma t}.
\end{equation}
In general, the rate $\Gamma$ can be obtained from the numerical solution of the saddle-point equations (see Sec.~\ref{sec:q4} for $q=4$), but also analytically in special cases (see Sec.~\ref{sec:q2} for $q=2$ and Sec.~\ref{sec:qinfty} for the large-$q$ limit).

\subsection{The Euclidean approach}
\label{sect:euclid}

The Liouvillian of Eq.~(\ref{doublelindbrad})
can also be interpreted as a two-site non-Hermitian but PT-symmetric SYK Hamiltonian with intersite coupling \cite{maldacena2018,garcia2021,garcia2022,garcia2021a}. Ignoring the irrelevant constant term, in this section we study
the Hamiltonian 
\be
\label{eq:2-Ham}
H = i H_L^{\rm SYK} - i(-1)^{q/2} H_R^{\rm SYK} + i \mu \sum_k \psi_L^k \psi_R^k.
\ee
Note that we have set $H=-\mathcal{L}$, because of the different conventions in the time evolution operator: $\exp(\mathcal{L}t)$ for real-time Liouvillian evolution and $\exp(-H \tau)$ for Euclidean evolution.
The ground state of this Hamiltonian is the thermofield-double (TFD) state at $\beta = 0$,
\be
\label{groundstate}
|0\rangle = \sum_k |k\rangle \otimes U K |k \rangle,
\ee
with the sum running over the complete basis of eigenstates of $H^\mathrm{SYK}_{L,R}$, $K$ the complex conjugation operator, and the unitary operator $U$ determined by the conditions (see Appendix~\ref{app:choi} for an explicit construction of $U$):
\be
i \mu  \sum_k\psi_L^k \psi_R^k|0\rangle = -\frac {N\mu}2 |0\rangle,
\qquad UK\,H_R^\mathrm{SYK}=(-1)^{q/2} H_R^\mathrm{SYK} UK.
\ee
Because
\be
H_L^{\rm SYK} |k\rangle = \lambda_k |k\rangle, \qquad  H_R^{\rm SYK} |k\rangle = \lambda_k |k\rangle,
\ee
we thus have that
\be
H|0\rangle = -\frac {N\mu}2 |0 \rangle,
\ee
and $|0\rangle$ is also the ground state of the coupling term.
Naturally, the $\beta=0$ TFD is the vectorization of the (unnormalized) steady-state density matrix~(\ref{steady}), which, as mentioned before, is the maximally entangled state.

The Euclidean action related to the Hamiltonian (\ref{eq:2-Ham})
is given by [with partition function $Z= \exp (-I)$]
\be
I =  \int_0^{\beta} d\tau \Big(\frac{1}{2} \sum_i \psi^i_L \partial_\tau \psi^i_L+\frac{1}{2} \sum_i \psi^i_R \partial_\tau \psi^i_R +H \Big).
\ee
The only difference with the real-time action is that the integral is over $[0,\beta]$ instead of the real axis.
The so-called $\Sigma G$ action is obtained in two steps~\cite{maldacena2016}. First, the integration over the Gaussian disorder results in a bilocal fermionic action. 
The original fermionic fields are then integrated by introducing two auxiliary fields: the Green's function,
\begin{equation}
G_{ab}(\tau_1,\tau_2)=\frac{1}{N}\sum_i \langle \psi^i_a (\tau_1)\psi^i_b(\tau_2)\rangle,
\end{equation} 
and the self-energy $\Sigma_{ab}(\tau_1,\tau_2)$, a Lagrange multiplier that fixes the above definition of $G_{ab}$ in the path integral, where $a,b=L,R$. The resulting Euclidean action is given by
\be
I/N&=& -\frac{1}{2}\log\det(\delta_{ab}\partial-\Sigma_{ab}) 
+\frac{1}{2} \sum_{ab}\int\!\!\!\int \left[\Sigma_{ab}(\tau_1,\tau_2)G_{ab}(\tau_1,\tau_2) +\frac{1}{q}t_{ab}J^2 s_{ab}G_{ab}(\tau_1,\tau_2)^q \right]d\tau_1 d\tau_2  \nn \\ && +\frac{i \mu}{2}\int \left[G_{LR}(\tau,\tau)-G_{RL}(\tau,\tau)\right]d\tau
\ee
with $a,b\in\{L,R\}$, $s_{LL}=s_{RR}=1$, $s_{LR}=s_{RL}=(-1)^{q/2}$, $t_{LL}=t_{RR}=1$, $t_{LR}=t_{RL}=-1$. In the large-$N$ limit, assuming time translational invariance, a saddle-point analysis results in the following SD equations:
\be
\label{eq:SDE_Eucl_1}
&&-(i\omega+\Sigma_{LL}(\omega))G_{LL}(\omega)+\Sigma_{LR}(\omega)G_{LR}(\omega)=1
,\nn\\
&&-(i\omega+\Sigma_{LL}(\omega))G_{LR}(\omega)-\Sigma_{LR}(\omega)G_{LL}(\omega)=0
\\
\label{eq:SDE_Eucl_2}
&&\Sigma_{LL}(\tau)=-J^2 G_{LL}^{q-1}(\tau)
,\qquad
\Sigma_{LR}(\tau)=i^q J^2 G_{LR}^{q-1}(\tau)-i\mu\delta(\tau),
\ee
where the first two equations are in Fourier space and
$\tau = \tau_1 -\tau_2$.
The details of the derivation of the Euclidean SD equations are given in Appendix~\ref{app:Euclidean_time}. If the system is gapped, naively, we expect that $G_{LR}(\tau)$ decays exponentially at a rate given by the spectral gap $\Delta$, namely, the gap between the ground state energy and the first excited state. However, we will see that, in the weak-coupling limit $\mu \ll 1$, the decay rate $E_g$, also referred to as the gap, that governs the exponential decay of $G_{LR}$ is much larger than the spectral gap $E_g \gg \Delta$.

In the time domain, at $\tau \ne 0$, the saddle-point equations
in the low-temperature limit admit solutions satisfying
$G_{LL} = \alpha G_{LR}$, which upon substitution gives
$\alpha^2 = -1$. 
In order to reproduce the delta function we thus find the solution
\be
 G_{LL}(\tau) = -i\,\sign(\tau)\,  G_{LR}(\tau),
\label{caus}
   \ee
  where we have used that, in the limit of zero temperature,
  \be
  \partial_\tau G_{LL}(\tau) = -2 i\,\delta(\tau)\, G_{LR}(0)=\delta(\tau).
 \ee
 
Using a suggestive notation, the saddle-point equations can be simplified by introducing the combinations \cite{plugge2020}
\be
 G^\rmR &\equiv& -i  G_{LL} - G_{LR},\qquad  \Sigma^\rmR \equiv i \Sigma_{LL}+ \Sigma_{LR},
 \label{greenE_1}\\
G^\rmA &\equiv& -i G_{LL} + G_{LR},\qquad  \Sigma^\rmA \equiv i\Sigma_{LL} -\Sigma_{LR}.
\label{greenE}
\ee
This results in the saddle-point equations \eref{gra} we found for the
real-time approach:
\be
G^\rmR(\omega)& =& \frac {1}{\omega - \Sigma^\rmR(\omega)},\qquad
  G^\rmA(\omega) = \frac {1}{\omega - \Sigma^\rmA(\omega)},
    \ee
and
    \be
    \Sigma^\rmR(\tau)+ \Sigma^\rmA(\tau) &=&
    - i^{q} 2^{2-q}{J^2} \left[G^\rmR(\tau) +G^\rmA(\tau)\right]^{q-1}, \\
    \Sigma^\rmR(\tau)- \Sigma^\rmA(\tau) &=&
    - i^q 2^{2-q}{J^2} \left[G^\rmR(\tau) -G^\rmA(\tau)\right]^{q-1}
    -2i \mu \delta(\tau).
    \ee
For $q=4$, these equations simplify to~\cite{plugge2020}
\be
\Sigma^\rmR(\tau) &=& \frac{J^2}{4}\left[ \left(G^\rmA(\tau)\right)^3 + 3 G^\rmA(\tau) \left(G^\rmR(\tau)\right)^2\right] +i\mu \delta(\tau),
\\
\Sigma^\rmA(\tau) &=& \frac{J^2}{4}\left[\left( G^\rmR(\tau)\right)^3 + 3 \left(G^\rmA(\tau)\right)^2 G^\rmR(\tau)\right] -i\mu \delta(\tau).
\ee

Using the relation \eref{caus} we find that
\be
G^\rmR(\tau) =  -2 \Theta(\tau)G_{LR}(\tau), \\ 
G^\rmA(\tau) =  2 \Theta(-\tau)G_{LR}(\tau), 
\ee
and we recognize $G^\rmR$ and $G^\rmA$ as the retarded and advanced Green's functions, introduced in the Keldysh approach of Sec.~\ref{sec:keldysh}.

Importantly, we note that in the limit of zero temperature, and assuming no memory of the initial state, both actions, the Euclidean one for the Hamiltonian and the Lorentzian one for the Liouvillian are identical, upon setting $\tau=t$.
We also have seen that the Green's functions satisfy the same Schwinger-Dyson equations, and at low temperature the solutions will be the same. More discussion of this equivalence
can be found in Appendix~\ref{app:relation_euclidean_real}.

We have thus established a complete equivalence between the two problems: the real-time path integral of an SYK coupled to a bath close to the steady state and the Euclidean path integral of a two-site non-Hermitian SYK related to wormhole configurations in the low-temperature limit.
For instance, the value of $E_g$ in the Euclidean problem will be the same as the decay rate $\Gamma$ that
describes the typical relaxation time to a steady state in the real-time problem. We have shown that the retarded Green's function in the real-time problem is equal to
$-i G_{LL}- G_{LR}$ in the Euclidean problem. We would like to remind the reader that, although we use $\psi_L$, $\psi_R$ in both Liouvillian and Euclidean discussions, they are not exactly the same: the former is for real time and $L$, $R$ represent different branches of the Keldysh contour, while the latter is for imaginary time and $L$, $R$ denotes two sites of the system. Rather than using a Keldysh contour, we could also have solved the real-time problem as a two-site system.

In the remainder of the paper, we discuss the solutions of the saddle-point equations which illustrate the equivalence of the Keldysh approach and the Euclidean approach at low temperature.
We will start with a study of the $q=4$ SYK model, whose dynamics is quantum chaotic, by solving numerically the saddle-point Schwinger-Dyson equations. We will then continue with analytical studies both for $q = 2$, whose dynamics is integrable, and in the large-$q$ limit, whose dynamics is quantum chaotic.
 
Finally, we will explore the possible existence of a gravity dual and its relevance in the late-time dynamics of our real-time SYK model. We are motivated by the recent result that, in the low-temperature limit, a similar non-Hermitian SYK model, with additional real couplings, is dual to a Euclidean wormhole \cite{garcia2021}. This possibility was originally noticed in Ref.~\cite{maldacena2018} and later formulated in detail in Ref.~\cite{garcia2021}, see also Refs.~\cite{garcia2022,garcia2021a,garcia2022b,garcia2022c}.
We shall show that qualitatively similar results are found in our model and we will dub the possible gravity dual a Keldysh wormhole. 
Interestingly, this identification between strongly correlated dissipative quantum matter and quantum gravity strongly suggests that the observation of a gap or decay rate, even if there is no coupling to the bath, is closely related to the existence of a wormhole gravity dual.
 
\section{Comparison between the Euclidean two-site SYK and the real-time open SYK for $q = 4$} 
\label{sec:q4}

We now proceed with the explicit calculation of $\Gamma$, $E_g$ that control the exponential decay of Green's functions in the real-time and Euclidean problem respectively, by solving the Schwinger-Dyson equations of both the Euclidean problem related to the non-Hermitian $q=4$ SYK model, or the Lorentzian problem related to Liouvillian dynamics, which, as we have seen in the previous section, give the same results.
For that purpose, we will compute the relevant Green's functions by solving the SD equations~(\ref{eq:SDE_Eucl_1}) and (\ref{eq:SDE_Eucl_2}) at $\beta=500$ and
the SD equations (\ref{sdreal_rho}) and (\ref{sdreal_sigma}), respectively.

Similar calculations have been carried out recently. In Ref.~\cite{kulkarni2022}, the real-time calculation was carried out but no results were presented in the small-$\mu$ region of interest. In this limit, we shall see that the large-$q$ calculation shows some deviations from the result obtained from the explicit solution of the SD equations for $q =4$. The Euclidean calculation for a similar model, where the SYK couplings are not purely imaginary but also have a real part, was carried out in Ref.~\cite{garcia2022c}. However, we shall see that $E_g$ and other observables are qualitatively different. 

\begin{figure}[t!]
	\centering
	\includegraphics[scale=.53]{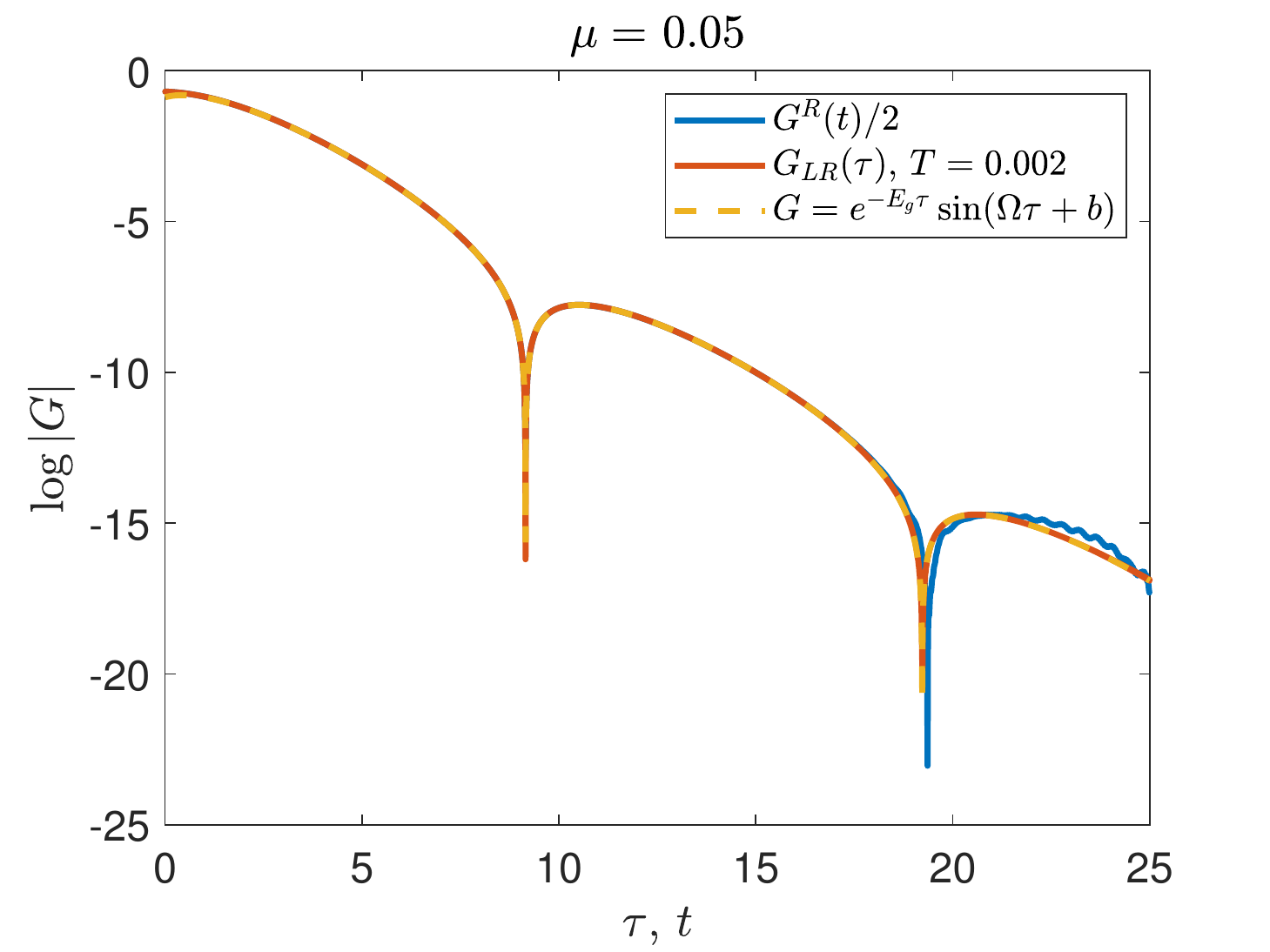}
	\includegraphics[scale=.53]{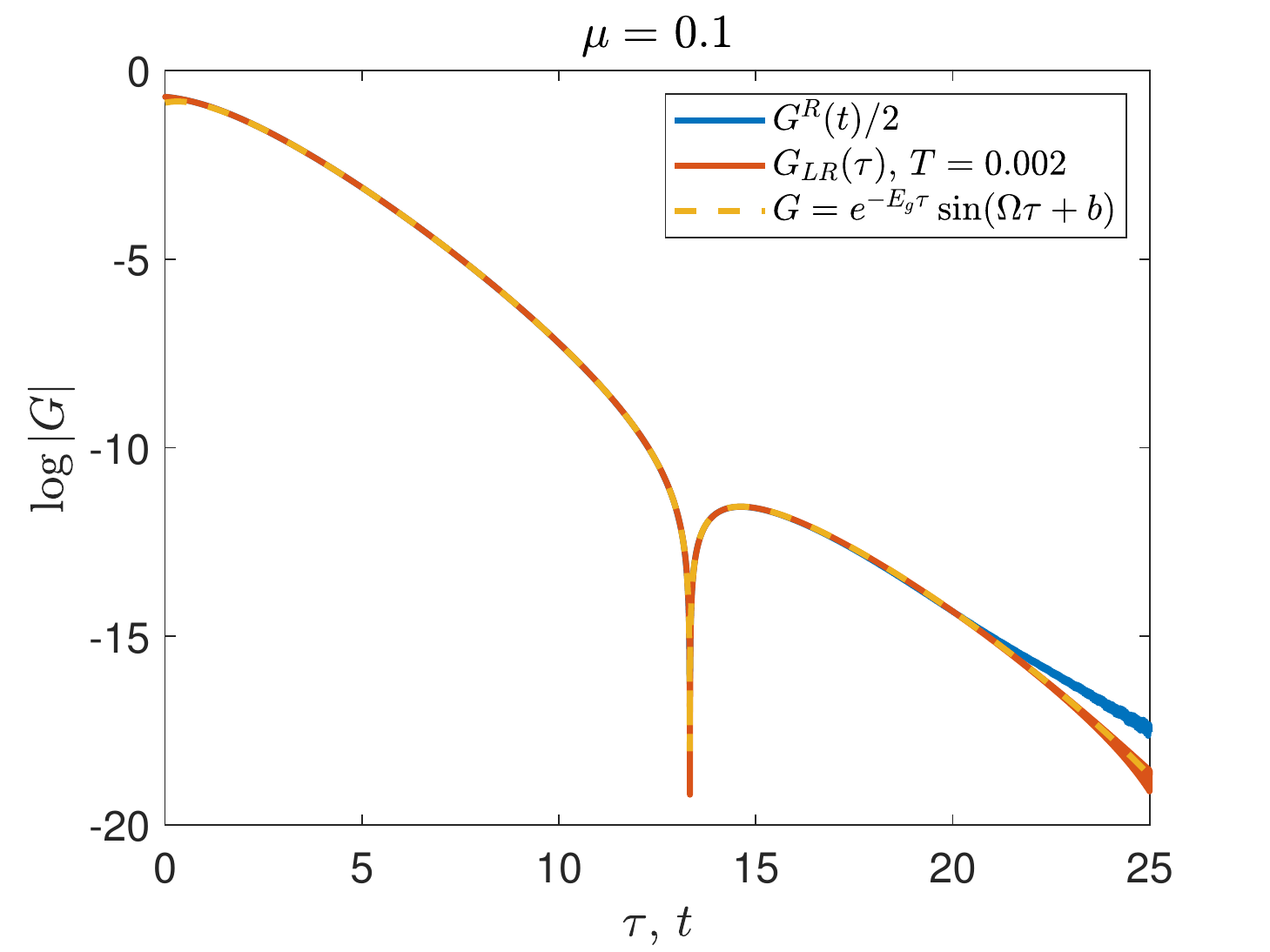}
	\includegraphics[scale=.53]{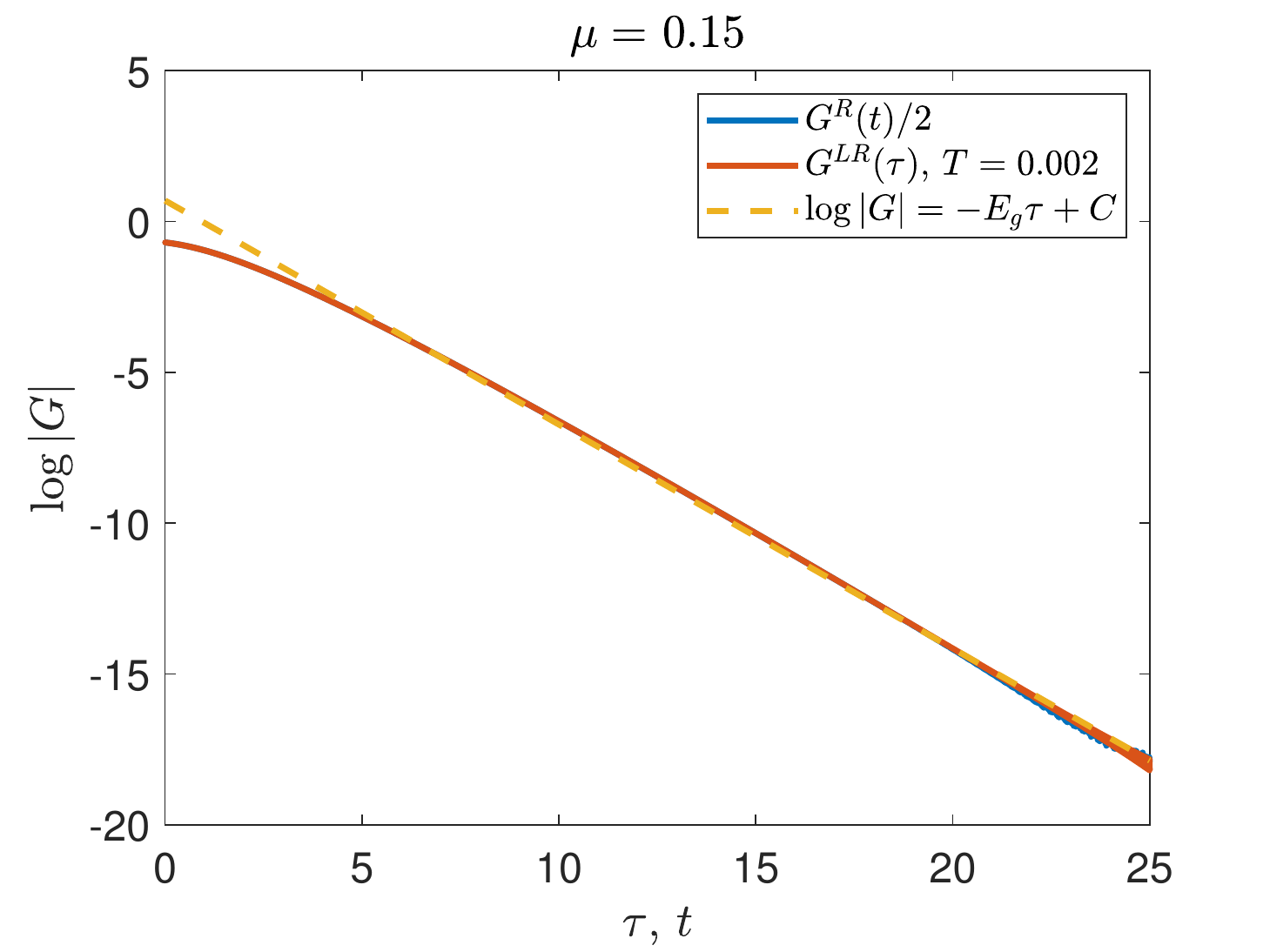}
	\includegraphics[scale=.53]{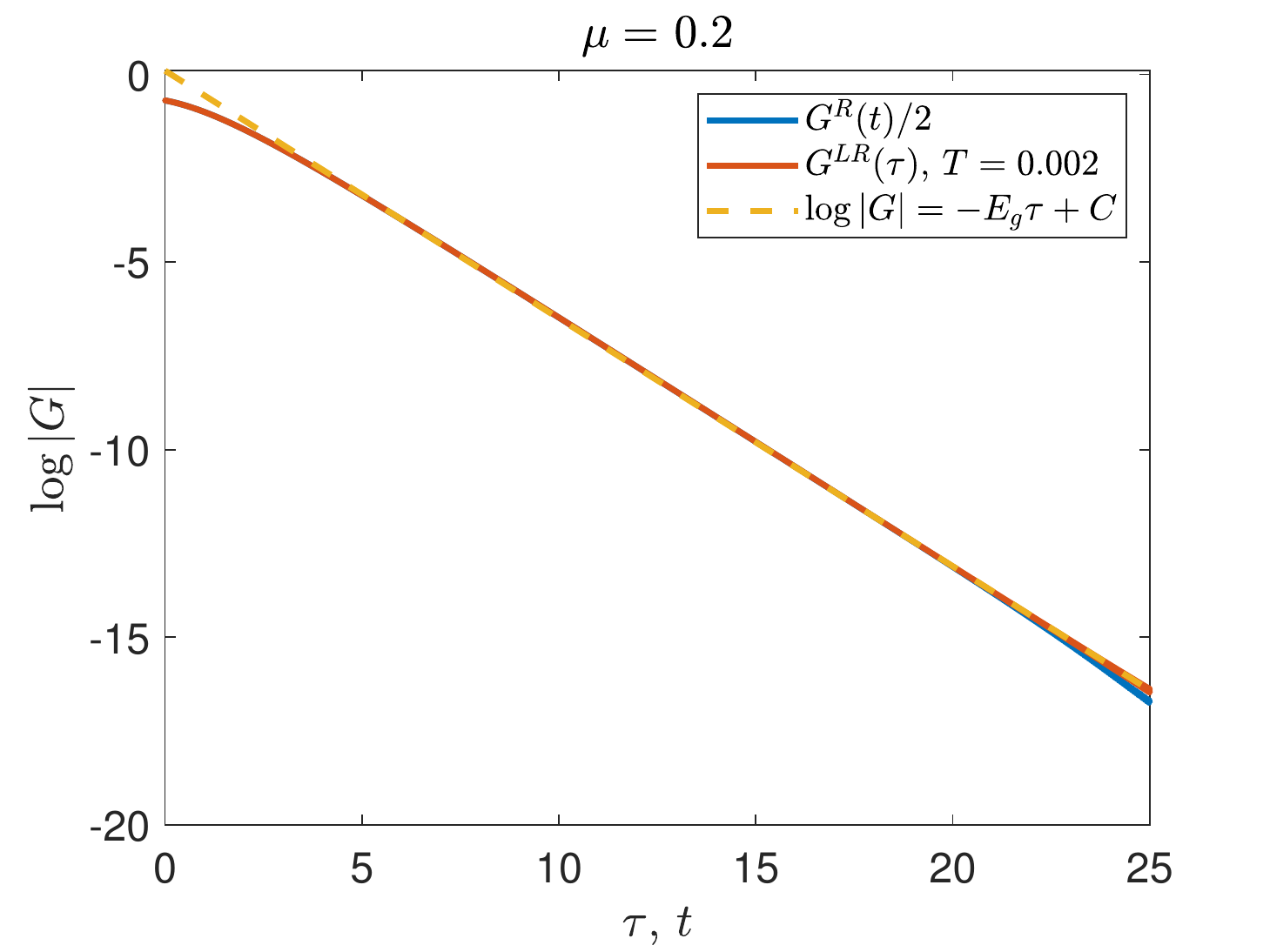}
	\caption{Green's function $G^\mathrm{R}(t)/2$ for the real-time problem and $G_{LR}(\tau)$ for the Euclidean problem, for several values of $\mu$. The two calculations coincide. We observe oscillations in both cases for small $\mu$ that stop for $\mu \sim 0.15$.
	\label{fig:Greens}
	}
\end{figure}
  
In Fig.~\ref{fig:Greens}, we depict the Euclidean, $G_{LR}(\tau)$, and retarded Lorentzian, $G^\mathrm{R}(t)/2$, Green's functions for different values of the coupling to the environment $\mu$. As argued above, the two computations coincide.
We stress that we are not making any analytical continuation
but just compare Euclidean and Lorentzian times. For small $\mu$, we observe oscillations in the Green's functions and a fast decay for long times. In both cases, as shown in the figure, we find an excellent agreement with a fit to $A e^{-E_g t}\sin{(\Omega t+\phi)}$ with $A$, $E_g$, $\Omega$, and $\phi$ fitting parameters. The frequency of the oscillations, $\Omega$, vanishes for sufficiently large $\mu > 0.15$.
In the Lorentzian case, similar oscillations \cite{sa2022,kulkarni2022} have been found in a model with a random jump operator involving two Majorana fermions.

\begin{figure}[t]
	\centering
	\includegraphics[scale=.54]{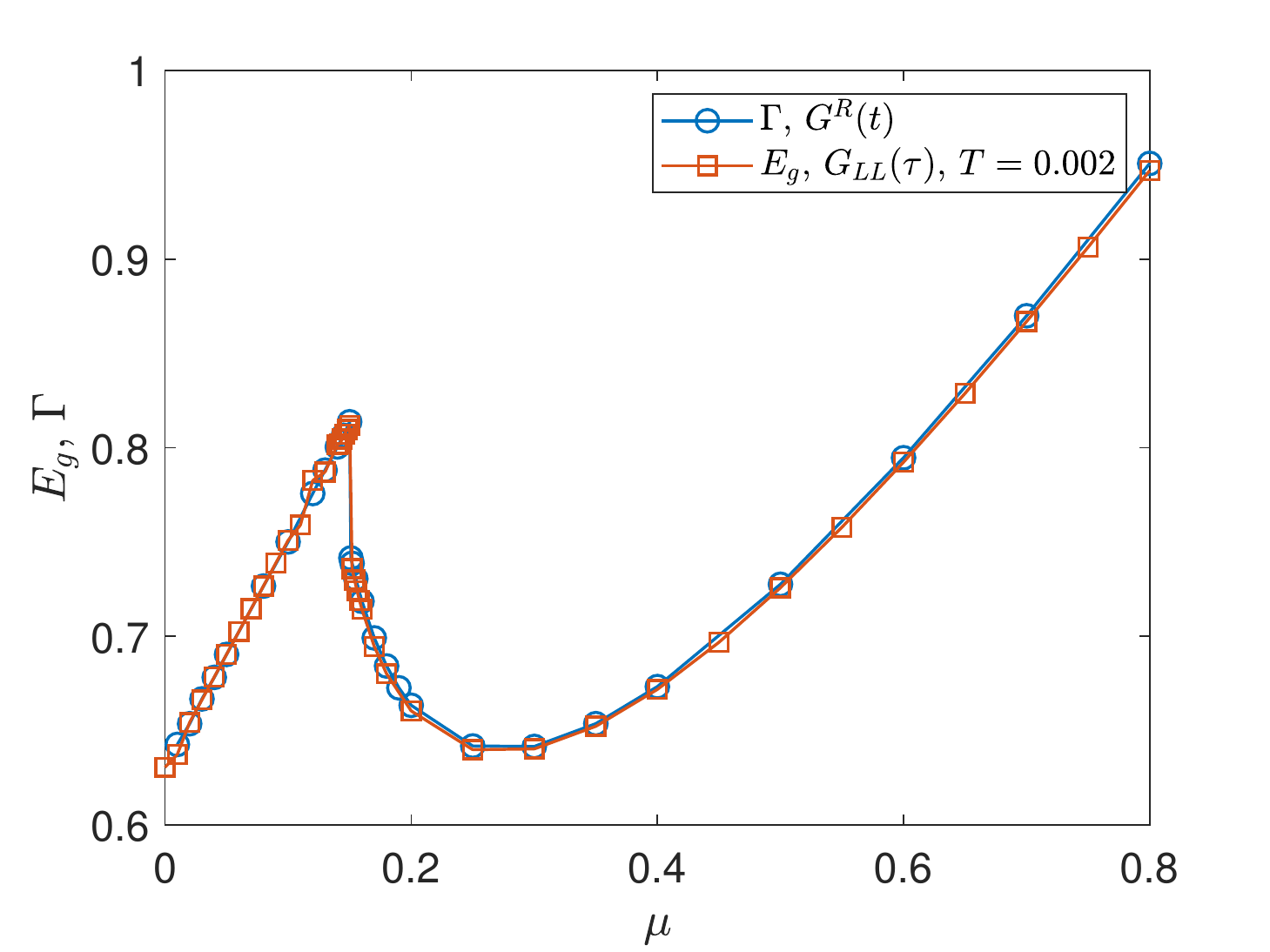}
	\includegraphics[scale=.54]{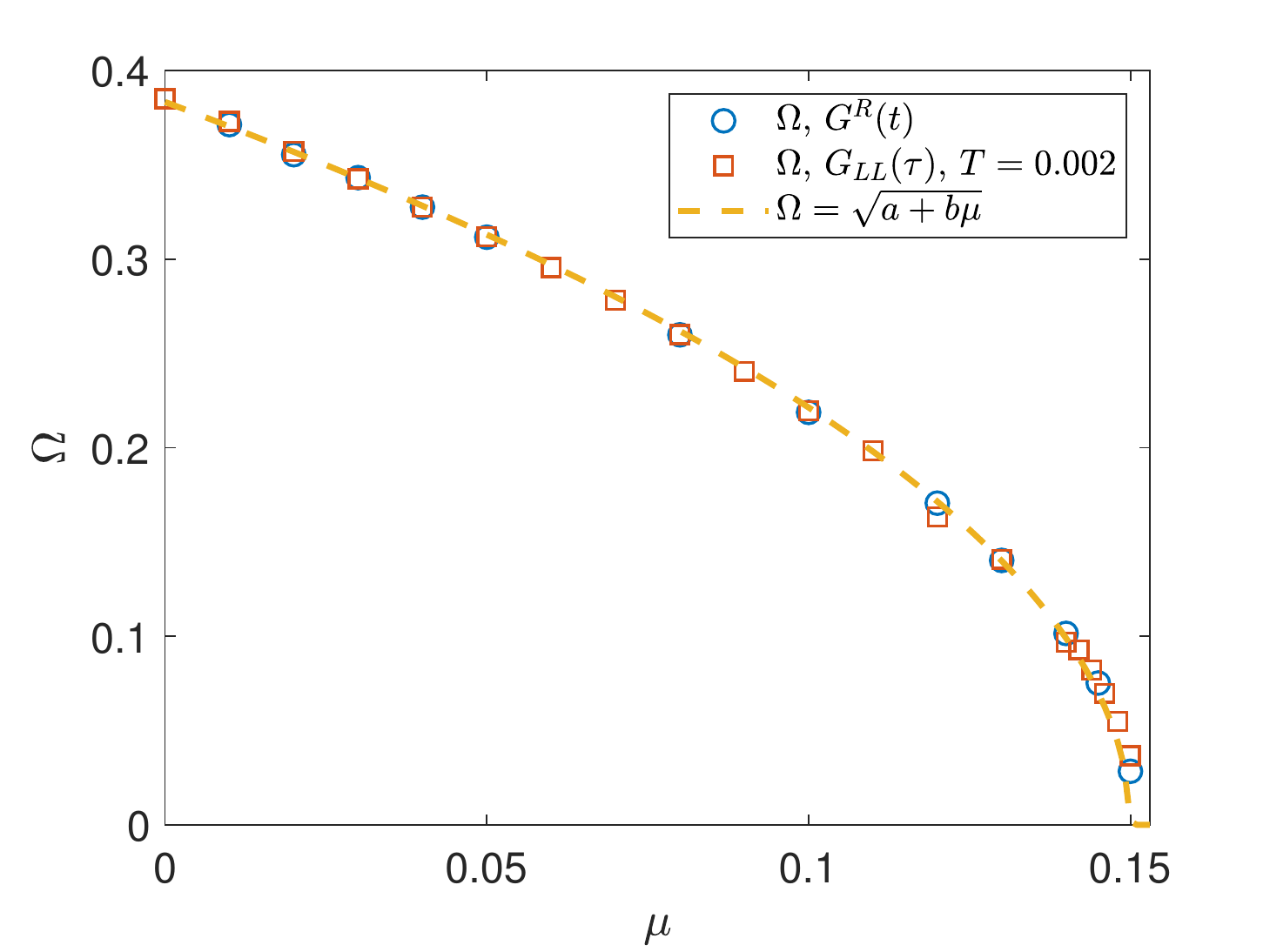}
	\caption{Left: gap $E_g$ (Euclidean time) or decay rate $\Gamma$ (real time) as a function of $\mu$. Right: frequency of the oscillations $\Omega$ of the Green's function, $G^\mathrm{R}$ (real) and $G_{LL}$ (Euclidean), as a function of $\mu$. The $\mu$ dependence of the frequency, right plot, is very well fitted (dashed curve) by $\Omega(\mu)\approx 0.99\sqrt{\mu_c-\mu}$, with $\mu_c \approx 0.15$. The oscillations in the Green's function terminate with a second-order phase transition at $\mu = \mu_c$.
	}
	\label{fig:comEuLor}
\end{figure}

The exponential decay, controlled by $E_g$, persists for any coupling strength $\mu$ provided that temperature is sufficiently low. Following previous literature \cite{maldacena2018,garcia2019,garcia2021,garcia2022c}, we have denoted the decay rate as $E_g$ in reference to a gap. However, we stress that, especially for zero or small $\mu$, this is not related to a gap in the spectrum between the ground state and the first excited state. It is rather the order parameter that, in the gravity dual, characterizes the wormhole phase. Therefore, it is a feature of the model resulting after ensemble averaging. A heuristic explanation of its origin is that the low-energy excitations in the partition function are nullified because the spectrum is complex and PT-symmetric. The ground state is not affected because it is real. As $\mu$ increases, $E_g$ approaches the spectral gap $\Delta$. 
In Fig.~\ref{fig:comEuLor}, we show the gap $E_g$ (left), or decay rate $\Gamma$ for the real-time calculation, and the frequency of the oscillations (right) of the Green's function as a function of the coupling $\mu$. The curves for $E_g(\mu), \Gamma(\mu)$ and $\Omega(\mu)$ are barely distinguishable which further supports that, in the $T \to 0$ limit, the Hamiltonian~(\ref{eq:2-Ham}) is identical to the Liouvillian~(\ref{doublelindbrad}) that represents an SYK coupled to a Markovian bath. Both $E_g$ and $\Omega$ have an intriguing $\mu$ dependence that we now analyze in detail.

For $\mu = 0$, the gap is finite and the frequency of the oscillations in the Green's function is largest. Physically, this corresponds to a complex order parameter whose real part controls the exponential decay and the imaginary part the frequency of oscillation.
As $\mu$ increases, the imaginary part of the excited eigenvalues becomes smaller. Therefore, a larger $\mu$ is expected to reduce the value of $\Omega(\mu)$. Indeed, we find that the frequency decreases monotonically with $\mu$. 
Interestingly, $\Omega(\mu)$ shows excellent agreement
with a simple ansatz $\Omega(\mu) \approx \sqrt{\mu_c- \mu}$ with $\mu_c \approx 0.15$. For $\mu >\mu_c$, the decay of Green's function is purely exponential, which implies the vanishing of the frequency of oscillations. This is consistent with the existence of a second-order phase transition at $\mu = \mu_c$.
 
The gap or decay rate in the small-$\mu$ limit, 
which has not been investigated before, has also a rather unexpected behavior. The decay rate (or gap) is not zero even in the limit $\mu \to 0$ of the coupling to the bath. We stress that, as mentioned earlier, $E_g$ at $\mu=0$ is not related to the spectral gap. On the Euclidean side, an analogous feature has already been predicted \cite{maldacena2018} by noticing that the existence of a finite $E_g$, which is a signature of the dual wormhole configurations, only requires that the variance of the left-right coupling is stronger than the left-left or right-right ones. This can be easily achieved with complex couplings in the SYK model. Indeed, the purely imaginary couplings of our model are sufficient since the variance of the left-left, or right-right, couplings are negative while the left-right ones are positive \cite{maldacena2018}. Strictly speaking, the gravity dual cannot be a Euclidean wormhole because the variance of real couplings (zero in our case) is not larger than the imaginary ones \cite{maldacena2018}. However, we shall see that our model still has wormhole-like features similar to those of a Euclidean wormhole.
Its counterpart in the real-time calculation is the existence of a finite decay rate even if there is no coupling with the environment, which is expected provided that the SYK model is quantum chaotic ($q > 2$)~\cite{kitaev2015,maldacena2016,garcia2016,cotler2016}. 
We term the wormhole configuration in real time and $\mu \ll 1$, responsible for the finite decay rate, a {\it Keldysh wormhole}. We shall give a more precise characterization of this possible gravity dual in Sec.~\ref{sec:dual}.

For a small but finite $\mu/\mu_c < 1$, $E_g(\mu) - E_g(0) \sim \mu$ (or $\Gamma(\mu) - \Gamma(0) \sim \mu$) increases linearly. This makes sense physically since a stronger coupling to the bath results in a larger decay rate. 
We think this linear behavior is related to the fact that the energy of the first excited state is always real and equal to $4\mu$. However, the overall effect on $E_g$ is relatively small compared with the $\mu = 0$ contribution, namely, the spectral gap $\Delta \equiv E_1-E_0$, where $E_0$ and $E_1$ are the energies of the ground and first excited state, is only a small contribution to $E_g$. 
The linearity suggests that the slope can be obtained by a perturbative treatment in $\mu$. This is in contrast with the $\mu^{2/3}$ behavior observed for a two-site SYK with $\mu \ll 1$ whose gravity dual is a traversable wormhole \cite{maldacena2018}. Therefore, we can rule out any quantitative relation between our setting and this gravity dual. 

At a critical value $\mu_c\approx 0.15$, corresponding to the vanishing of the oscillation frequency, the linear increase of $E_g$ stops abruptly.
A further increase in the coupling $\mu$ leads to a sharp, likely continuous, drop in $E_g$ or $\Gamma$. This is rather unexpected as an increase in the coupling to the bath should always lead to an increase in the decay rate as the approach to equilibrium should occur faster. Similarly, a stronger explicit left-right coupling in the Euclidean problem is expected to give a larger gap because the hopping probability increases. However, we clearly observe a drop as $\mu \gtrsim 0.15$ increases. A possible explanation of this nonmonotonic behavior is that the oscillations become imaginary, $\Omega \to i \Omega$, for $\mu > \mu_c$, so that they effectively become a negative contribution to the gap.
More generally, we believe that this sharp decrease is directly related to the effective weakening of the strength of the imaginary couplings in each SYK due to the left-right term which is Hermitian in our model. 
For a sufficiently large $\mu$, in agreement with Ref.~\cite{kulkarni2022}, the decay rate $\Gamma$, or $E_g$, approaches a linear dependence $E_g\sim \mu$, which is the expected result \cite{sieberer2016} for a strong coupling to the environment. In this region, the spectrum is mostly real and $E_g$ is close to the spectral gap $\Delta$.

In order to gain a more complete understanding of the physical meaning of $E_g$ and $\Omega$, we computed the finite-$N$ Green's functions, see Appendix~\ref{app:ED}, by exact diagonalization (ED) techniques. Most of our finite-$N$
results have been obtained for $N=9$, and at this point we are satisfied
with a qualitative agreement between the finite-$N$ results and the SD results.
For $\mu <0.15$, we find a large difference between the average of the
Green's function and the average of the absolute value of the Green's function.
In both cases, we find an initial fast exponential decrease followed by a slower
exponential decrease at longer times. However, the range of the initial exponential decay increases substantially upon ensemble averaging because of cancellations in the average of the Green's function. We expect that in the large-$N$ limit, and in the
limit of a very large number of realizations, where these cancellations become
manifest, the initial exponential decrease will become dominant and will converge to
the result obtained from the solution of the Schwinger-Dyson equations.
Indeed, 
we find qualitative agreement with the results of this section, see Fig. \ref{fig:gap}
of Appendix \ref{app:ED}. In particular,
we find a finite gap at $\mu =0$ which increases linearly for small $\mu$.
The long-time tail remains dominant for the average of the absolute value
of the Green's function. For small $\mu$ its decay rate converges to the result
given by the spectral gap, which is equal to $\Delta=E_1- E_0= 4\mu$, while for large
$\mu$ the decay rates of the initial exponential decay and the long-time
exponential decay become equal.
At finite $N$, we observe oscillations 
realization by realization but, because their period fluctuates, they are
mostly averaged out after ensemble averaging.

The detailed nature of the cancellation mechanism of the long-time tail
is not clear at this point. For $\mu>0.175$ no cancellations happen, but the
initial decay rate is still different from the long-time decay rate. A better
understanding of this difference requires a detailed
finite-size scaling analysis.
Results on the finite-$N$ scaling of the Green’s function are relatively scarce in the SYK literature. They have been studied to some extent for out-of-time-order correlators (OTOC)~\cite{kobrin2020} but in that case the problem is that the exponential behavior is barely visible at finite $N$, while in the present problem the exponential decay is very clear for the long-time tail. Also the free energy, which is studied in Sec.~\ref{sec:dual}, only shows a weak $N$ dependence at low temperatures, much weaker than the one observed in a different non-Hermitian SYK model \cite{garcia2021a}. One of the reasons is that, in the present case, the finite-$N$ ground state is reproduced identically by ED techniques. The free energy also has finite-$N$ issues that are not well understood: for small $\mu$, it has oscillations that are not observed in the solutions of the SD equations. Another potential source for the absence of the long-time tail in the SD equations is the averaging procedure. In the exact diagonalization, the average is done directly on the individual Green’s function while in the SD analysis the average over disorder is taken at earlier stages of the calculation. At present, we cannot rule out completely that this is in part related to other solutions of the saddle-point equations. In any case, progress in this problem requires substantially larger values of $N$ than those discussed in Appendix~\ref{app:ED} and a more precise understanding of the relation between the spectral gap $\Delta$ and the decay rate $E_g$.

Finally, a natural question to ask is to what extent the anomalous behavior we have found in the weak-coupling limit is universal or whether it is a particularity of the SYK model. More specifically, is it related to the fact that the system is strongly interacting and quantum chaotic? In order to answer this question we turn to the study of a nonchaotic $q=2$ SYK coupled to a Markovian bath described by the same Lindblad operator as for $q=4$.

\section{Analytical comparison between the Euclidean two-site SYK and the real-time open SYK for $q = 2$}
\label{sec:q2}

The $q=2$ SYK model is not quantum chaotic and it is not dual to any wormhole background. Therefore, it is an ideal candidate to investigate whether the anomalous behavior observed for weak coupling to the environment is related to quantum chaos and field theories with gravity duals.

\begin{figure}[t]
	\centering
	\subfigure[]{\includegraphics[scale=.5]{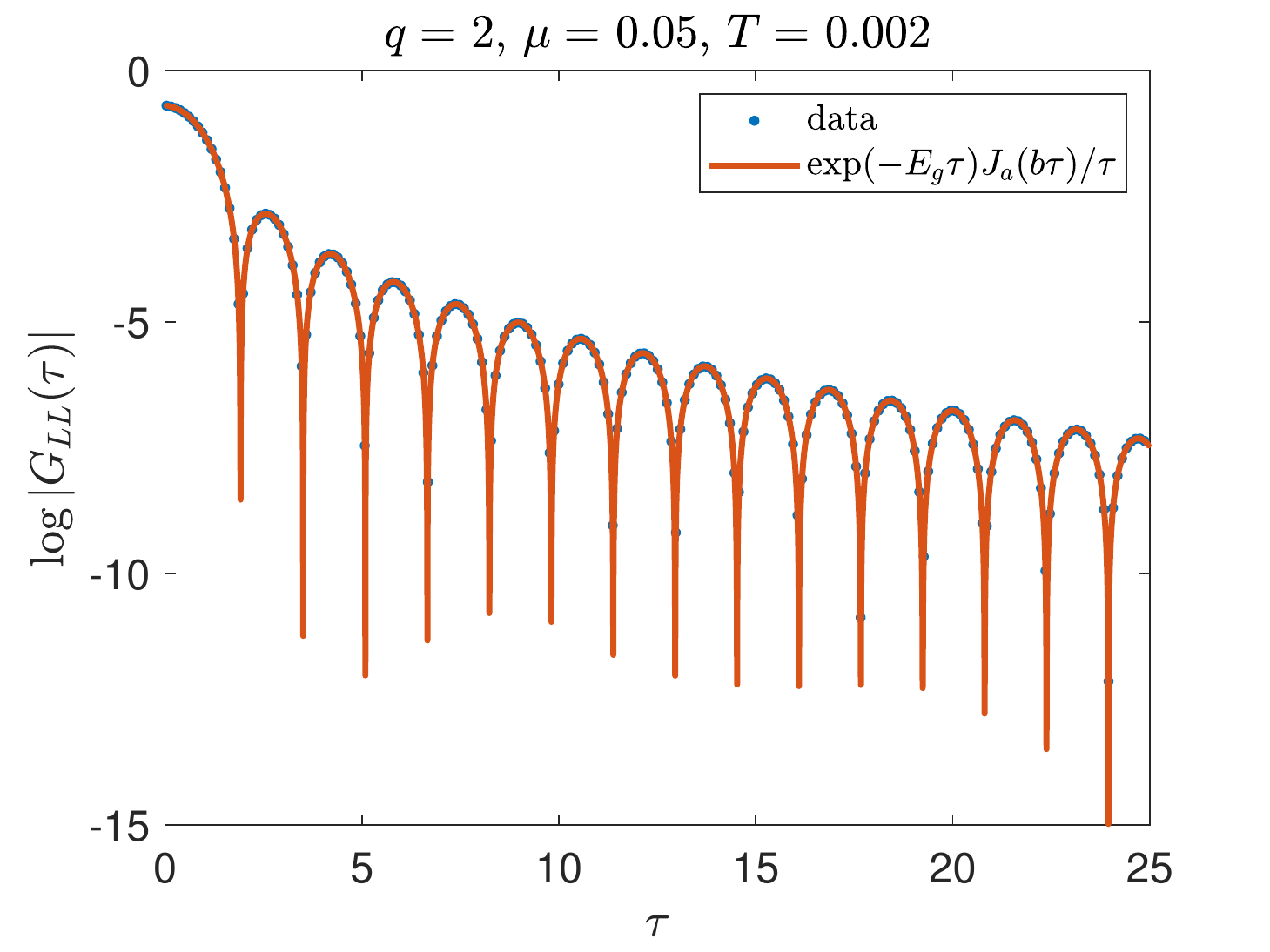}} 
	\subfigure[]{\includegraphics[scale=.5]{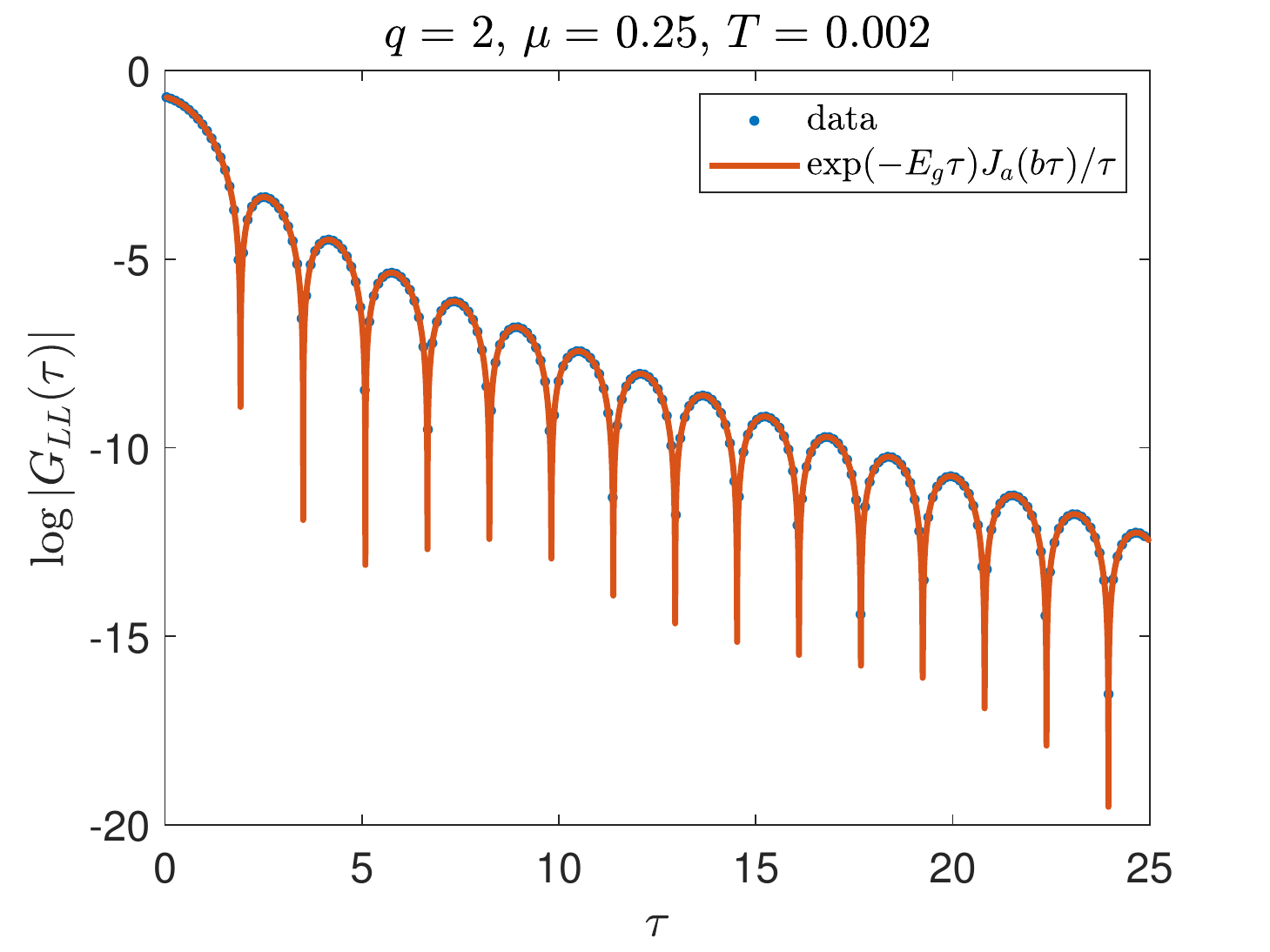}} 
	\subfigure[]{\includegraphics[scale=.5]{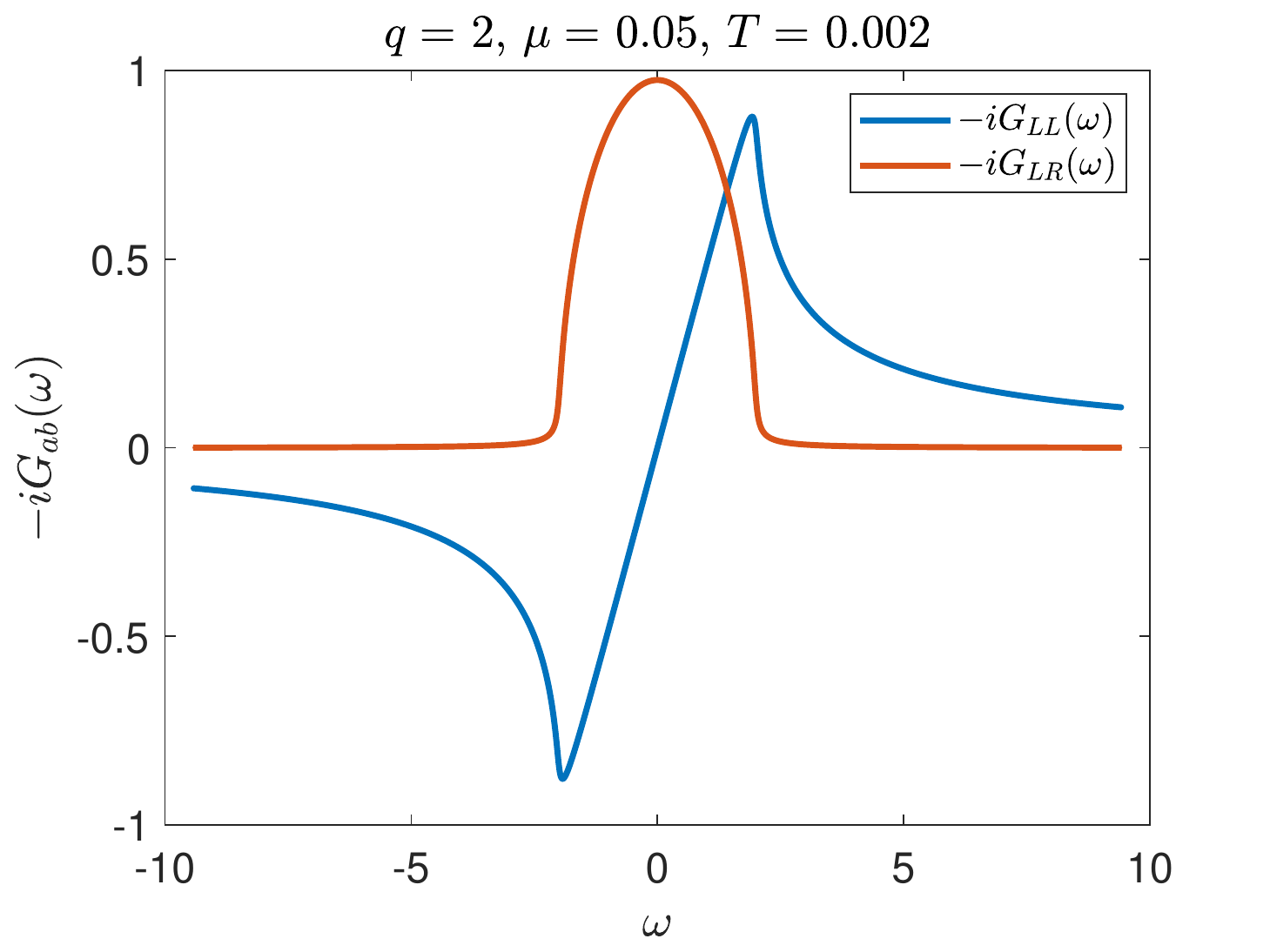}} 
	\subfigure[]{\includegraphics[scale=.5]{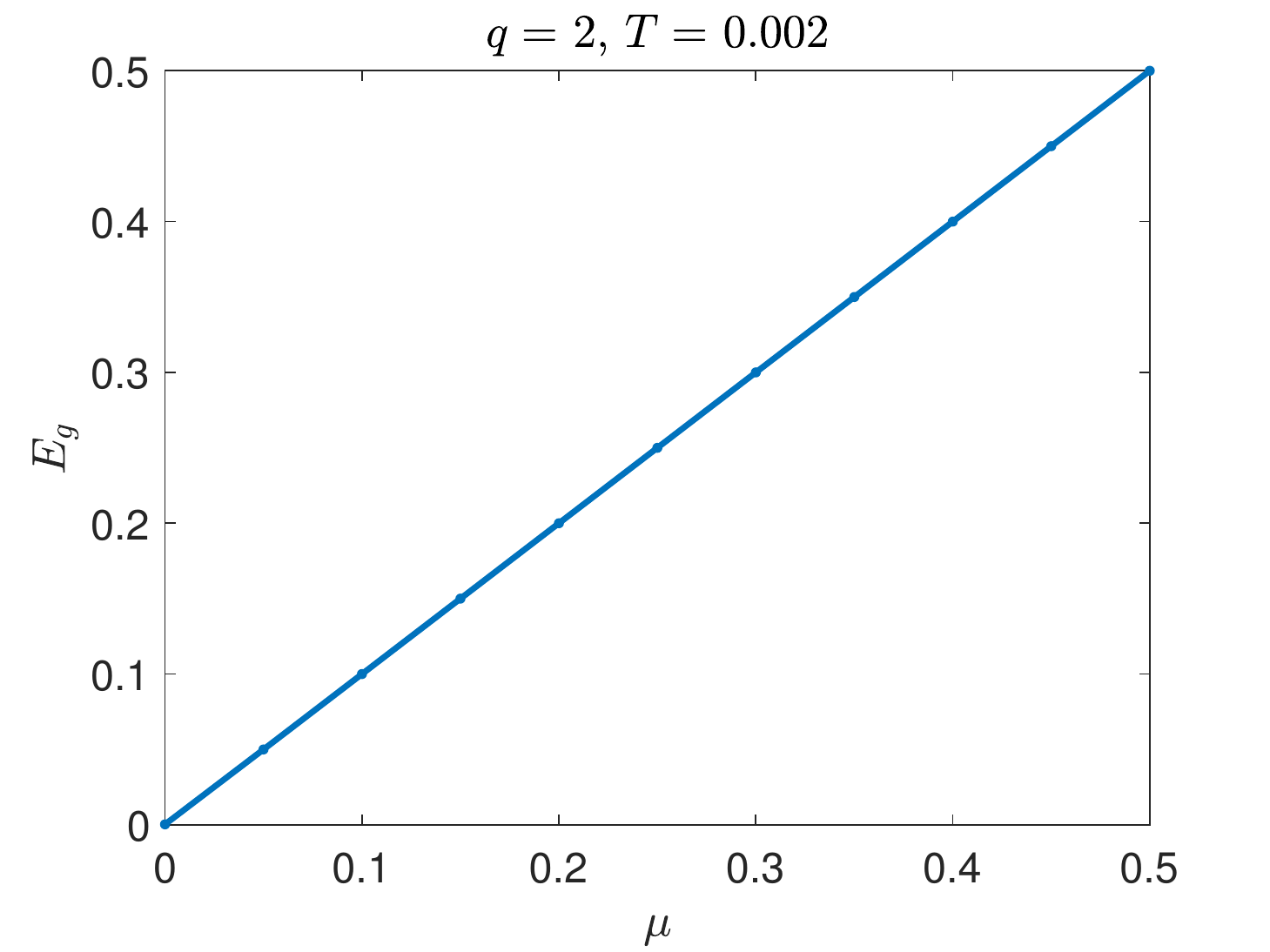}}
	\caption{Top row: $G_{LL}(\tau)$ for $q=2$ and (a) $\mu=0.05$, (b) $ \mu = 0.25$. We compare the numerical solution of the	 Euclidean SD equations with the analytic prediction $\frac{1}{2}e^{-E_g\tau}J_1(2\tau)/\tau$ with $E_g = \mu$. The analytical result for the real-time problem is identical with $\tau = t$. Bottom left: $-iG_{LL} (\omega)$ and $-iG_{LR} (\omega)$ in Fourier space. The Green’s function $i G_{LR} (\omega) /\pi$ coincides with the spectral function of the real-time problem, given by Eq.~(\ref{eq:rho-2}). Bottom right: the gap $E_g$ as a function of $\mu$. We find $E_g = \mu$ for all values of $\mu$ which, in the real-time problem, is an indication that the process of relaxation is always dissipation-driven.
	}\label{fig:gllq2com}
\end{figure}

We start with the Euclidean problem at $\mu = 0$, for which it is possible to find analytical solutions of the SD equations in Fourier space given by
\begin{equation}\begin{aligned}
		& 1 = (-i\omega +J^2 G_{LL})G_{LL} -J^2 G_{LR}G_{LR},\\ 
		& 0 = (-i\omega +J^2 G_{LL})G_{LR} +J^2 G_{LR}G_{LL}. 
\end{aligned}\end{equation}
In terms of the $G^{\rmR/\rmA}$ variables introduced in Eqs.~(\ref{greenE_1}) and (\ref{greenE}), the equations simplify to
\be
{J^2} \left[G^\rmR(\omega)\right]^2 - \omega G^\rmR(\omega) + 1 =0,
\ee
and the same equation for $G^\rmA$. In the real-time Liouvillian formulation, the same equation for the retarded Green's function can be obtained combining Eqs.~(\ref{gra})--(\ref{eq:sigmaR_needs}) and (\ref{sdreal_sigma}) with $q=2$.
The solution that vanishes at $\omega \to \pm \infty$ is given by
\be
G^\rmR(\omega) = -\frac \omega {2 J^2} +\frac{\sign(\omega)}{2J^2} \sqrt{(\omega+i\epsilon)^2-4J^2}
\ee
with $\epsilon\to0^+$, and the advanced Green's function is equal to
\be
G^\rmA(\omega) = -\frac \omega {2J^2}
+\frac{\sign(\omega)}{2J^2} \sqrt{(\omega-i\epsilon)^2-4J^2}.
\ee
Taking linear combinations of
$G^\rmR$ and $ G^\rmA$, we have
\begin{alignat}{99}
&G_{LR}(\omega) = i \frac{1}{J}\sqrt{1-\frac{\omega^2}{4J^2}},
&&\qquad
G_{LL}(\omega)=\frac{i\omega}{2J^2}, 
&\qquad 
\text{if }|\omega|<2J,
\\
&G_{LR}(\omega) = 0, 
&&\qquad 
G_{LL}(\omega) = i \frac{1}{J}\left(\frac{\omega}{2J} - {\rm sign}(\omega) \sqrt{\frac{\omega^2}{4J^2}-1} \right),
&\qquad
\text{if }| \omega| >2J.
\end{alignat}
In the time domain, the retarded and advanced Green's functions are given by
\be
G^\rmR(\tau) = -i \Theta(\tau) e^{-\epsilon \tau}\frac{J_1(2J\tau)}{J\tau}, \qquad  G^\rmA(\tau) =
i \Theta(-\tau)e^{\epsilon \tau} \frac{J_1(2J\tau)}{J\tau},
\ee
where $J_1$ is a Bessel function of the first kind.
Note that these solutions are also well defined for finite $\epsilon>0$.
Then, letting $\epsilon\to0$, we find
\be
G_{LR}(\tau)&=&- \frac 12 \(G^\rmR(\tau) -G^\rmA(\tau)\) =
i \frac{J_1(2J\tau) }{2J\tau},
\nn\\
G_{LL}(\tau)&=&\frac i2\(G^\rmR(\tau) + G^\rmA(\tau)\) = \sign(\tau)
\frac{J_1(2J\tau) }{2J\tau},
\ee
in agreement with the fundamental identity \eref{caus}.
The same result is obtained for real time.
We observe, see Fig.~\ref{fig:gllq2com}, that oscillations persist as in the $q = 4$ case, but, in stark contrast to it, the decay of Green's function is power law, not exponential. Therefore, the gap vanishes for $\mu = 0$.

We now turn to the region of nonzero $\mu$. In fact, we already solved this problem because $\mu$ plays the role of $\epsilon$ at $\mu > 0$. In this case, the saddle-point equations for $G^{\rm R/\rm A}$ are given by
\be
{J^2} \left[G^\rmR(\omega)\right]^2 - (\omega+i\mu) G^\rmR(\omega) + 1 &=&0, \nn \\
{J^2} \left[G^\rmA(\omega)\right]^2 - (\omega-i\mu) G^\rmA(\omega) + 1 &=&0.
\ee
This results in the Green's functions
\be
G^\rmR(\omega) &=& \frac{ (\omega+i\mu) }{2 J^2} -\frac{\sign(\omega)}{2J^2} \sqrt{(\omega+i\mu)^2-4J^2},\nn\\
G^\rmA(\omega) &=& \frac{( \omega-i\mu)} {2J^2}
-\frac{\sign(\omega)}{2J^2} \sqrt{(\omega-i\mu)^2-4J^2}.
\ee
Fourier-transforming to the time domain, we find
\be
  G^\mathrm{R}(\tau)&=&-i\Theta(\tau)\, e^{-\mu \tau}\,\frac{J_1(2J\tau)}{J\tau},
  \nn\\
  G^\mathrm{A}(\tau)&=&i\Theta(-\tau)\, e^{\mu \tau}\,\frac{J_1(2J\tau)}{J\tau}.
\ee
For completeness, we also write down the single-particle spectral function of the model:
\be
\label{eq:rho-2}
\rho^-(\omega)=-\frac{1}{\pi}\mathrm{Im}G^\mathrm{R}(\omega)
=\frac{-\mu+\frac{1}{\sqrt{2}}\sqrt{
		4J^2+\mu^2-\omega^2+\sqrt{(4J^2+\mu^2-\omega^2)^2 +4\omega^2\mu^2
}}}{2\pi J^2}.
\ee
As before, the saddle-point equations for the real-time calculation are the same, resulting in the same solutions provided that boundary/initial conditions are not important (i.e., at low temperature in the Euclidean calculation and for times such that the system is close to the steady state in the real-time calculation). 

In Fig.~\ref{fig:gllq2com}, we compare the analytical solution discussed above with the exact numerical solution of the SD equations for both the real-time and Euclidean-time problems. We find that both are indistinguishable from the analytical result without using any fitting parameters which confirms that also for $q=2$ the real-time calculation agrees with the Euclidean calculation in the low-temperature limit.
However, unlike the $q = 4$ case, we do not observe an anomalously large value for the gap or the decay rate for small values of $\mu$. For $q=2$, $E_g = \Gamma = \mu$ for all values of
$\mu$, so the approach to equilibrium is dissipation-driven.
The reason is the integrable nature of the $q = 2$ SYK, which requires a coupling to the environment to reach a steady state. Therefore, in the absence of any coupling to the environment, 
(many-body) quantum chaos and strong interactions seem necessary conditions to reach a steady state.

Finally, we note that, for $q = 2$, the analytic prediction for $E_g$ is in agreement with the spectral gap. In this case, the finite-$N$ 
spectrum is given by (for $k =0, 1, \dots, N$)
\be
\lambda_{k,n} = -\mu \left (\frac N2 - k \right ) + i e_{k,n}, {\qquad}
n= 1, \dots, {N\choose k}
\ee
with $e_{k,n}$ real, and it is realization dependent except for
the single real eigenvalue for each value of $k$. The gap between the first excited state energy with zero
imaginary part and the ground state energy is thus equal to $\mu$.
So, unlike $q = 4$, the resulting spectral gap
is equal to the decay rate of the Green's functions, $E_g =\Delta=\mu$. The observed oscillations in the Green's functions are determined by the imaginary parts of $\lambda_{k,n}$.

\section{Analytical comparison between the Euclidean two-site SYK and the real-time open SYK in the large-$q$ limit}
\label{sec:qinfty}

We now show that the equivalence between Euclidean and Lorentzian problems also holds in the large-$q$ limit \cite{maldacena2016,maldacena2018}. In order to carry out the large-$q$ analysis for the two-site non-Hermitian SYK model, we follow closely the method introduced in Ref.~\cite{maldacena2016} for the one-site SYK model and applied in Ref.~\cite{maldacena2018} to a two-site SYK model dual to a traversable wormhole.

The Green's function in the large-$q$ limit can be written as
\be\label{eq:largeqG1}
&& G_{LL}=\frac{1}{2}{\rm sign}(\tau)e^{\frac{1}{q}g_{LL}}=\frac{1}{2}{\rm sign}(\tau)(1+\frac{1}{q}g_{LL}+\dots), \\
&& G_{LR}=\frac{i}{2}e^{\frac{1}{q}g_{LR}}=\frac{i}{2}(1+\frac{1}{q}g_{LR}+\dots).
\ee
To leading order in $1/q$, the SD equations simplify to
\be
&& \sign(\tau)  ( \partial_{\tau}^2 g_{LL}(\tau) + 2\mathcal{J}^2 e^{g_{LL}(\tau)}) = 0, 
\label{eq:EoM_gll} \\
&&\partial_{\tau}^2 g_{LR}(\tau) + 2\mathcal{J}^2e^{g_{LR}(\tau)} +2\hat{\mu}\delta(\tau) = 0, 
\label{eq:EoM_glr}
\ee
where $J^2 = 2 ^{q-1}\mathcal{J}^2/q$ and $\mu =\hat{\mu}/q$.
To satisfy the causality relation \eref{caus}
at low temperature we need that
\be
G_{LL}(\tau) =-i\, \sign(\tau)\, G_{LR}(\tau),
\ee
which requires that $g_{LL}(\tau) = g_{LR}(\tau)$. In Eq.~\eref{eq:EoM_gll}, the $\hat \mu \delta(\tau) $ term does not contribute because of the $\sign(\tau)$ factor.
The boundary conditions are \cite{maldacena2018} $g_{LL}(0) =0$, $\partial_{\tau}g_{LR}(\tau=0) = - \hat{\mu}$, and $g_{LL}(\tau) - g_{LR}(\tau) \to 0$ as $\tau \to \infty$. We note that the above SD equations and boundary conditions coincide with those of the real-time problem of the open single-site SYK model \cite{kulkarni2022}. In the low-temperature limit, the solution is given by
\be\label{eq:glargeq}
e^{g_{LL}}=e^{g_{LR}}=\frac{\alpha^2}{\mathcal{J}^2 \cosh^2(\alpha|\tau|+\gamma)}
\ee
with ${\alpha^2}/{\mathcal{J}^2 \cosh^2 \gamma} =1$, $2\alpha\tanh\gamma=\hat{\mu}$.
Using Eqs.~(\ref{eq:largeqG1}) and (\ref{eq:glargeq}), the gap is given by
\be
E_g = \frac{2\alpha}{q}= \frac{2\mathcal{J}}{q}\sqrt{\Big(\frac{\hat{\mu}}{2\mathcal{J}}\Big)^2 + 1},
\ee
which is also the result for the decay rate of the real-time one-site SYK coupled to a bath \cite{kulkarni2022}. The $\mu = 0$ limit of the decay rate was derived earlier in Ref.~\cite{maldacena2016}. 

We know that it is expected \cite{maldacena2018} that this large-$q$ limit is qualitatively different from the $q = 4$ case, which is also quantum chaotic. For instance, Green's functions have no oscillations and the gap is monotonic with $\mu$. Still, some similarities remain, and the gap tends to a constant in the $\mu \to 0$ limit, which indicates that for weak coupling the gap and the decay rate are fully controlled by internal dynamics and not by the coupling to the bath. Likewise, in the $\mu \to \infty$ limit, the decay rate is fully controlled by the environment.

A similar calculation, see Appendix \ref{app:large_q}, leads to the free energy in the large-$q$ limit at low temperature $T$:
\be
\beta F =-\frac 12 \beta \mu,
\ee
which is the value that is obtained from the ground-state energy of the Hamiltonian discussed in the next section, see Fig.~\ref{fig:feuc}. 

\section{Gravity dual of a dissipative SYK: Keldysh wormhole}
\label{sec:dual}

\begin{figure}[!ht]
	\centering
	\includegraphics[width=7cm]{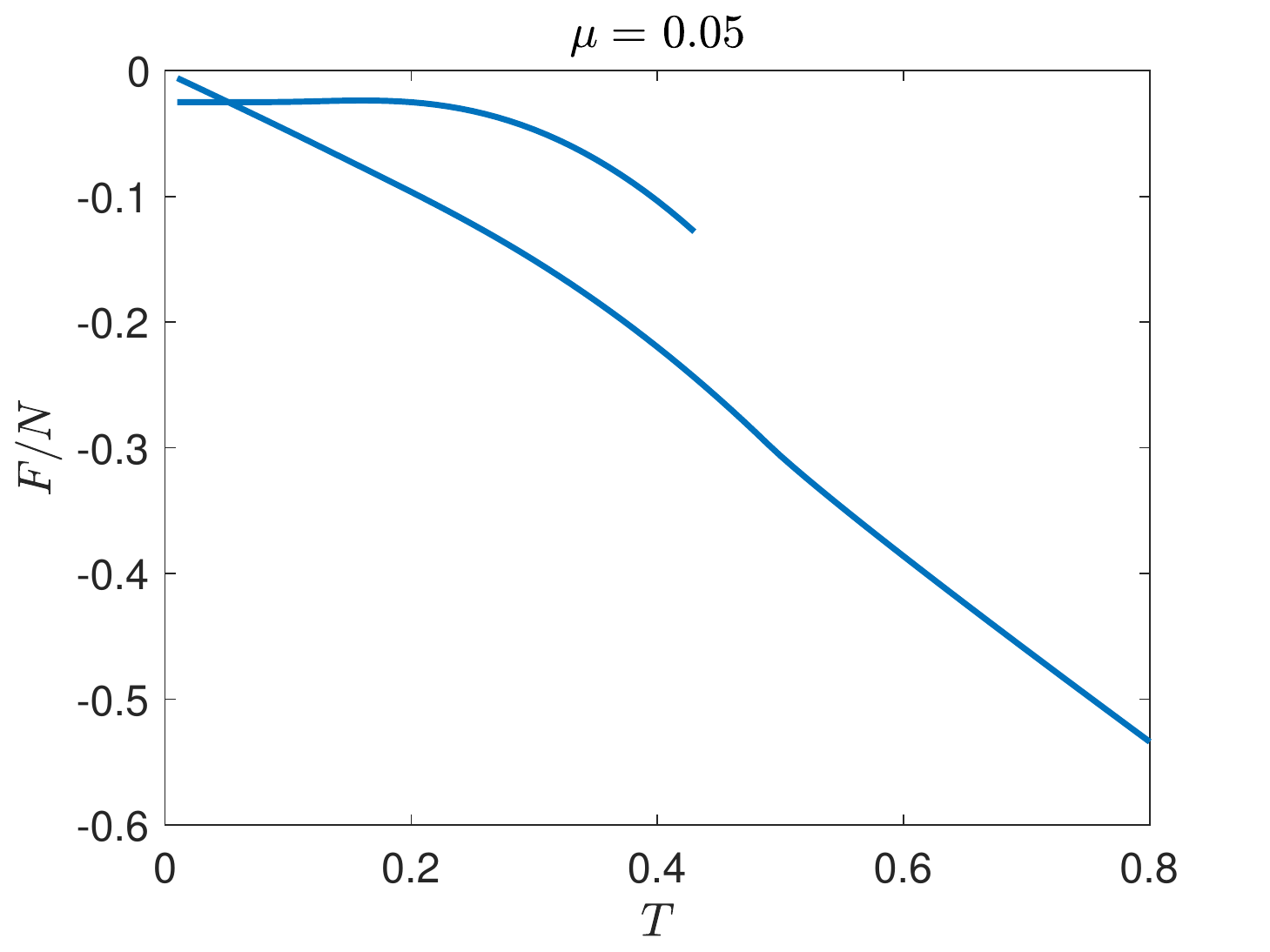}
	\includegraphics[width=7cm]{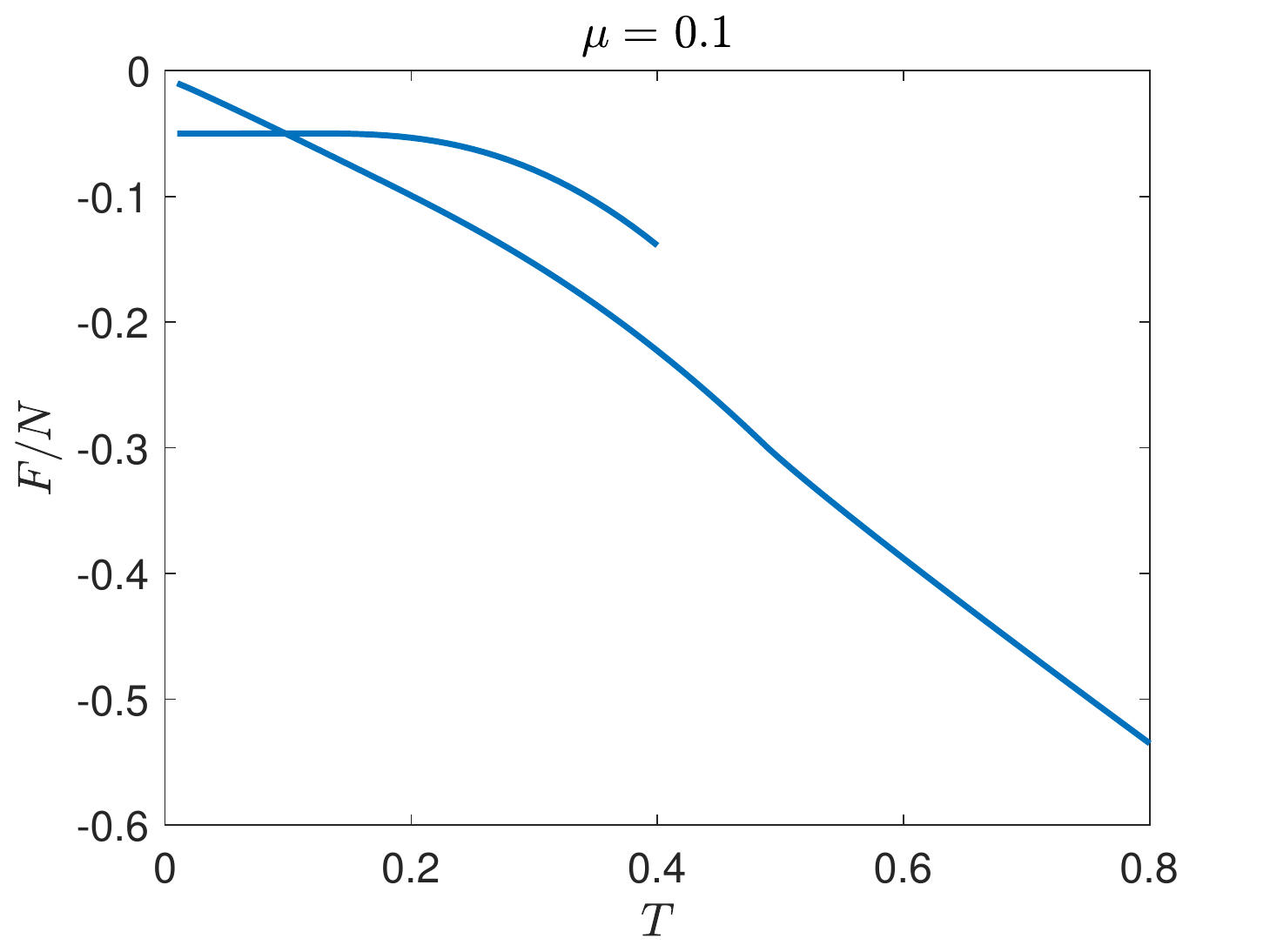}
	\includegraphics[width=7cm]{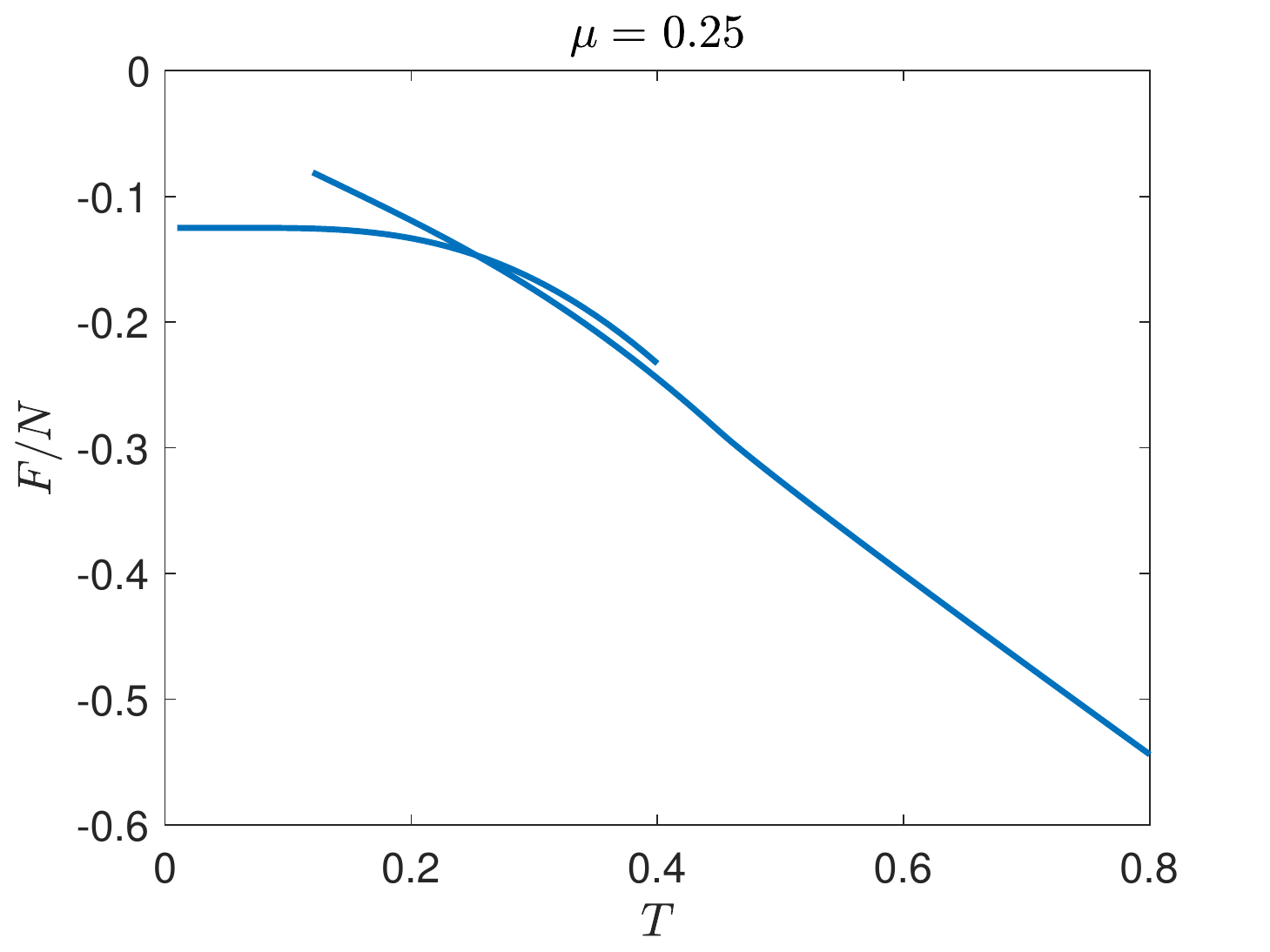}
	\includegraphics[width=7cm]{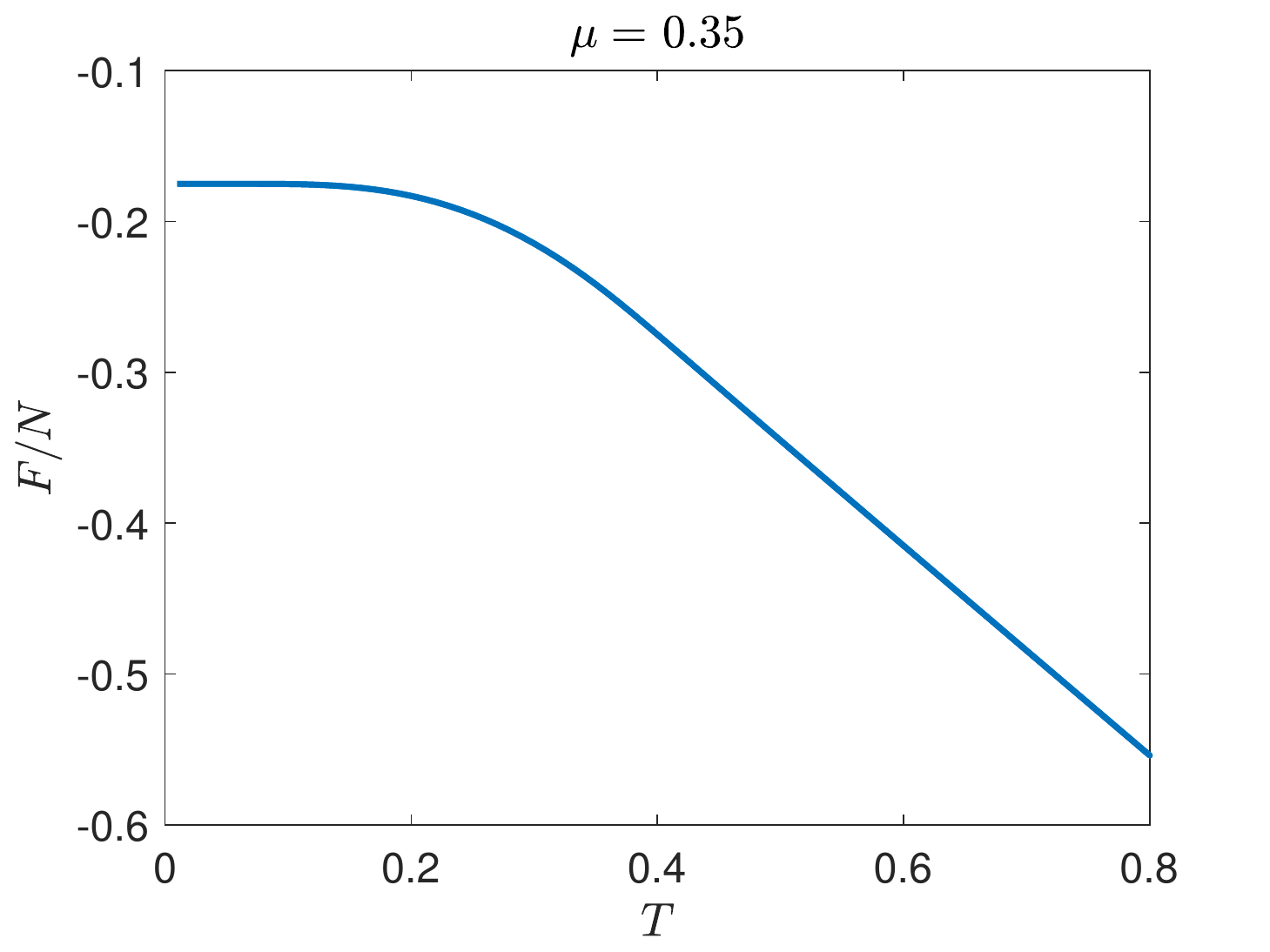}
	\includegraphics[width=7cm]{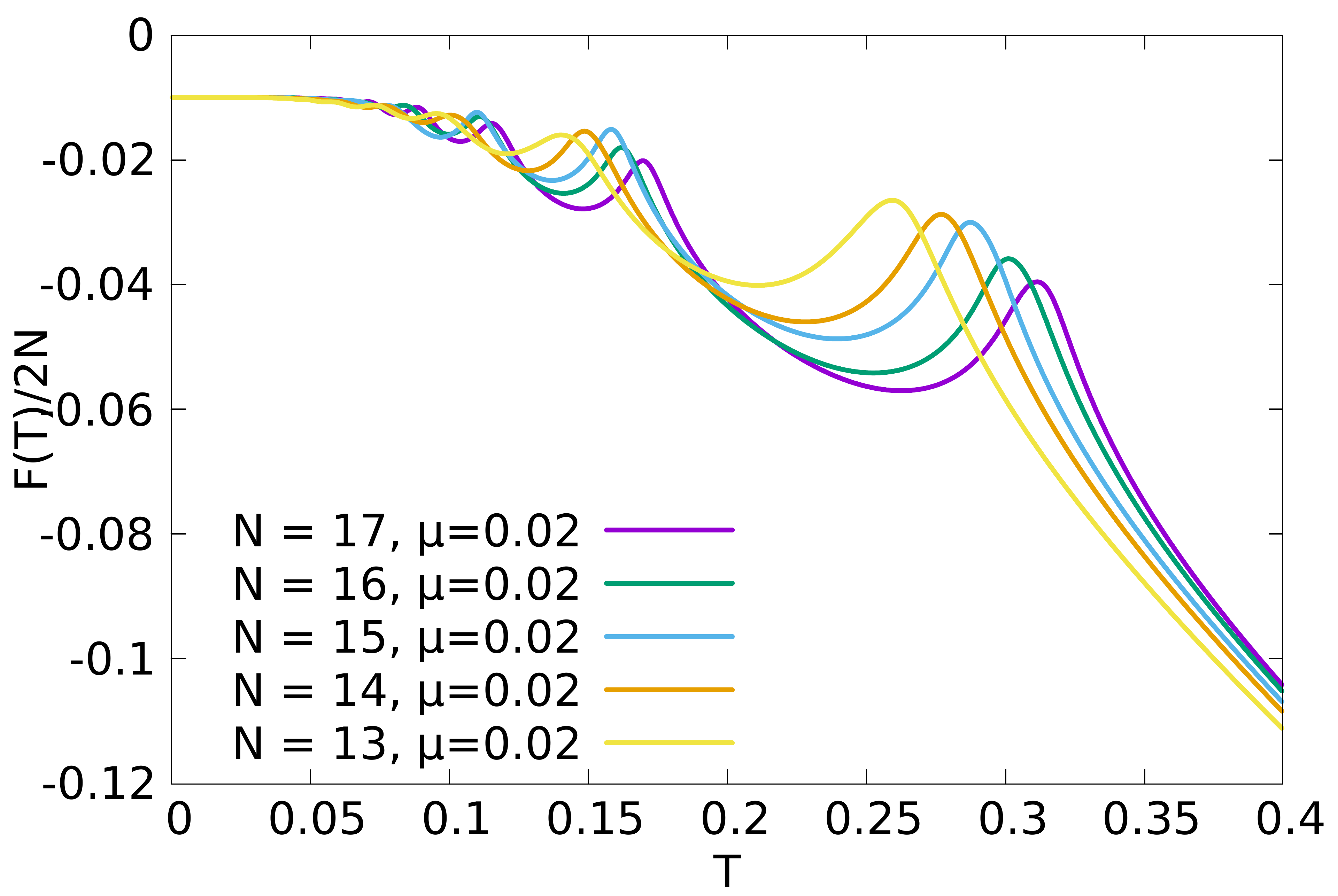}
	\includegraphics[width=7cm]{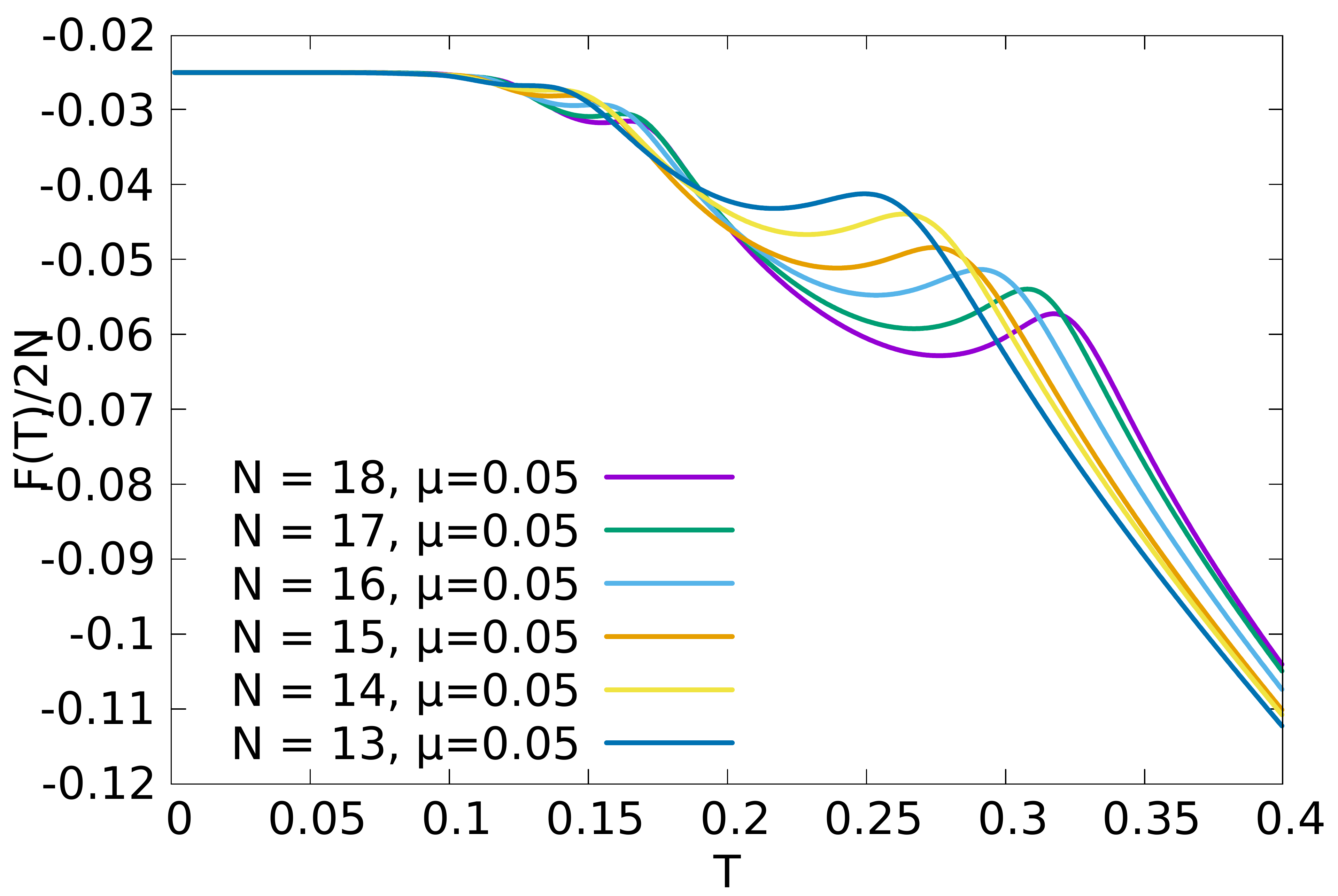}
	\includegraphics[width=7cm]{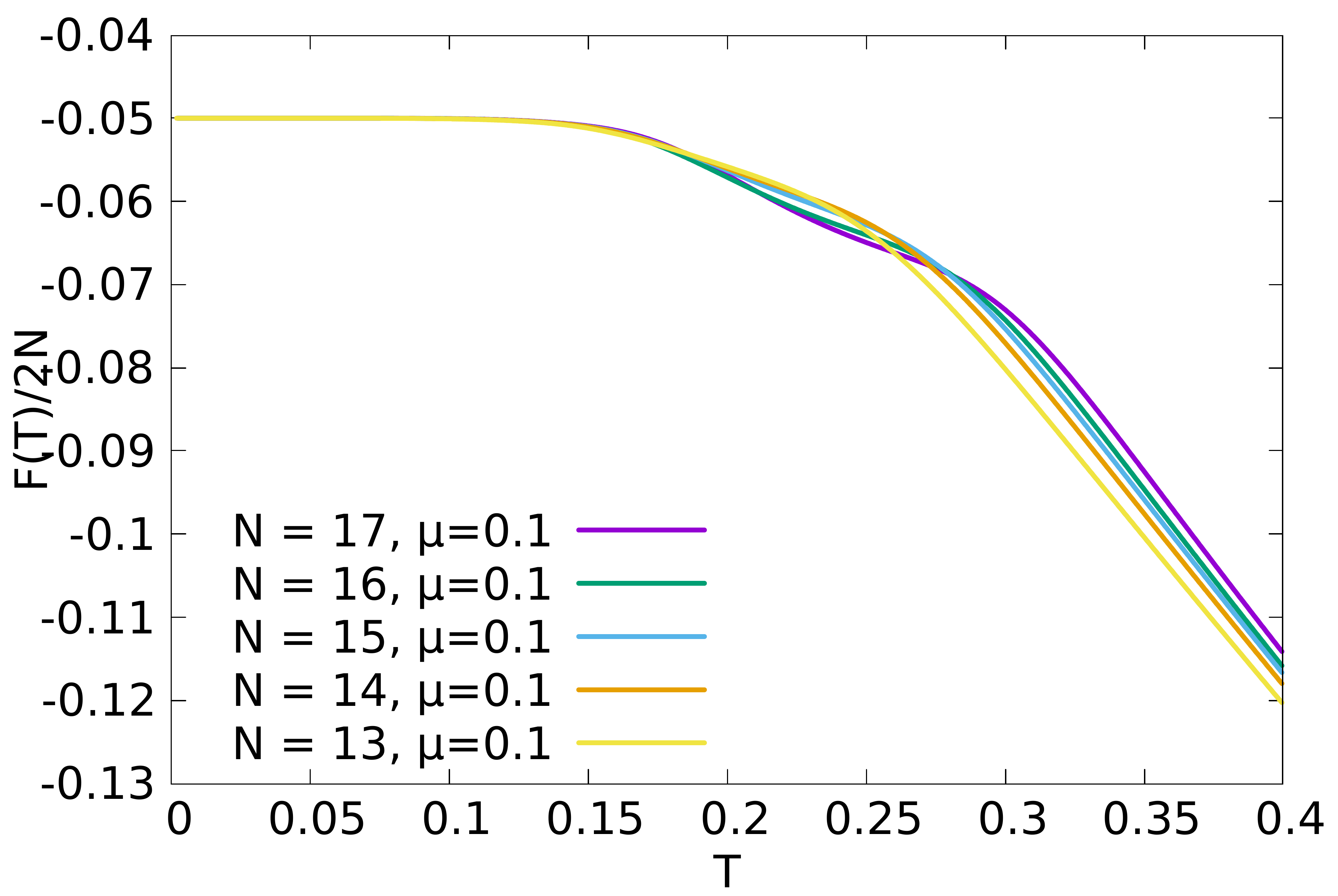}
	\includegraphics[width=7cm]{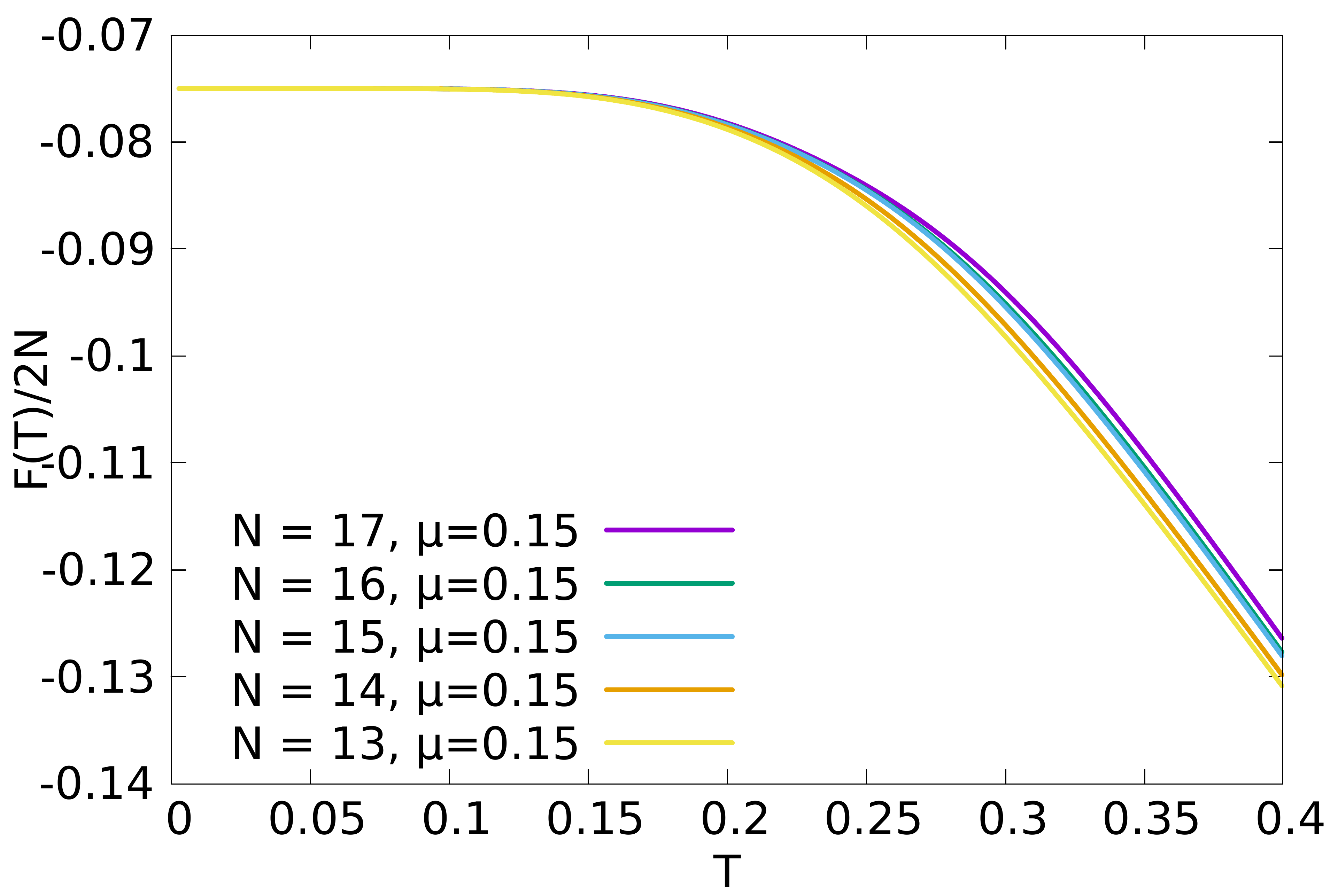}
	\caption{Free energy of the Euclidean problem as a function of temperature $T$ for $q = 4$ from the solution of the SD equations (first two rows) and from exact diagonalization of the Hamiltonian for $13\le N \le 18$ (last two rows). For the latter, we use a different convention than for the SD calculation: $\{\psi^i,\psi^j\}=2\delta_{ij}$ and $\langle J_{ijkl}^2\rangle =6 J^2/(2N)^3$. 
          For small $\mu$, we identify a first-order phase transition where the low-temperature phase is characterized by a flat free energy. For large $\mu$, in line with the result observed for Euclidean \cite{garcia2021} and traversable \cite{maldacena2018} wormholes, the transition becomes a crossover. For intermediate temperatures above the first-order transition, we observe oscillations only in the exact diagonalization calculation which are suppressed as $\mu$ increases. The change in slope observed for intermediate $\mu$ at a temperature above the first-order transition, which likely signals a higher-order transition, is reproduced by both calculations.}\label{fig:feuc}
\end{figure}

We now turn to the tentative gravitational interpretation of the results obtained above. 
As was mentioned earlier, it is well established \cite{maldacena2018,garcia2021} that, in the low-temperature limit, the dual field theory of two-dimensional anti-de Sitter 
(AdS$_2$) wormholes is a two-site SYK. Depending on the setting, the gravity dual can be a traversable or a Euclidean wormhole. Since the path integral describing the dissipative SYK is identical to the path integral of the two coupled non-Hermitian SYK models,
its gravity dual will be the same. Therefore, the findings of this section could shed light on the gravity dual of a single-site SYK coupled to an environment in real time.

Our two-site SYK model is qualitatively similar to that recently investigated in Ref.~\cite{garcia2022c}, which 
reported a Euclidean-to-traversable wormhole transition by tuning the intersite coupling. 
However, there is an important difference: the random couplings in our model are purely imaginary while those of the model of Ref.~\cite{garcia2022c} are complex with the variance of the real part larger than the variance of the imaginary part. Therefore, the low-energy effective action of our model cannot be easily identified with a Euclidean or a traversable wormhole. 
The reason is that in our model the coupling $J$ in the Schwarzian term of the low-energy effective action becomes complex, $J_L \to iJ$ and $J_R \to -i J$, with respect to that in the Euclidean and traversable cases. This is a direct consequence of performing the ensemble average over a two-site SYK Hamiltonian that is purely imaginary after vectorization. 
Intriguingly, this complexification of couplings is precisely the necessary prescription \cite{cotler2020} to obtain the correct Schwarzian action describing a double-trumpet configuration in a near-de Sitter background in two dimensions (dS$_2$)~\cite{cotler2020} in Lorentzian time from the equivalent in a near-AdS$_2$ in Euclidean time \cite{jackiw1985,teitelboim1983}. If this is the case, our setting would correspond to that of a wormhole in this near-dS$_2$ geometry. Indeed, the so-called bra-ket wormholes \cite{chen2021,page1986}\footnote{A.\ M.\ G.\ G.\ thanks Victor Godet for suggesting bra-ket wormholes as a possible gravity dual of our SYK setting.} with a similar double-trumpet geometry have been identified in near-dS$_2$ backgrounds coupled to conformal matter. These wormholes are connected geometries between bra (ket) states prepared forward (backward) in time following the Keldysh protocol. Finally, it has been speculated that the SYK dual of a one-boundary near-dS$_2$ background \cite{turiaci2020} in Lorentzian time, not related to wormholes, would require the same coupling complexification $J \to iJ$. 
Beyond the explicit comparison of the low-energy effective action, a potential way to identify the gravity duality more explicitly is to compare the gravitational quasinormal modes in a near-dS$_2$ setting and the decay rate and frequency that characterizes the approach to a steady state in the dissipative SYK. 
%We note that since the real time SYK is formally at infinite temperature, and therefore in the weak coupling region, the quasinormal modes is a natural observable to test the duality.  
For an interpolating AdS$_2$-to-dS$_2$ geometry, the study of quasinormal modes was carried out in Ref.~\cite{anninos2018,anninos2019}. More recently \cite{rahman2022,susskind2022}, quasinormal modes leading to a purely exponential decay---with no oscillations---of the retarded Green's function have been computed in a dS$_2$ static patch. It has been conjectured \cite{rahman2022,susskind2022} that the field theory dual may be a certain Hermitian single-site SYK in the double scaling limit \cite{erdos2014}. We note that this is qualitatively different from our non-Hermitian two-site SYK model whose retarded Green's function in the small $\mu$ limit also has oscillations and an exponential decay that does not require large $q$ even for $\mu = 0$. Therefore, it is expected that the gravity dual of both models is qualitatively different as well.

We now turn to confirm the wormhole features of the low-temperature limit of the two-site non-Hermitian model by the study of the free energy, where wormhole configurations
are characterized by an almost
flat free energy at low temperature that ends abruptly at a certain temperature where a first-order phase transition takes place. 
In the first two rows of Fig.~\ref{fig:feuc}, we present results for the free energy of our model by solving the SD equations for different values of the coupling $\mu$. We note that the free energy is obtained \cite{maldacena2016} by inserting the solutions of the SD equations back in the action.
The free energy has two branches from different solutions of the SD equations. For very low temperatures, and finite but not very strong coupling, we observe a flat part typical of wormhole
configurations. This branch eventually crosses a fast decreasing branch, signaling the presence of the first-order transition consistent with a wormhole-black hole transition \cite{maldacena2018,garcia2019}.
For a stronger coupling $\mu$, the transition becomes a crossover, a feature that has been observed in the free energy corresponding to a two-site SYK model dual to wormhole
configurations \cite{garcia2019,garcia2022c}. For intermediate temperatures above the first-order transition, where the two branches coexist, the free energy undergoes a close to linear decrease. At a higher temperature, close to the end of the wormhole branch, the slope of $F(T)$ in the high-temperature phase
changes rather abruptly. This is consistent with the existence of a higher-order phase transition separating this linear temperature behavior from the black hole phase. 
We note that around the same temperature at which this change in slope occurs, there is also a qualitative change in the Green's functions: $G_{LL}(\tau)$ and $G_{LR}(\tau)$ have finite real and imaginary parts in this range of temperatures, while they are, respectively, purely real and imaginary elsewhere. These results for the free energy and the ones of Appendix~\ref{app:large_q} strongly overlap with those of Ref.~\cite{kawabata2022}, which was posted on the arXiv a few days after this paper.

Motivated by this intriguing intermediate phase, 
we have also carried out a numerical calculation of the free energy by computing the eigenvalues of the Hamiltonian using ED techniques. The results depicted in the last two rows of Fig.~\ref{fig:feuc} reproduce well the flat part for low temperatures and the crossover for larger values of $\mu$. For intermediate temperatures and weak intersite coupling, corresponding to the region of coexistence of two solutions in the large-$N$ calculation above, we observe oscillations that are suppressed as $\mu$ increases. This is an indication of a pathological behavior characterized by a negative specific heat and a negative entropy. For $\mu \gtrsim 0.125$, there are no oscillations and the free energy of the large-$N$ calculation is reproduced qualitatively including the change in slope in the high-temperature limit mentioned above.
Further research will be needed to find out both the physical mechanism causing the change of slope and the origin of the observed oscillations in the finite-$N$ free energy. Overall, the first-order transition is rather similar to that reported in other SYK settings \cite{garcia2022c} where the wormhole gravity dual is explicitly known. 

Another distinct feature of wormhole configurations \cite{sahoo2020,plugge2020,garcia2022c} are oscillations of Euclidean Green's functions after analytic continuation to real time. The frequency of the oscillations is related to $E_g$. For Euclidean wormholes, $G_{LR}$ and $G_{LL}$ are in phase \cite{garcia2022c}, while for traversable wormholes, they are out of phase \cite{plugge2020,sahoo2020}. Since we have that $G_{LL}(\tau)=- i\,\sign(\tau)G_{LR}(\tau)$, the Green's functions of our Euclidean model in real time will oscillate in phase. 

Although we cannot identify precisely the dual gravity background, we summarize, based on results in this and previous sections, its main features: a gapped ground state, with a gap that has a nontrivial dependence on the couplings but which is different from that of Euclidean or traversable wormholes; oscillating Green's functions in real time with $|G_{LL}(t)| = |G_{LR}(t)|$, as observed in Euclidean wormholes; and a flat free energy in the low-temperature limit that terminates in a first-order phase transition (for higher temperatures, we observe a change of trend in the free energy consistent with a higher-order phase transition). Moreover, as mentioned above, the Schwarzian action of our model should correspond to that of bra-ket wormholes \cite{chen2021} in a near-dS$_2$ background.
Since these features are similar to those corresponding to Euclidean or traversable wormholes, but with important qualitative differences, we have termed \emph{Keldysh wormholes} the field configurations with wormhole-like features that govern the path integral describing the time evolution of an SYK model coupled to an environment.

\section{Conclusions}
\label{sec:conclusions}
We have shown that in the near-zero-temperature limit, the dynamics of a two-site non-Hermitian SYK model in Euclidean time
are equal to the real-time evolution of the Liouvillian describing a one-site SYK model coupled to an environment close to the infinite-temperature steady state. 
The two-site model has features, like Green's functions that decay exponentially with a decay rate that is finite even if there is no coupling to the environment and a free energy with a first-order phase transition, consistent with those found earlier in SYK models with a wormhole gravity dual. Indeed, the low-energy effective action is that of a Euclidean wormhole but
with the coupling $J$ in the Schwarzian term complexified, $J \to iJ$. This complexification of the coupling has been related \cite{cotler2020,turiaci2019} to wormholes in a near de-Sitter background and we have termed it a Keldysh wormhole. However, we stress that a precise identification of the gravity dual requires further research.

We have also shown that the decay rate of the open SYK is finite even if the coupling to the environment is zero provided that the dynamics is quantum chaotic (for $q=4$). Only in the limit of strong coupling to the bath, the approach to the steady state is controlled by the environment. Interestingly, only in the region of weak or vanishing coupling do Keldysh wormholes control the path integral. 
An analytical study of the large-$q$ limit, which is also quantum chaotic, confirms that the decay rate is finite even if there is no coupling to the environment. 
Another distinct feature of these wormhole configurations is that Green's functions show damped oscillations whose amplitude decreases as the coupling to the bath increases. For a sufficiently strong coupling, the oscillations vanish and, later, the decay rate becomes a monotonic function of the coupling.
In contrast, for a dissipative integrable SYK model ($q=2$), the decay rate, which we have obtained analytically, is always dominated by the coupling to the environment. Indeed, contrarily to the quantum chaotic case, it vanishes if there is no coupling to the bath, showing no anomalous relaxation. Also unlike the quantum chaotic case, the pattern of oscillations is observed for any coupling.

Natural extensions of this work are a more detailed analysis of the low-energy effective action in Euclidean signature in order to establish a precise relation to wormhole configurations and de Sitter holography \cite{turiaci2019,cotler2020}. We also plan to study the relation between the decay rate and the spectrum of the Hamiltonian and the nature of the large-$N$ limit. Another interesting avenue of research is to investigate more general Lindblad jump operators in order to characterize the conditions for the observation of a phase at weak or zero coupling to the environment where the decay rate is not controlled by the environment and investigate whether it is related to the presence of Keldysh wormholes.

\acknowledgments{
AMGG acknowledges illuminating correspondence with Victor Godet. LS thanks Pedro Ribeiro and Toma\v{z} Prosen for insightful discussions and collaboration in related projects. JJMV thanks Joshua Leeman and Derek Teaney for useful discussions. AMGG and JPZ were partially supported by the National Natural Science Foundation of China (NSFC) (Grant number 11874259), by the National Key R\&D Program of China (Project ID: 2019YFA0308603). We acknowledge partial support from a Shanghai talent program (AMGG).
LS is supported by Funda\c{c}\~ao para a Ci\^encia e a Tecnologia (FCT-Portugal) through Grant No.\ SFRH/BD/147477/2019. This project was funded within the QuantERA II Programme that has received funding from the European Union’s Horizon 2020 research and innovation programme under Grant Agreement No 101017733. JJMV acknowledges partial support from U.S. DOE grant No.\ DE-FAG-88FR40388. 
}

\appendix

\section{Choi-Jamiolkowski isomorphism}
\label{app:choi}

The idea of the Choi-Jamiolkowski (CJ) transformation is to write an operator as a vector. In
bra-ket notation
\be
A = \sum_{ij} \langle i| A |j\rangle |i\rangle \langle j|
\to \sum_{ij} \langle i| A |j\rangle (|i\rangle \otimes | j \rangle).
\ee
If the operator $A$ acts on the space $V$, then, after the CJ transformation, $A$ is represented as a vector in $V\otimes V$. The first space will be labeled by $L$ and the second one
by $R$.

The Lindblad operator ${\cal L}$ acts on the density matrix $\rho$. After the CJ transformation, $\rho$ becomes a vector in $V\otimes V$, so that ${\cal L}$ maps to an operator acting on $V \otimes V$.
The basis of $V$ will be denoted by $|k \rangle \otimes |l \rangle $ or, in short, $|k \rangle |l \rangle $. After the CJ transformation, we thus have
\be
   {\cal L}(\rho) \to \langle k | \langle m| {\cal L} |l\rangle |j\rangle
   \langle l |\rho |j\rangle.
   \label{lindblad-matrix}
   \ee
To work out the Lindblad operator after the CJ transformation, we first need to derive what is known as operator reflection.
\subsection{Operator reflection}

The idea of operator reflection is to find, for a given
$A_L$, the operator $A_R$ satisfying
\be
A_L\sum_j \kj UK \kj = \sum_j \kj A_R UK \kj,
\label{refl}
\ee
where $K$ is the complex conjugation operator and $U$ is unitary.
For the system of interest $A_L = \psi_L$ and $A_R =\alpha \psi_R$ with
$\alpha $ a constant, and we determine $U$ and $\alpha$ such that
\eref{refl} is satisfied. The representation of $\gamma $ matrices
is taken to be
\be
\psi_L^k = \gamma_k \otimes 1,\qquad
\psi_R^k = \gamma_c \otimes \gamma_k
\ee
with $\gamma_k$ the gamma matrices for the left SYK model, and $\gamma_c$
the chirality matrix. Note that with this choice, $A_R$ in \eref{refl} also
acts on the first $\kj$ in \eref{refl}.
This leads to
the requirement (repeated indices summed)
\be
\bl\gamma_k \kj \bm UK \kj= \alpha \bl \gamma_c \kj \bm \gamma_k UK\kj
\ee
for all $l$ and $m$.
This can be simplified to
\be
\bl \gamma_k U^T K \km&=& \alpha  \bm \gamma_k UK \gamma_c\kl\nn\\
&=& \alpha  \bl( \gamma_k U\gamma_c)^T K\km.
\ee
Therefore, we must have that
\be
\gamma_k U^T = \alpha \gamma_c U^T \gamma_k^T
\ee
or
\be
\gamma_k  = \alpha \gamma_c U^T \gamma_k^T {U^T}^{-1}.
\ee
Since the charge conjugation matrix $C$ satisfies
\be
C^{-1} \gamma_k C = \gamma_k^T,
\ee
we choose
\be
U^T = e^{i\pi\gamma_c/4} C.
\ee
This gives the condition
\be
\alpha \gamma_c e^{i \pi \gamma_c/2} =1,
\ee
resulting in $\alpha = -i$.

We conclude that
\be
\psi_L^k|0\rangle = - i\psi_R^k|0\rangle
\label{reflec}
\ee
with 
\be
|0\rangle = \sum_j \kj e^{\pi i\gamma_c/4}CK \kj.
\ee
The overall sign from $C^T=\pm C$ is irrelevant and has been ignored.

\subsection{The vectorized Lindblad operator}
\label{app:lind}

\noindent
\paragraph{The MQ term in the Lindblad operator.}
Let us consider the $L_n \rho L_n^\dagger $ term in the Lindblad
evolution. We take $L_n = \sqrt \mu \psi^n$ with $\psi^n$ a
Majorana operator. The CJ transformation in this case is given by
(repeated indices are summed)
\be
\mu \bi \psi^n \rho \psi^n \kj \ki \bj |
\to \mu \bi \psi_{L}^{n} \rho \psi_{L}^{n} \kj \ki \kj,
\ee
where the factor $UK$ has been absorbed in the 
second $|j\rangle$ of $|j \rangle |j\rangle$. Using completeness and the reflection property \eref{reflec}, we obtain
 \be
\mu \bi \psi_{L}^{n} \rho \psi_{L}^{n} \kj \ki \kj
 &=& -i \mu  \psi_{L}^{n} \rho  \kj  \psi_{R}^{n}\kj  \nn \\
 &=& -i \mu  \bk \psi_{L}^{n} \kl \bl  \rho \kj  \bm \psi_{R}^{n}\kj  \kk  \km  \nn \\
 &=& -i \mu  \bm  \psi_{L}^{n} \kl  \bk \psi_{R}^{n} \kj  \km \kk \bl  \rho \kj.
\ee
Comparing to Eq.~\eref{lindblad-matrix}, we can read off the operator acting on $V\otimes V$:
\be
-i \mu \,  \psi_{L}^{n} \psi_{R}^{n}  \equiv -i \mu   \psi_{L}^{n} \otimes \psi_{R}^{n}.
\ee

\noindent
\paragraph{The $-i[H^\mathrm{SYK}, \rho]$ term.}
After the CJ transformation, the $-iH^\mathrm{SYK}\rho$ term becomes
\be
  -i \bi H^\mathrm{SYK} \rho \kj \ki\kj & = & -i  H^\mathrm{SYK}_L \rho \kj\kj \nn\\
   & = & -i  \bk  H_L^\mathrm{SYK} \kl \bm \mathbb{1}_R \kj  \kk \km \bl \rho  \kj.  
\ee
This gives $-i H^\mathrm{SYK}_L\equiv-i H^\mathrm{SYK}_L \otimes \mathbb{1}_R$ for the vectorized operator.

After the CJ transformation, the $i\rho H^\mathrm{SYK}$ term becomes (for even $q$)
  \be
  i \bi \rho  H^\mathrm{SYK}  \kj \ki\kj & = & i   \rho H_L^\mathrm{SYK} \kj\kj \nn \\
  & = & i(-i)^q \bm \mathbb{1}_L  \kl  \bk  H_R^\mathrm{SYK} \kj \km \kk \bl  \rho \kj.
  \ee
This gives the term $(-1)^{q/2} i H_R^\mathrm{SYK}\equiv (-1)^{q/2} (\mathbb{1}_L \otimes i H_R^\mathrm{SYK})$ in the SYK operator.

  \noindent
  \paragraph{The constant term.}
  For $-\frac 12 \mu \({\psi^n}^\dagger \psi^n\rho+\rho{\psi^n}^\dagger \psi^n\)$ we use that (note the summation convention)
  \be
  \frac 12 {\psi^n}^\dagger \psi^n = \frac N4
  \ee
  with $N$ the number of $L$ (or $R$) Majorana fermions.
  Then, we obtain
  \be
  -\frac{N\mu}2\bi \rho \kj \ki \kj &=& -N\mu \rho \kj \kj\nn\\
  &=& -\frac{N\mu}2 \bk \mathbb{1}_L \kl \bm \mathbb{1}_R\kj \kk \km \bl\rho \kj.
  \ee
  This gives $-(N\mu/2)  \mathbb{1}_L\otimes  \mathbb{1}_R$ in the SYK operator. 
  
Combining all terms gives the operator \eref{doublelindbrad} obtained in Ref.~\cite{kulkarni2022}.

\section{Details on the real-time evolution of the open SYK model}
\label{app:real_time}

The partition function is
\be
\label{eq:app_path_integral}
Z 
= \int \mathcal{D}\psi_L\mathcal{D}\psi_R e^{iS[\psi_L,\psi_R]},
\ee
with action
\be
\label{eq:app_action}
i S =  \int_{-\infty}^{+\infty} dt \Big[ -\frac{1}{2} \sum_i \psi^i_L \partial_t \psi^i_L-\frac{1}{2} \sum_i \psi^i_R \partial_t \psi^i_R +\mathcal{L} \Big]
\ee
and Liouvillian  
\be
\label{eq:app_doublelindbrad}
\mathcal{L} = - iH_{{\rm SYK}}^L + i  (-1)^\frac{q}{2} 
H_{{\rm SYK}}^R    -i \mu \sum_i\psi_L^i\psi_R^i - \frac{1}{2}\mu N.  
\ee
As discussed in the main text, we now move to the closed-time contour representation ($\sC=\sC^+\cup\sC^-$), identifying $\psi^i(t^+)=\psi^i_L(t)$, with $t^+\in\sC^+$, and $\psi^i(t^-)=i\psi^i_R(t)$, with $t^-\in\sC^-$. On the Keldysh contour the action reads as (note that $dz=dt$ on $\sC^+$ and $dz=-dt$ on $\sC^-$)
\begin{equation}
\begin{split}
iS=&-\int_\sC dz \,\frac{1}{2}\sum_{i=1}^N\psi^i(z)\partial_z \psi^i(z)
-i\int_\sC d z\, i^{q/2} \sum_{i_1<\cdots<i_q}^N J_{i_1\cdots i_q}\psi^{i_1}(z)\cdots \psi^{i_q}(z)
\\
&+\mu\int_\sC d z\, d z'\, K(z,z')\sum_{i=1}^N \psi^i(z)\psi^i(z'),
\end{split}
\end{equation}
with dissipative kernel
\begin{equation}\label{eq:app_kernel}
K(t_1^+,t_2^+)=K(t_1^-,t_2^-)=K(t_1^-,t_2^+)=0,
\qquad
K(t_1^+,t_2^-)=\delta(t_1-t_2).
\end{equation}
Here, we have dropped any dependence on the initial conditions, since we are interested in the long-time relaxation, and the immaterial constant $-N\mu/2$.
Performing the disorder average over the couplings $J_{i_1\cdots i_q}$ we obtain
\begin{equation}
\begin{split}
iS=&-\int_\sC dz\, \frac{1}{2}\sum_{i=1}^N \psi^i(z)\partial_z \psi^i(z)
-\frac{i^q}{2}\int_\sC d z\, d z'\, \frac{(q-1)!J^2}{N^{q-1}}\sum_{i_1<\cdots<i_q}^N 
\psi^{i_1}(z)\cdots \psi^{i_q}(z)
\psi^{i_1}(z')\cdots \psi^{i_q}(z')
\\
&+\mu\int_\sC d z\, d z'\, K(z,z')\sum_{i=1}^N \psi^i(z)\psi^i(z').
\end{split}
\end{equation}
Introducing the collective field
\begin{equation}
G(z,z')=-\frac{i}{N}\sum_{i=1}^N \psi(z)\psi(z'),
\end{equation}
and its Lagrange multiplier $\Sigma(z,z')$, the action reads as
\begin{equation}
\label{srt}
\begin{split}
iS=\frac{N}{2}\Bigg\{
&\Tr\log(i\partial_z-\Sigma)-
\int_\sC d z\, d z'\, \Sigma(z,z')G(z,z')-\frac{i^q J^2}{q}\int_\sC d z\, d z'\, [G(z,z')]^q
\\
&+2i\mu\int_\sC d z\, dz'\, K(z,z')G(z,z')
\Bigg\}.
\end{split}
\end{equation}
The Schwinger-Dyson (saddle-point) equations on the contour are given by
\begin{align}
\label{eq:C_SD_Sigma}
\left(i\partial-\Sigma\right)\cdot &\,G=\mathbb{1}_{\sC},
\\
\begin{split}\label{eq:C_SD_G}
\Sigma(z,z')=&-i^qJ^2\left[G(z,z')\right]^{q-1}+i\mu\left[K(z,z')-K(z',z)\right].
\end{split}
\end{align}
We now return to real times $(t_1,t_2)$. Different components of the real-time Green's function are obtained by restricting $(z,z')$ to the two branches $\sC^+$ and $\sC^-$:
\begin{equation}
G_{ab}(t_1,t_2)=-\frac{i}{N}\sum_{i=1}^N \psi^i (t_1^a)\,\psi^i(t_2^b),
\end{equation}
with $a,b=+,-$ corresponding to the respective contours. At the saddle point, we introduce the greater, lesser, time-ordered, anti-time-ordered, retarded, advanced, and Keldysh components, respectively, as
\begin{align}\label{eq:SM_G^>}
&G^>(t_1,t_2)=
\left\langle G_{-+}(t_1,t_2)\right\rangle=
-\frac{i}{N}\sum_{i=1}^N \left\langle\psi^i(t_1^-) \psi^i(t_2^+)\right\rangle,
\\
\label{eq:SM_G^<}
&G^<(t_1,t_2)=
\left\langle G_{+-}(t_1,t_2)\right\rangle=
-\frac{i}{N}\sum_{i=1}^N\left\langle \psi^i(t_1^+) \psi^i(t_2^-)\right\rangle
=-G^>(t_2,t_1),
\\
&G^\rmT(t_1,t_2) =
\heav{t_1-t_2}G^>(t_1,t_2) - \heav{t_2-t_1} G^>(t_2,t_1),
\label{eq:SM_G^T}\\
&G^{\bar{\rm T}}(t_1,t_2) =
\heav{t_1-t_2}G^<(t_1,t_2) - \heav{t_2-t_1} G^<(t_2,t_1),
\label{eq:SM_G^Tb}\\
&G^\rmR(t_1,t_2)
=\heav{t_1-t_2}\left(G^>(t_1,t_2)-G^<(t_1,t_2)\right)
=\heav{t_1-t_2}\left(G^>(t_1,t_2)+G^>(t_2,t_1)\right),
\label{eq:SM_G^R}
\\
\label{eq:SM_G^A}
&G^\rmA(t_1,t_2)
=\heav{t_2-t_1}\(G^<(t_1,t_2)-G^>(t_1,t_2)\)
=-\heav{t_2-t_1}\(G^>(t_1,t_2)+G^>(t_2,t_1)\),
\\
\label{eq:SM_G^K}
&G^\rmK(t_1,t_2)
=G^>(t_1,t_2)+G^<(t_1,t_2)
=G^>(t_1,t_2)-G^>(t_2,t_1)
=-G^\rmK(t_2,t_1).
\end{align}
The components of the contour self-energy $\Sigma(z,z')$ satisfy the same relations as the Green's functions.
In the long-time limit $\mathcal{T}=(t_1+t_2)/2\to\infty$, the system loses any information about the initial condition as it relaxes to the steady state. Time-translational invariance in $\mathcal{T}$ emerges and the Green's functions depend only on the relative time $t=t_1-t_2$. Moving to Fourier space with continuous frequencies $\omega$ using the convention
\begin{equation}
\begin{split}
A(t)&=\int_{-\infty}^{+\infty} \frac{d\omega}{2\pi}\, A(\omega)\,e^{-i \omega t},
\\
A(\omega)&=\int_{-\infty}^{+\infty} d t\, A(t)\,e^{i \omega t},
\end{split}
\end{equation}
we define the real quantities
\begin{align}
\label{eq:SM_rho^+-_def}
\rho^{\pm}(\omega)
&=-\frac{1}{2\pi i}\(G^>(\omega)\pm G^<(\omega)\)
=-\frac{1}{2\pi i}\(G^>(\omega)\mp G^>(-\omega)\),
\\
\label{eq:SM_sigma^+-_def}
\sigma^{\pm}(\omega)
&=-\frac{1}{2\pi i}\(\Sigma^>(\omega)\pm \Sigma^<(\omega)\)
=-\frac{1}{2\pi i}\(\Sigma^>(\omega)\mp \Sigma^>(-\omega)\),
\end{align}
and their Hilbert transforms
\begin{align}
\label{eq:SM_rhoH_def}
\rho^\rmH(\omega)=-\frac{1}{\pi}\mathcal{P}\!\int_{-\infty}^{+\infty} d \nu\, \frac{\rho^-(\nu)}{\omega-\nu},
\qquad\text{and}\qquad
\sigma^\rmH(\omega)=-\frac{1}{\pi}\mathcal{P}\!\int_{-\infty}^{+\infty} d \nu\, \frac{\sigma^-(\nu)}{\omega-\nu}.
\end{align}
The spectral function $\rho^-(\omega)$ is symmetric and normalized as $\int d \omega\, \rho^{-}(\omega)=1$.
Because the jump operators are Hermitian, the system relaxes to the maximally mixed state.
Close to the steady state, $\rho^+(\omega)$ and $\rho^-(\omega)$ are related by a fluctuation-dissipation-like relation, $\rho^+(\omega)=\tanh(\beta_\mathrm{TFD}\, \omega/2)\, \rho^-(\omega)$, where $\beta_\mathrm{TFD}$=0 is the parameter of the infinite-temperature TFD steady state. Hence, $\rho^+(\omega)$ vanishes identically. In terms of the two-site non-Hermitian SYK model, the vanishing of $\rho^+$ can be traced back to the PT symmetry of the model~\cite{garcia2022b}. Note that for more general jump operators that lead to finite-temperature equilibrium or nonequilibrium steady states we have, in general, $\rho^+(\omega)\neq0$.

In terms of $\rho^\pm$, the components of the Green's function read as [in the last equality of each line we use the special property $\rho^+(\omega)=0$ of our model]
\begin{align}
\label{eq:SM_G>_rho}
G^{>}(\omega)
&=-\pi i\(\rho^+(\omega)+\rho^-(\omega)\)
=-\pi i \rho^-(\omega),
\\
\label{eq:SM_G<_rho}
G^{<}(\omega)
&=-\pi i\(\rho^+(\omega)-\rho^-(\omega)\)
=\pi i \rho^-(\omega),
\\
\label{eq:SM_GT_rho}
G^{\rm T}(\omega) &=-\pi\(\rho^{\rmH}(\omega) + i \rho^+(\omega)\)
=-\pi \rho^\rmH(\omega), \\
G^{\rm \bar T}(\omega) &=\pi\(\rho^{\rmH}(\omega) - i \rho^+(\omega)\)=\pi \rho^\rmH(\omega), \\
\label{eq:SM_GR_rho}
G^\rmR(\omega)&=-\pi\(\rho^\rmH(\omega)+i \rho^-(\omega)\),
\\
G^\rmA(\omega)&=-\pi\(\rho^\rmH(\omega)-i \rho^-(\omega)\),
\\
G^\rmK(\omega)&=-2\pi i\rho^+(\omega)=0.
\end{align}
Exactly the same relations hold for the self-energies $\sigma^\pm$. 

In terms of the quantities $\rho^-(\omega)$ and $\sigma^-(\omega)$, the Schwinger-Dyson equations \eref{eq:SD_real_I} and \eref{eq:SD_real_II}
can be rewritten as~\cite{sa2022}
\begin{align}
\rho^-(\omega)&=\frac{\sigma^-(\omega)}{
	\left[\omega+\pi \sigma^\rmH(\omega)\right]^2
	+\left[\pi \sigma^-(\omega)\right]^2},
\label{sdreal_rho}
\\
\sigma^-(\omega)&=\frac{J^2}{2^{q-2}}\left(\rho^-\right)^{*(q-1)}(\omega)+\frac{\mu}{\pi},
\label{sdreal_sigma}
\end{align}
where $(\rho^-)^{*n}(\omega)$ denotes the $n$-fold convolution of the spectral function with itself.
In real time $t$, our main quantity of interest is $G^\rmR(t)$ which can be obtained from the spectral function by
\begin{equation}\label{eq:GR_rho-}
i G^\rmR(t)=\heav{t}\int_{-\infty}^{+\infty} d \omega\,  \rho^-(\omega)\,e^{-i\omega t}.
\end{equation}

\section{Details on the Euclidean evolution of the two-site SYK model}
\label{app:Euclidean_time}

The partition function can be expressed as a path integral 
\begin{equation}\begin{aligned}
Z=\int \mathcal{D}\psi_a e^{-I[\psi_a]}
\end{aligned}\end{equation}
with
\begin{equation}\begin{aligned}
    I=& \int d\tau\left[\frac{1}{2}\sum_{a,i}\psi_{a}^i\partial_\tau\psi_{a}^i +i^{q/2}\sum_{a,i_1<\cdots<i_q}
      \left(i J^L_{i_1\cdots i_q} \psi_{L}^{i_1} \cdots \psi_{L}^{i_q}
      -i J^R_{i_1\cdots i_q} \psi_{R}^{i_1} \cdots \psi_{R}^{i_q}\right)
      +i\mu \sum_i \psi_{L}^{i} \psi_{R}^{i}   \right] \\
\end{aligned}\end{equation}
Integrating out the random couplings $J_{i_1,\cdots,i_q}$ satisfy the normal distribution 
\begin{equation}\begin{aligned}
\langle J^a_{i_1\cdots i_q}\rangle = 0, \quad \langle {J^a_{i_1\cdots i_q}}^2\rangle = \frac{(q-1)!J^2}{N^{q-1}} ,\qquad J^L_{i_1\cdots i_q}=(-1)^{q/2}J^R_{i_1\cdots i_q},
\end{aligned}\end{equation}
introducing $G_{ab}(\tau)$ and $\Sigma_{ab}(\tau)$ with identities 
\begin{equation}\begin{aligned}
\mathbbm{1}\sim \int \mathcal{D}G_{ab} \mathcal{D}\Sigma_{ab} \exp\left( -\frac{N}{2}\int\!\!\!\int d\tau d\tau'\Sigma_{ab}(\tau,\tau')\left[G_{ab}(\tau,\tau')-\frac{1}{N}\sum_{i}\psi_{a}^{i}(\tau)\psi_{b}^{i}(\tau')\right]  \right),
\end{aligned}\end{equation}
and finally integrating out the fermions, we obtain the
partition function in terms of $G_{ab}$ and $\Sigma_{ab}$ variables,
\begin{equation}\begin{aligned}
Z=\int \mathcal{D}G_{ab}\mathcal{D}\Sigma_{ab}e^{-I[G_{ab},\Sigma_{ab}]},
\end{aligned}\end{equation}
and the effective action
\begin{equation}\begin{aligned}
I/N=& -\frac{1}{2}\log\det(\delta_{ab}\partial_\tau -\Sigma_{ab}) \\
+\frac{1}{2}&\sum_{ab}\int\!\!\!\int [\Sigma_{ab}(\tau,\tau')G_{ab}(\tau,\tau') +\frac{1}{q}t_{ab}J^2 s_{ab}G_{ab}^q(\tau,\tau') ]d\tau d\tau' + \frac{i \mu}{2}\int [G_{LR}(\tau,\tau)-G_{RL}(\tau,\tau)]d\tau,
\end{aligned}\end{equation}
where $a,b\in\{L,R\}$ and $s_{LL}=s_{RR}=1$, $s_{LR}=s_{RL}=(-1)^{q/2}$, $t_{LL}=t_{RR}=1$, $t_{LR}=t_{RL}=-1$.
Changing variables according to
\begin{alignat}{99}
&G_{LL} \to i  G_{LL}, &\qquad& G_{RR} \to -i  G_{RR},&\qquad&
G_{LR} \to   G_{LR},&\qquad& G_{RL} \to   G_{RL},\nn\\ 
&\Sigma_{LL} \to -i  \Sigma_{LL}, &\qquad& \Sigma_{RR} \to i  \Sigma_{RR},&\qquad&
\Sigma_{LR} \to   -\Sigma_{LR},&\qquad& \Sigma_{RL} \to   -\Sigma_{RL}, 
\end{alignat}
the time derivative in this action appears as $i\partial_\tau$ and it directly maps to the real-time action \eref{srt}. This will be discussed in more detail in the next appendix.

The equations of motion, the Schwinger-Dyson equations, follow from the variation with respect to $G_{ab}$ and $\Sigma_{ab}$ and are
\begin{equation}\begin{aligned}
& \partial_\tau G_{LL}(\tau) -\Sigma_{LL}*G_{LL}(\tau) -\Sigma_{LR}*G_{RL}(\tau)=\delta(\tau), \qquad \partial_\tau G_{LR} -\Sigma_{LL}*G_{LR} -\Sigma_{LR}*G_{RR}=0,  \\
& \Sigma_{LL}(\tau)=-J^2 G_{LL}^{q-1}(\tau), \qquad \text{and} \qquad
\Sigma_{LR}(\tau)=i^q J^2 G_{LR}^{q-1}(\tau)-i\mu\delta(\tau),
\end{aligned}\end{equation} 
where $*$ denotes convolution. 
\section{Relation between the Euclidean and real-time calculations}
\label{app:relation_euclidean_real}

As established above, the relation between the real-time Majoranas on the Keldysh contour and the Euclidean-time Majoranas is given by
\begin{equation}
\label{eq:app_identifi_psi}
\psi^i(t^+)\leftrightarrow \psi_L^i (\tau)
\qquad \text{and} \qquad
\psi^i(t^-)\leftrightarrow i\psi_R^i (\tau).
\end{equation}

The real-time and the Euclidean Green's functions are defined by
\begin{equation}
\label{eq:app_identifi_defG}
G_{ab}(t_1,t_2)=-\frac{i}{N}\sum_{i=1}^N \psi^i(t_1^a) \psi^i(t_2^b)
\qquad \text{and} \qquad
G_{ab}(\tau_1,\tau_2)=\frac{1}{N}\sum_{i=1}^N \psi_a^i(\tau_1) \psi_b^i(\tau_2),
\end{equation}
respectively, where $a,b=-,+$ in real time and $a,b=L,R$ in Euclidean time. Using the identification (\ref{eq:app_identifi_psi}), we find the following relations between the components (with time-translation invariance)
\begin{equation}
\label{eq:app_identifi_G}
G^<(t)=\langle G_{+-}(t)\rangle\leftrightarrow G_{LR}(\tau)
\qquad \text{and} \qquad
G^\mathrm{T}(t)=\langle G_{++}(t)\rangle \leftrightarrow -i G_{LL}(\tau).
\end{equation}
These identifications are consistent with the symmetries of the Green's functions in both real and Euclidean time: $G^<(t)$ and $G_{LR}(\tau)$ are imaginary and even (about $t=0$ and $\tau=0$, respectively); $G^\mathrm{T}(t)$ is imaginary and odd and $G_{LL}(\tau)$ is real and odd. In order to keep the term $\int \Sigma G$ in the action invariant, the self-energy components must be identified as
\begin{equation}
\label{eq:app_identifi_Sigma}
\Sigma^<(t) =\left\langle\Sigma^{+-}(t)\right\rangle \leftrightarrow -\Sigma_{LR}(\tau)
\qquad \text{and} \qquad
\Sigma^\mathrm{T}(t) =\langle \Sigma^{++}(t)\rangle \leftrightarrow i \Sigma_{LL}(\tau),
\end{equation}
where the minus sign in the first relation results from the backward time direction in the integration.

Let us now compare the Schwinger-Dyson equations.
In real time they are given by
\begin{equation}
\begin{split}
\label{eq:app_identifi_SDE_real}
(\omega-\Sigma_{++})G_{++}-\Sigma_{+-}G_{+-}=1
\qquad &\text{and} \qquad
(\omega-\Sigma_{++})G_{+-} -\Sigma_{+-}G_{++}=0,
\\
\Sigma_{++}(t)=-i^qJ^2 G_{++}^{q-1}(t)
\qquad &\text{and} \qquad
\Sigma_{+-}(t)=-i^q J^2G_{+-}^{q-1}(t)+i\mu\delta(t),
\end{split}
\end{equation}
where the equations in the first line are in frequency space and the ones in the second line are in the time domain.

In Euclidean time we have
\begin{equation}
\label{eq:app_identifi_SDE_Eucl}
\begin{split}
-(i\omega+\Sigma_{LL})G_{LL}+\Sigma_{LR}G_{LR}=1
\qquad &\text{and} \qquad
-(i\omega+\Sigma_{LL})G_{LR}-\Sigma_{LR}G_{LL}=0,
\\
\Sigma_{LL}(\tau)=-J^2 G_{LL}^{q-1}(\tau)
\qquad &\text{and} \qquad
\Sigma_{LR}(\tau)=i^q J^2 G_{LR}^{q-1}(\tau)-i\mu\delta(\tau).
\end{split}
\end{equation}
Using the identifications of Eq.~(\ref{eq:app_identifi_G}) in Eqs.~(\ref{eq:app_identifi_SDE_Eucl}) we recover the saddle-point equations \eref{eq:app_identifi_SDE_real}.
\section{Exact finite-$N$ results}
\label{app:ED}

In this appendix, we discuss exact diagonalization results obtained for
a finite number $N$ of Majorana particles. We expect that for large
$N$ these results will approach the results obtained from the SD equations.
Before discussing the numerical finite-$N$ results, we first discuss
the exact analytical results we have used as checks of the numerical finite-$N$
calculations, i.e., the eigenfunctions and eigenvalues of the ground state and the first excited state, and the large-$\omega$ behavior of $G_{LR}(\omega)$ in the low-temperature limit.

\subsection{Analytical finite-$N$ results}

The Hamiltonian is given by
\be
H = iH_L^\mathrm{SYK} - i (-1)^{q/2}H_R^\mathrm{SYK} +i\mu\sum_k\psi_{L}^{k}\psi_{R}^{k}.
\ee
It has the ground state (see Appendix \ref{app:choi})
\be
|0 \rangle = \sum_k |k\rangle e^{i\pi \gamma_c/4} C K |k\rangle,
\ee
which satisfies 
\be
(H_L^\mathrm{SYK} -(-1)^{q/2}H_R^\mathrm{SYK})|0\rangle =0,
\ee
and
\be
i\mu\sum_k\psi_{L}^{k}\psi_{R}^{k}|0\rangle =-\mu  \sum_k\psi_{L}^{k}\psi_{L}^{k}|0\rangle
=-\frac N2 \mu |0\rangle \equiv E_0|0\rangle.
\ee

The excited states of $i\mu  \sum_k\psi_{L}^{k}\psi_{R}^{k}$ are obtained by acting
with raising operators on the ground state. The lowest excited states
are given by
\be
i\mu  \sum_k\psi_{L}^{k}\psi_{R}^{k}\(\psi_{L}^{m} -i \psi_{R}^{m} \) |0\rangle
&=&\left(-\frac N2 +\frac 12 \right )\mu \(\psi_{L}^{m} -i \psi_{R}^{m}\) |0\rangle
+i(-i)\mu \( \frac 12 \psi_{L}^{m} -i\frac i2 \psi_{R}^{m} \) |0\rangle \nn \\
&=&\left (-\frac N2 + 1 \right ) \mu \(\psi_{L}^{m} -i \psi_{R}^{m}\) |0\rangle,
\ee
which can be shown by separation of the $k=m$ and $k\neq m$ terms.
The next excited state is obtained by acting twice with the raising operator on the ground
state, etc.
Since
\be
\(\psi_{L}^{m} -i \psi_{R}^{m}\) |0\rangle
= 2\psi_{L}^{m} |0\rangle,
\ee
we can take
\be
\prod_{i_1<i_2< \cdots <i_p} \psi_{L}^{i_1} \psi_{L}^{i_2}\cdots  \psi_{L}^{i_p}\ket{0}
\ee
as a basis for the eigenstates of the Hamiltonian.

Numerically we find that, for small $\mu$ and sufficiently large $N$ ($N >8$), the energy of the first excited state is given by
\be
E_1 = E_0 +4 \mu.
\ee
This excited state $|1\rangle$ is the next eigenstate of $i\mu\sum_k\psi_{L}^{k}\psi_{R}^{k}$ which can be similarly annihilated by $H_L^\mathrm{SYK} -(-1)^{q/2}H_R^\mathrm{SYK}$, i.e.,
\be
(H_L^\mathrm{SYK} -(-1)^{q/2}H_R^\mathrm{SYK})|1\rangle = 0,
\ee
suggesting that $|1\rangle$ is an entangled state composed of states
$|j\rangle  |j\rangle  $, or alternatively can be expressed as a linear combination of
\be  \prod_{i_1<i_2< i_3 <i_4} \psi_{L}^{i_1} \psi_{L}^{i_2}\psi_{L}^{i_3}  \psi_{L}^{i_4}
|0\rangle.
\ee
Indeed, this has been observed numerically. 

Next we discuss analytical checks for the Green's functions. As discussed in the main text, in the low-temperature limit we must have that 
$G_{LL}(\tau) =  -i G_{LR}(\tau)$
for $0<\tau\ll \beta/2$ and $\beta \to\infty$.
The symmetry of $G_{LL}(\tau)$ and $G_{LR}(\tau)$ 
gives the relation for $-\beta/2\ll \tau<0$. Combining the two we obtain
\be
G_{LL}(\tau) =  -i\,\sign(\tau)\,G_{LR}(\tau).
\label{caus_app}
\ee
The values at $\tau=0^{\pm}$ follow from the properties of the gamma matrices.
The result for $G_{LL}$ is given by
	\be\begin{aligned}
		G_{LL}(0^+) & =
		\frac 1N \frac 1Z \Tr\left( e^{-\beta H} T\sum_k \psi_{L}^{k}(0^+)\psi_{L}^{k}(0)\right) =\frac 12, \\
		G_{LL}(0^-) & =
		\frac 1N \frac 1Z \Tr\left( e^{-\beta H} T\sum_k \psi_{L}^{k}(0^-)\psi_{L}^{k}(0)\right) = \frac 1N \frac 1Z \Tr\left( e^{-\beta H} T\sum_k \psi_{L}^{k}(0)\psi_{L}^{k}(0^+)\right) \\
		& = -\frac 1N \frac 1Z \Tr\left( e^{-\beta H} \sum_k \psi_{L}^{k}(0^+)\psi_{L}^{k}(0)\right)= -\frac 12, 
	\end{aligned}\ee 
which are the Green's function in free limit, as is expected.
The time-ordering operator has been denoted by
$T$, which for fermionic operators satisfies $ T\psi(0) \psi(t) = -\psi(t)\psi(0)$ for $t<0$.
For the calculation of $G_{LR}(0)$ in the low-temperature limit
we use the property $(\psi_L^k+ i \psi_R^k) |0\rangle =0$ to obtain
\be
G_{LR} (0)  = \frac 1Z \frac 1N \Tr\left( e^{-\beta H} \sum_k \psi_{L}^{k}\psi_{R}^{k} \right) \approx  \frac 1Z \frac 1N \langle 0| e^{-\beta H} \sum_k \psi_{L}^{k}\psi_{R}^{k}  |0\rangle = \frac i 2 .
\ee
These values for $G_{LL}$ and $G_{LR}$ agree with what we observe in
finite-$N$ calculations and for solutions of the SD equations.

As a last check we discuss the asymptotic large-$\omega$ behavior of
$G_{LR}(\omega)$, the
Fourier transform of $G_{LR}(\tau)$.
The Green's function $G_{LR}(\tau)=(1/N)\sum_i\langle\psi_{L}^{i}(\tau)\psi_{R}^{i}(0)\rangle=(1/N)\sum_i G_{LR}^{ii}(\tau)$ is equal to the Green's function for the $i$th particle $G^{ii}_{LR}(\tau)$, which reads
\begin{equation}\begin{aligned}
		G^{ii}_{LR}(\tau) =  \langle\psi_{L}^{i}(\tau)\psi_{R}^{i}(0)\rangle =
		\frac{1}{Z} {\rm Tr}\left(e^{-\beta H}e^{\tau H}\psi_{L}^{i} e^{-\tau H}\psi_{R}^{i}\right).
		\label{E14}
\end{aligned}\end{equation}
In the low-temperature limit we have contributions from $\tau/\beta \ll 1$
and $(\beta-\tau)/\beta \ll1$. 
Using the approximation,  
\begin{equation}\begin{aligned}
		Z = \sum_i e^{-\beta E_i} \approx e^{-\beta E_0},
\end{aligned}\end{equation}we obtain for $\tau/\beta \ll 1$ 
\begin{equation}\begin{aligned}
		\label{eq:Gii}
		& G^{ii}_{LR}(\tau) = \langle 0| \psi_{L}^{i} e^{-\tau (H - E_0)}  \psi_{R}^{i} |0\rangle. 
\end{aligned}\end{equation}

Because $G_{LR}(\tau) = - G_{LR}(\beta-\tau)$ the contribution
for $(\beta-\tau)/\beta \ll 1$ is given by $-G_{LR}(\beta-\tau)$.
The Fourier transform $G^{ii}_{LR}(\omega)$
of both contributions adds up to
\begin{equation}\begin{aligned}
		& G^{ii}_{LR}(\omega) = \int_0^{\beta} (G^{ii}_{LR}(\tau)-G^{ii}_{LR}(\beta-\tau)) e^{i\omega\tau}
		\approx \langle 0| \psi_{L}^{i} \int_0^\beta d\tau \(e^{\tau (i\omega + E_0 - H )}+e^{\tau (-i\omega + E_0 - H )}\) \psi_{R}^{i} |0\rangle  .
\end{aligned}\end{equation}
We have used that for fermionic Matsubara frequencies, $\omega_n\beta= 2\pi (n+\frac{1}{2})$, so we have
$\exp(-i\beta\omega_n)=-1$. 
Therefore,
\begin{equation}\begin{aligned}
		& G^{ii}_{LR}(\omega) = -\langle 0| \psi_{L}^{i} \left (
		\frac{1}{i\omega + E_0 - H} -\frac{1}{-i\omega + E_0 - H} \right )\psi_{R}^{i} |0\rangle. 
\end{aligned}\end{equation}
We expand this result to second order in $1/\omega$. The first-order term
cancels and from the second-order term we obtain
\be
G^{ii}_{LR}(\omega) = \frac 2{\omega^2}
\langle 0| \psi_{L}^{i} (H-E_0) \psi_{R}^{i} |0\rangle.
\ee
To evaluate this expectation value we use that 
\be
H = i H_L^\mathrm{SYK} - (-1)^{q/2}i H_R^\mathrm{SYK} + i\mu\sum_{k}\psi_{L}^{k}\psi_{R}^{k},
\ee
and
\be\begin{aligned}
	& \langle 0|\psi_{L}^{i} H_L^\mathrm{SYK} \psi_{R}^i|0\rangle = \langle 0|\psi_{L}^{i} \psi_{R}^{i}H_L^\mathrm{SYK}|0\rangle = \frac{i}{2}\langle 0|H_L^\mathrm{SYK}|0\rangle , \\
	& \langle 0|\psi_{R}^{i} H_R^\mathrm{SYK} \psi_{R}^i|0\rangle = \langle 0|H_R^\mathrm{SYK}\psi_{L}^{i} \psi_{R}^{i}|0\rangle = \frac{i}{2}\langle 0|H_R^\mathrm{SYK}|0\rangle , \\
\end{aligned}\ee
as well as 
\be
\langle 0|\psi_{L}^{i}\left(i\mu\sum_{k}\psi_{L}^{k}\psi_{R}^{k}-E_0\right) \psi_{R}^{i}|0\rangle
= \frac{i\mu}{4} + \frac{i}{2}\langle 0|\left(i\mu
\sum_{k\neq i}\psi_{L}^{k}\psi_{R}^{k}\right )\ket{0}
+N\frac {i\mu}4= \frac{i\mu}{2}.
\ee
To derive this result we have used the commutation rules of the
fermion operators and properties of the ground state,
\be \begin{aligned}
	H|0\rangle = E_0|0\rangle &= -\frac {N}2 \mu |0\rangle,\\
	\psi_{L}^{i}\psi_{R}^{i} |0\rangle = \frac i2 |0\rangle ~~ \Rightarrow ~~  i\mu \sum_{k\neq i}\psi_{L}^{k}\psi_{R}^{k}|0\rangle &= -\frac {N-1}2 \mu
	|0\rangle.
\end{aligned}\ee
Our final result for $G^{ii}_{LR}(\omega)$
is given by
\be
G^{ii}_{LR}(\omega) =\frac{i\mu}{\omega^2} +O(1/\omega^3).
\label{eq:Glr_long_tau}
\ee

\subsection{Numerical finite-$N$ results for $q=4$}
\label{app:finite}

\begin{figure}[t!]
\includegraphics[width=8cm]{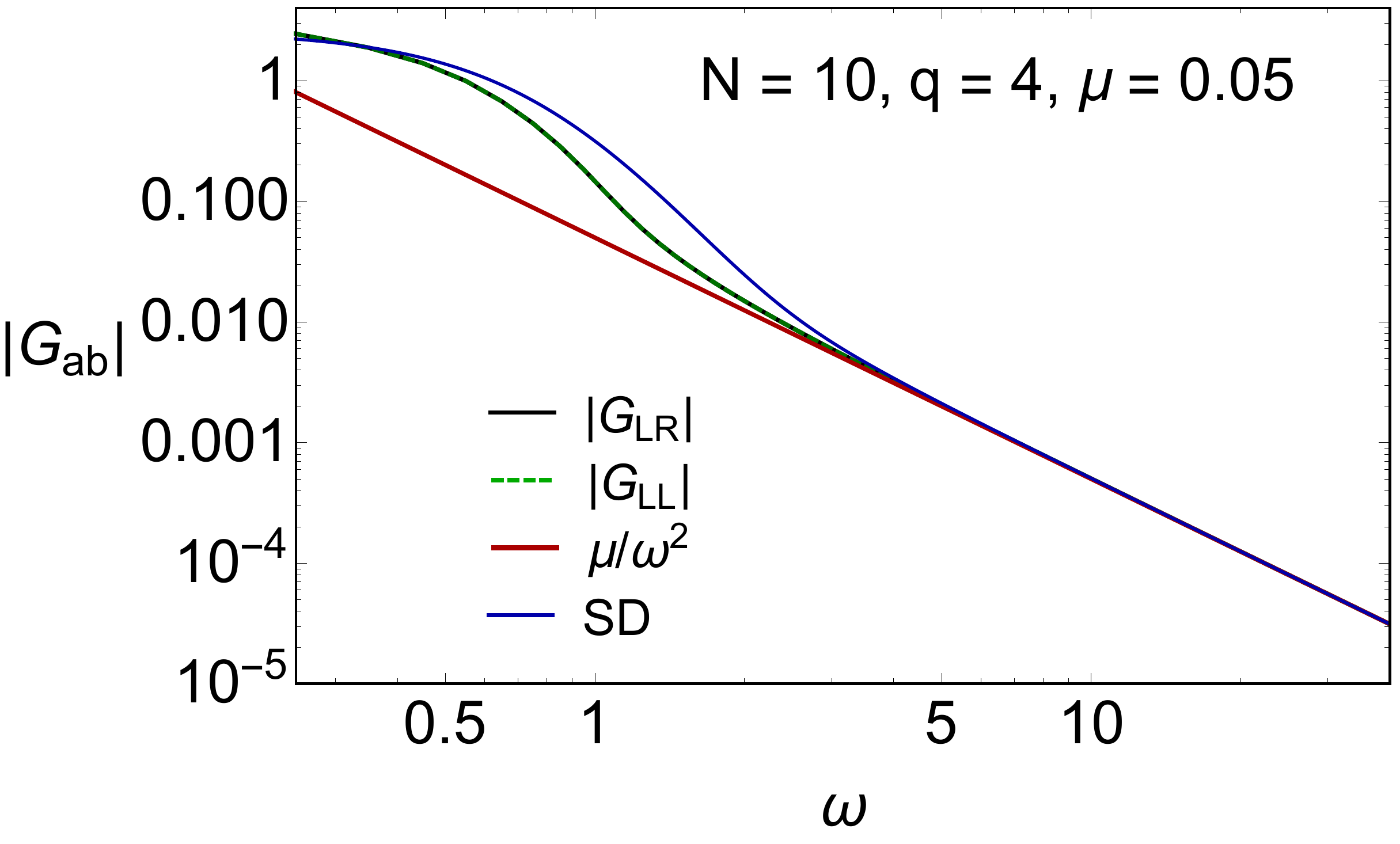}
\includegraphics[width=8cm]{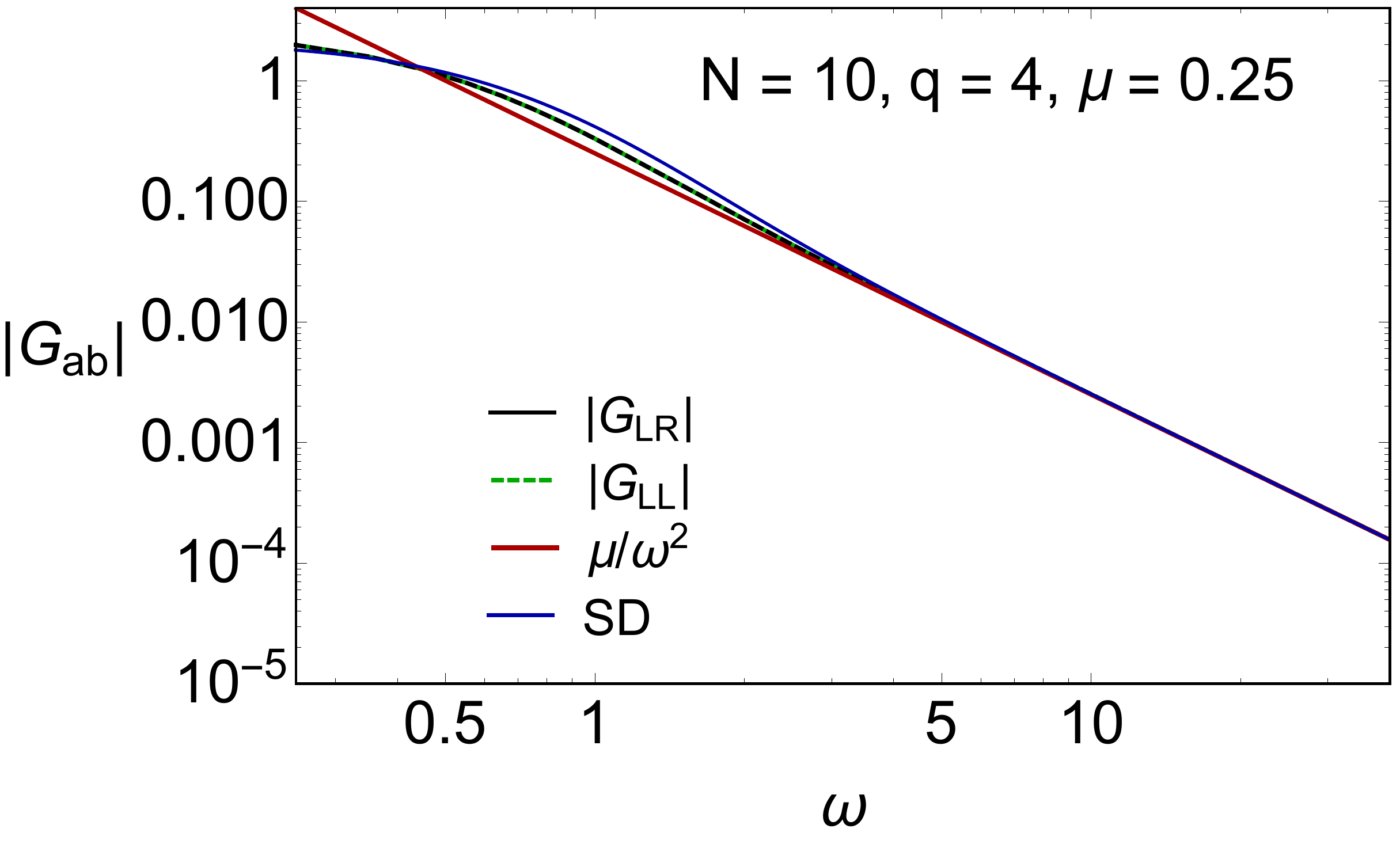}
\caption{Log-log plot of comparison of $|G_{LR}(\omega)|$ (black) and $|G_{LL}(\omega)|$ (green dashed) with the analytical asymptotic result of $\mu/\omega^2$ (red line) and the solution of the SD equations (blue curve). Results are given for $N=10$, $q=4$, and $\mu=0.05$ (left) and $\mu=0.25$ (right). 
}
\label{fig:comp}
\end{figure}

In this section, we discuss the low-temperature finite-$N$ results for $G_{LR}(\tau)$ obtained
by exact diagonalization.
The Green's functions can be evaluated using Eq.~\eref{E14} and
%This expression can be evaluated using 
the completeness relation of the eigenstates of
$H$. Care must be taken that the right eigenvectors are only
orthogonal to the left eigenvectors. The right eigenvectors satisfy
\be
H V^T = V^T D,
\ee
where $D$ is a diagonal matrix with the eigenvalues of $H$. Inserting $V^T {V^T}^{-1}$
after the exponential into Eq.~(\ref{E14}) within the region $\tau/\beta\ll 1$, we obtain
\be
G_{LR}^{ii}(\tau) =  \langle 0|\psi_{L}^{i}V^T e^{-(D-E_0)\tau}{V^T}^{-1} \psi_{R}^{i} |0\rangle.
\ee
This can be easily evaluated numerically for small values of $N$. 
As a check of our numerics, we show in Fig.~\ref{fig:comp} a comparison of the Green's functions in frequency space with the large-$\omega$ results derived in the previous subsection. Results are shown for $\mu =0.05$ (left) and
$\mu =0.25$ (right) and $N=10$.
The solutions of the SD equations (blue curve) show the same large-$\omega$ behavior but deviate from the finite-$N$ results at low and intermediate frequency.

\begin{figure}[t]
\centering
\subfigure[]{\includegraphics[width=8cm]{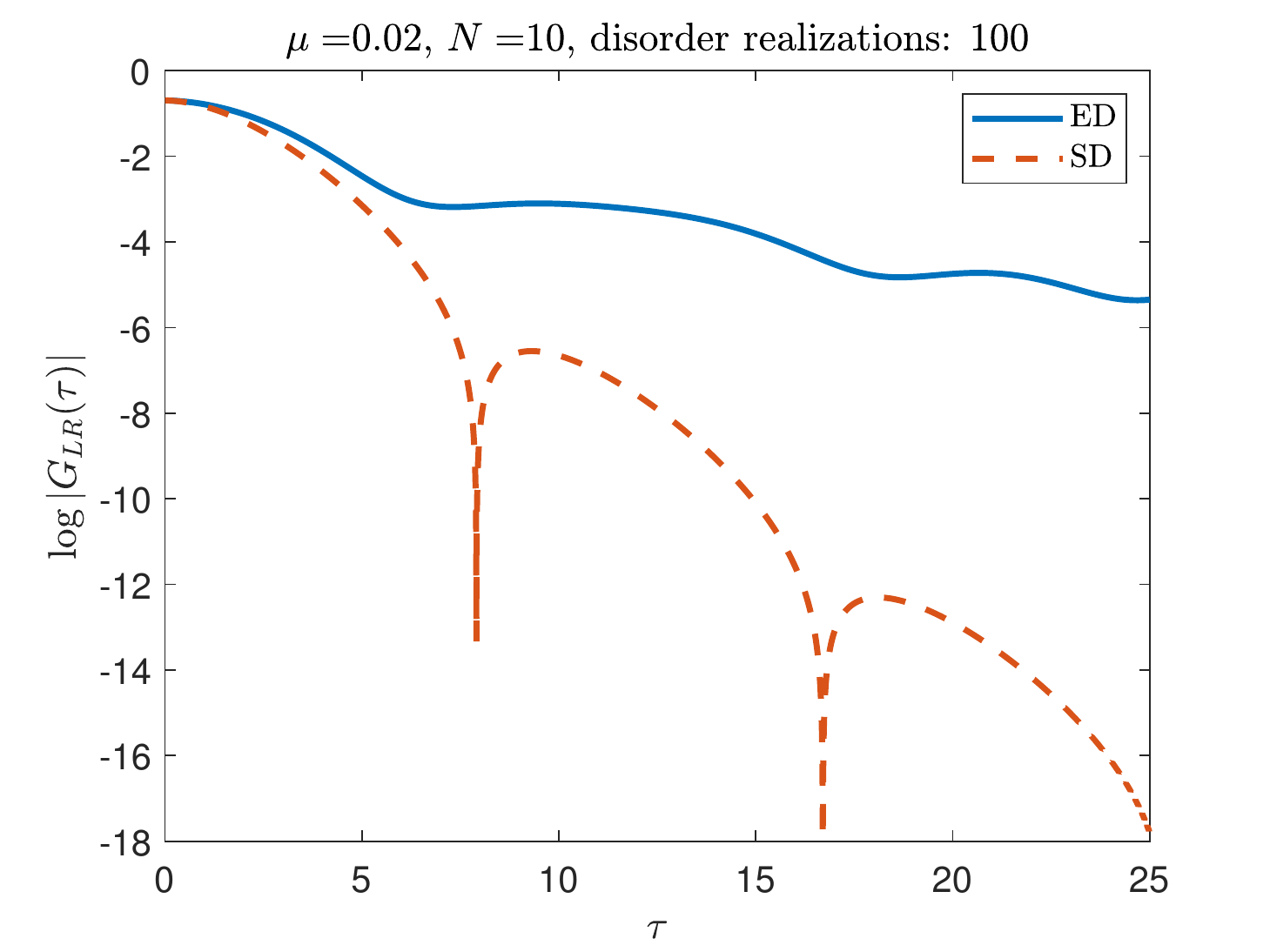}}
\subfigure[]{\includegraphics[width=8cm]{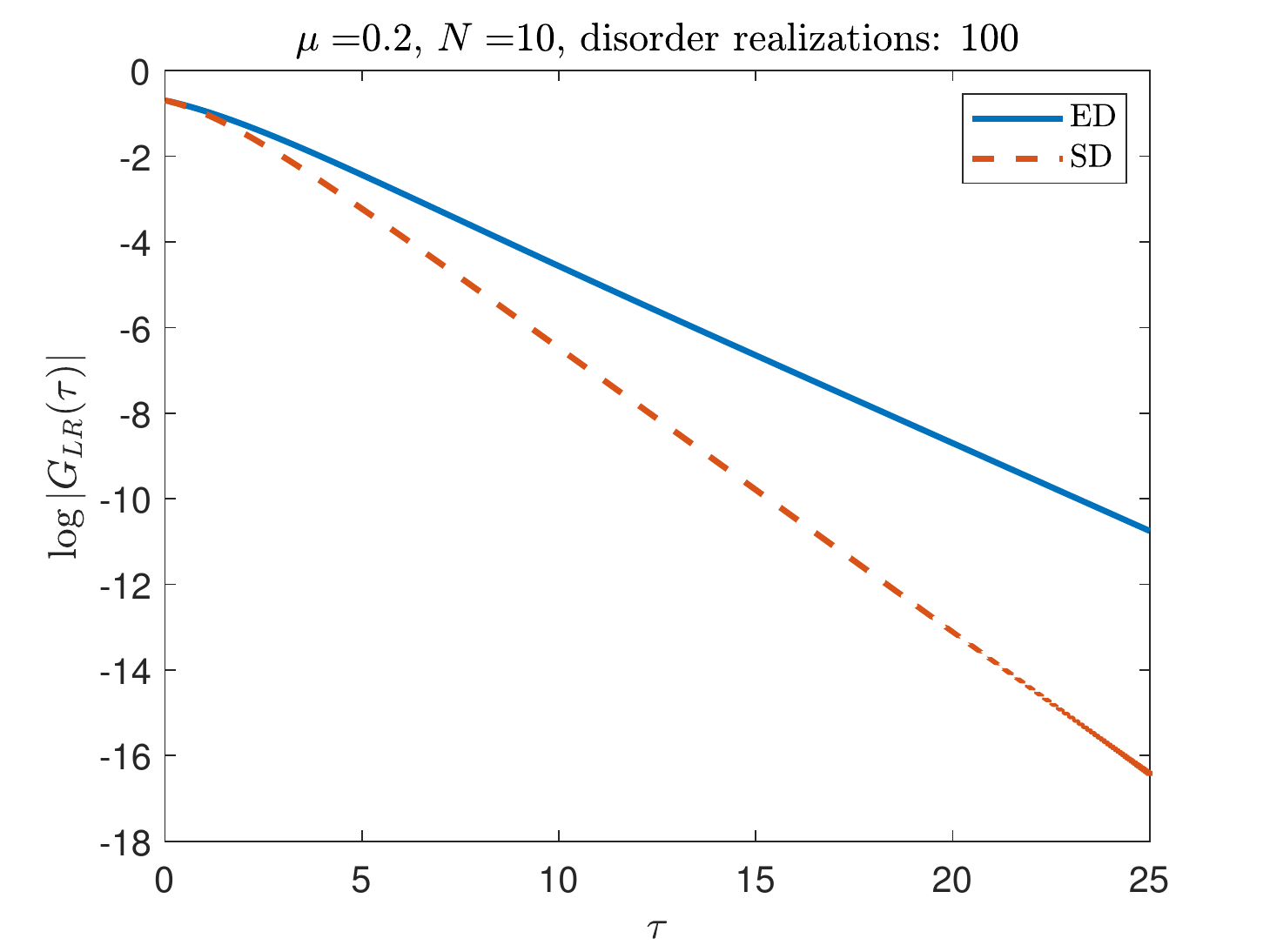}}
\caption{Plot of $\log|G_{LR}(\tau)|$ versus $\tau$ for $T=0.002$, $\mu=0.02$ (left), and $\mu=0.2$ (right), for $N=10$, with $100$ samples. The blue curve is the result calculated from exact diagonalization and the red dashed curve depicts the result obtained from
	the solution of the Schwinger-Dyson equations. 
}
\label{fig:GLR_ka_1_ld_02_p7_T_002_N_10_cp_30}
\end{figure}

\begin{figure}[t!]
\centering
{\includegraphics[scale=.4]{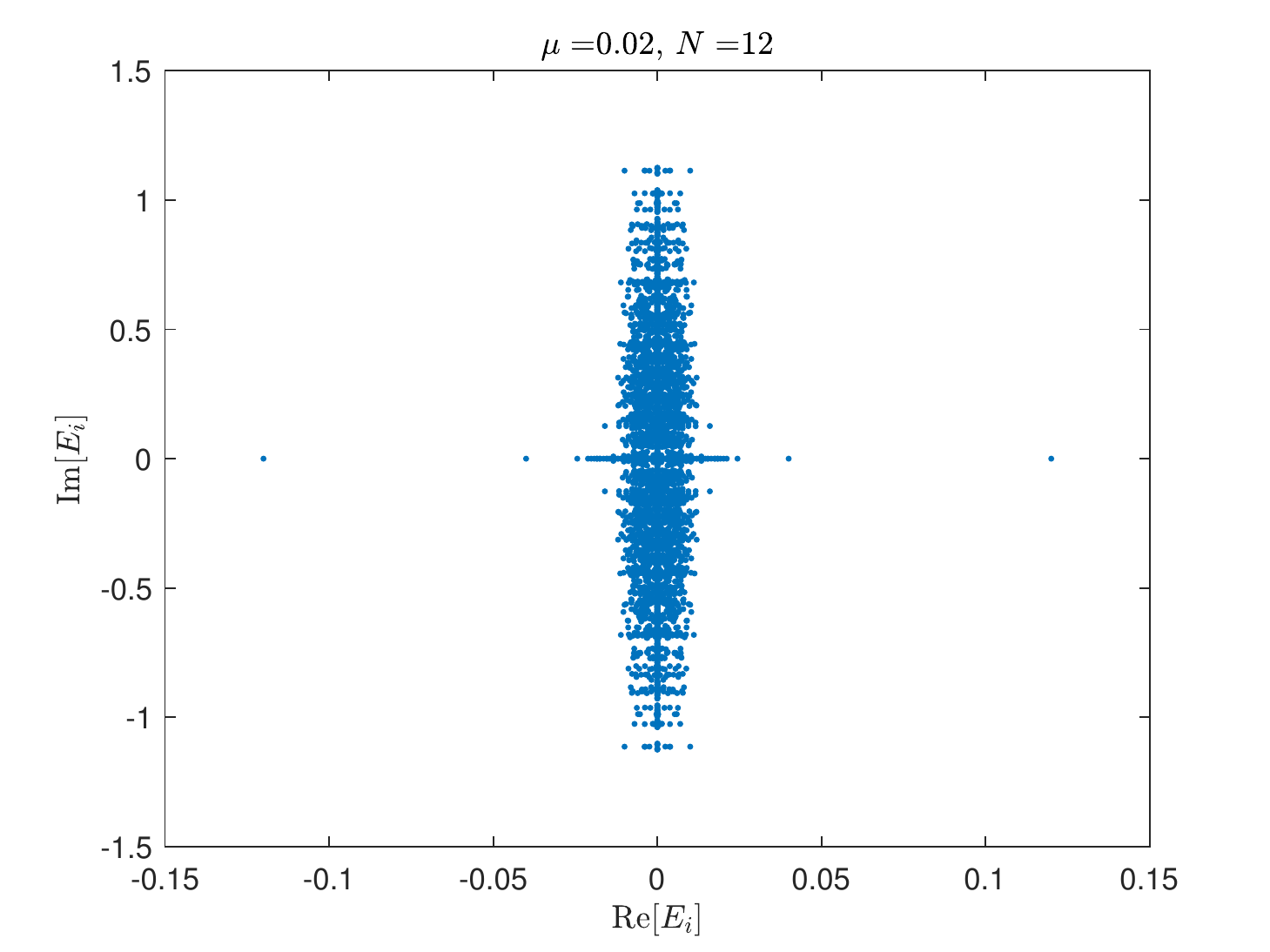}}
{\includegraphics[scale=.4]{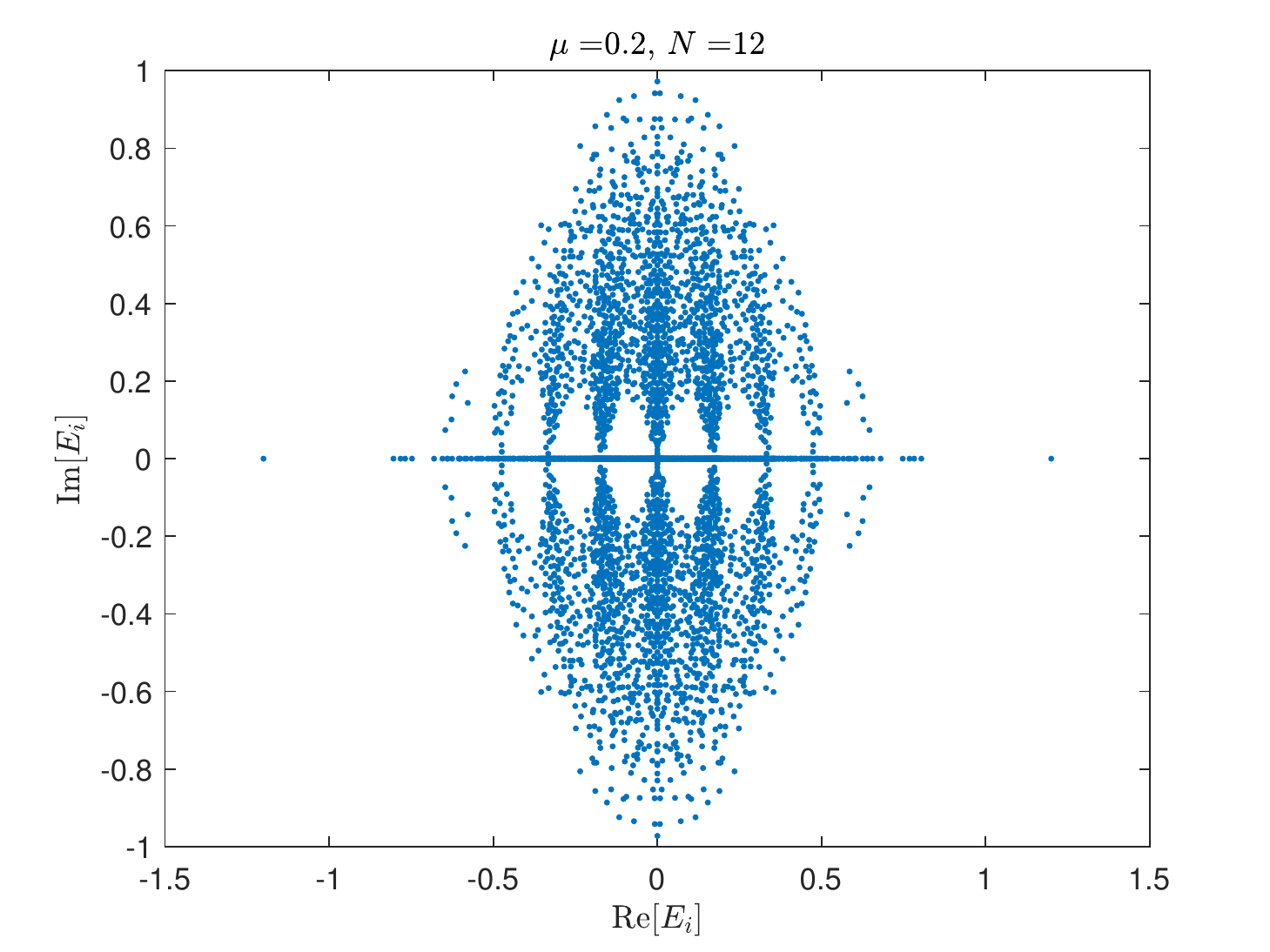}}
{\includegraphics[scale=.4]{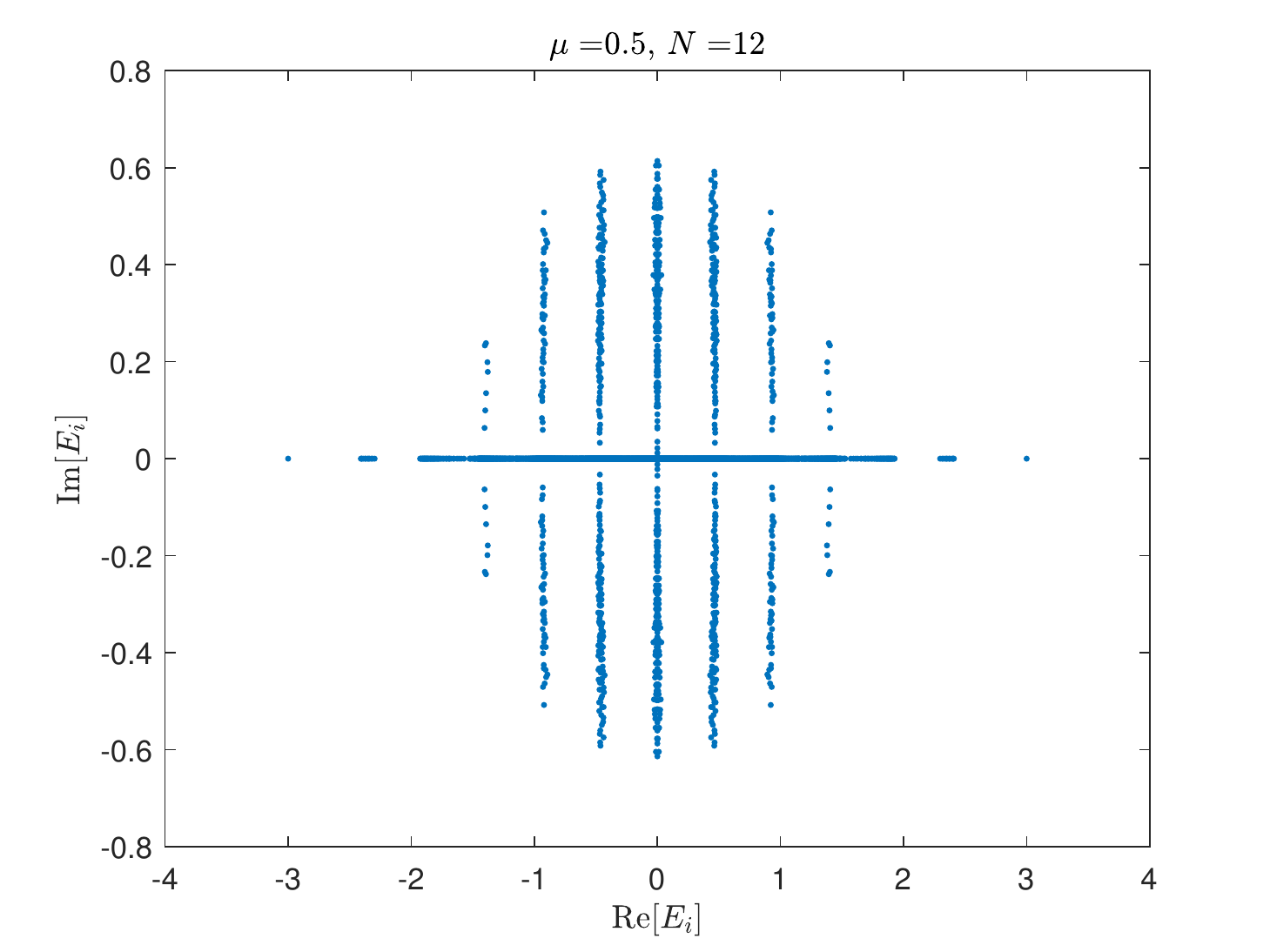}}
{\includegraphics[scale=.4]{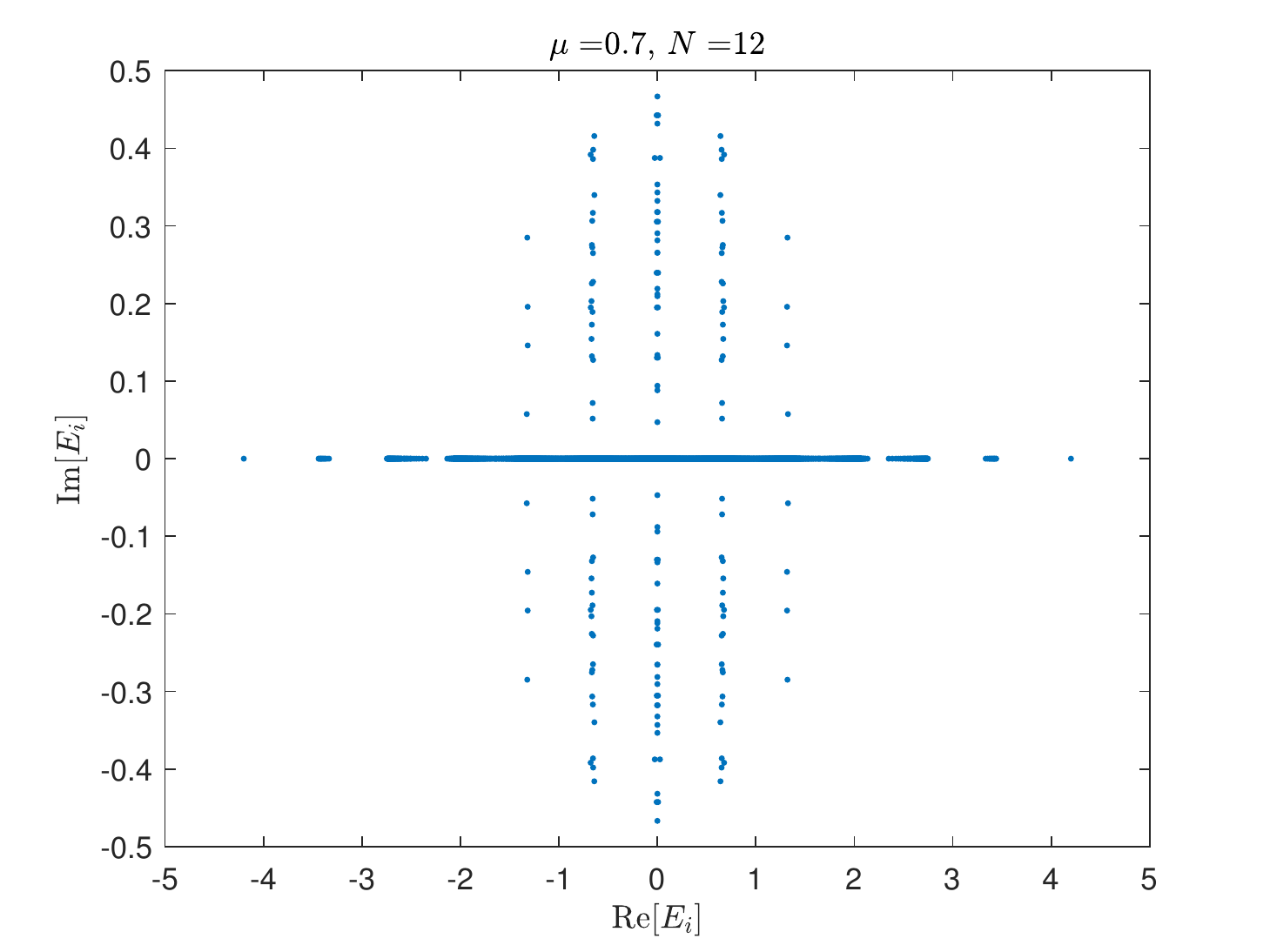}}
\caption{Spectrum for $\mu=0.02$, $\mu=0.2$, $\mu=0.5$, and $\mu=0.7$, $N=12$, and $q=4$. In the small-$\mu$ limit, most eigenvalues are complex and are located close to the imaginary axis, but low-lying eigenvalues remain real. When $\mu$ gets larger, the number of real eigenvalues increases,
	but the gap in units of $\mu$ between the first excited state
	and the bulk of the spectrum decreases. 
	For large $\mu$, we observe vertical stripes corresponding to the eigenvalues of $i\mu \sum_k \psi_{L}^{k}\psi_{R}^{k}$.}
\label{fig:sp}
\end{figure}

Next, we study the large-time behavior of $G_{LR}(\tau)$.
In Fig.~\ref{fig:GLR_ka_1_ld_02_p7_T_002_N_10_cp_30} we show results for $N=10$ and two values of $\mu$ (see legend) and compare them to the result from the SD equations. We observe a discrepancy between the SD results and the finite-$N$ behavior of the Green's function at long times, especially when $\mu$ is small. At this moment we do not have a complete understanding of this discrepancy. 
It might have its origin in the large-$N$ limit of the spectrum, but we do not see this from the finite-$N$ calculations. As an example, in Fig.~\ref{fig:sp} we give scatter plots of spectra for $N=12$ and various values of $\mu$. Next,
we will show that the discrepancy is mainly due to cancellations that become
manifest only for a large number of realizations.

\begin{figure}[t!]
\centering
\includegraphics[width=8cm]{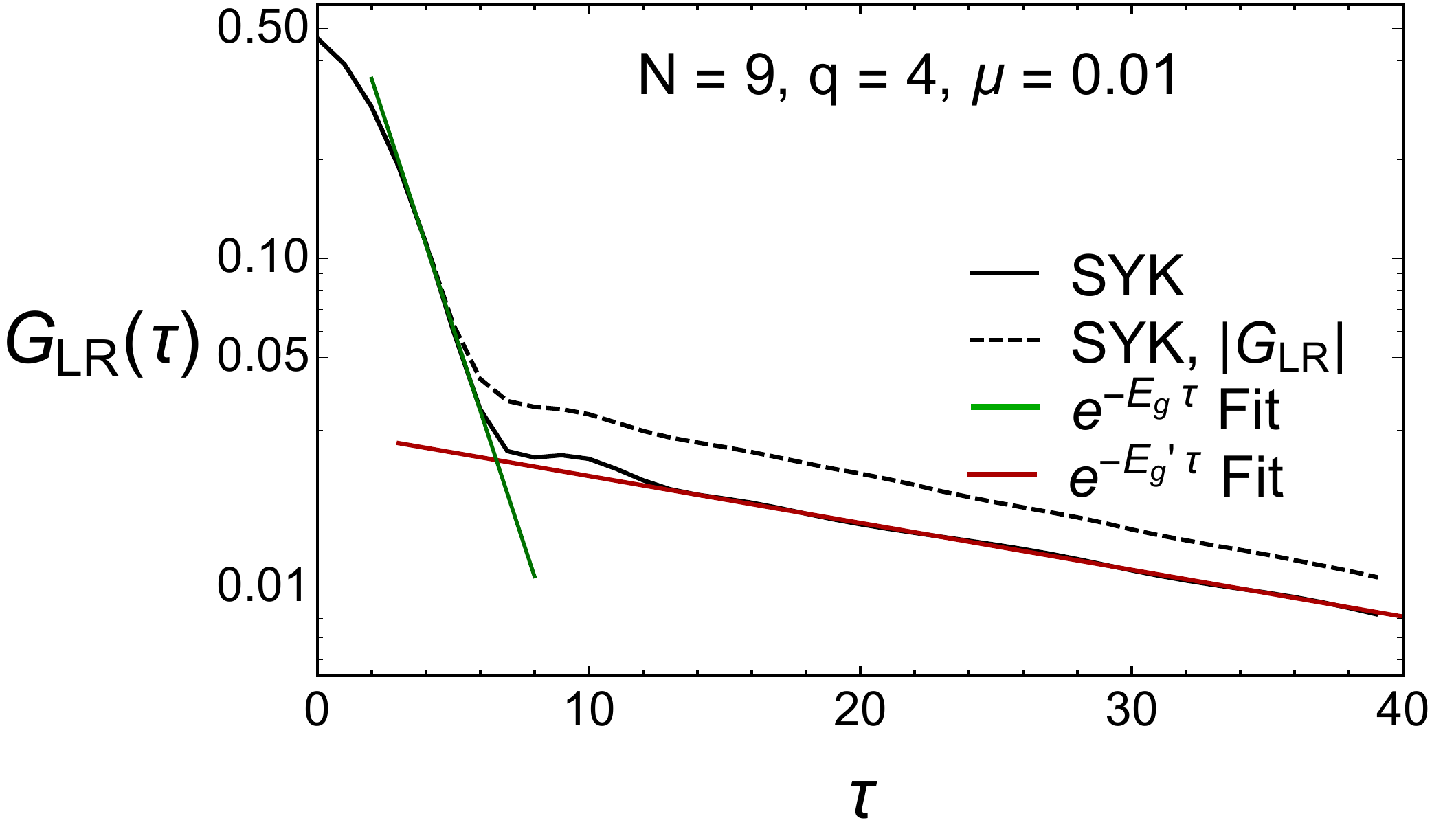}
\includegraphics[width=8cm]{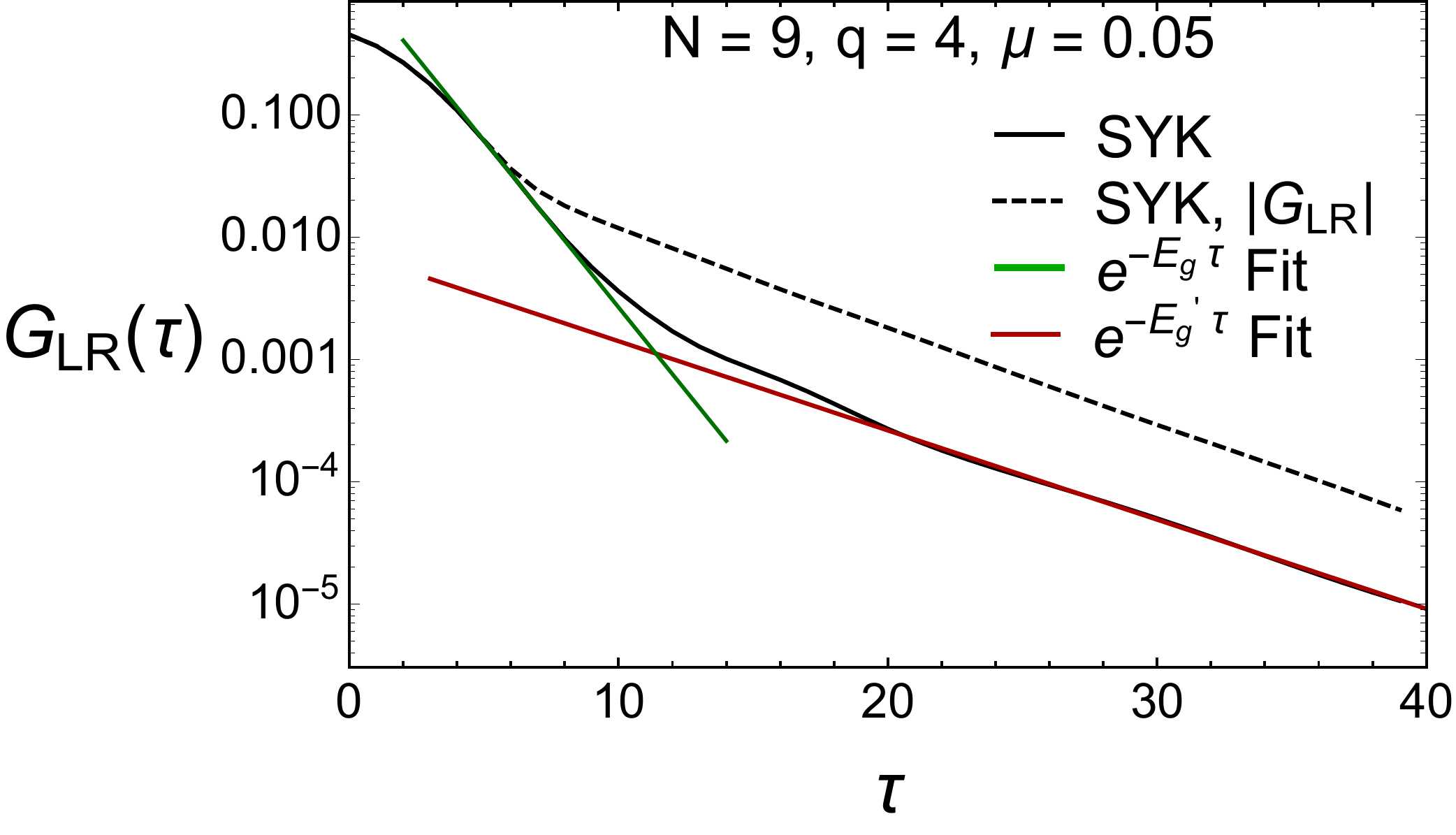}\\
\includegraphics[width=8cm]{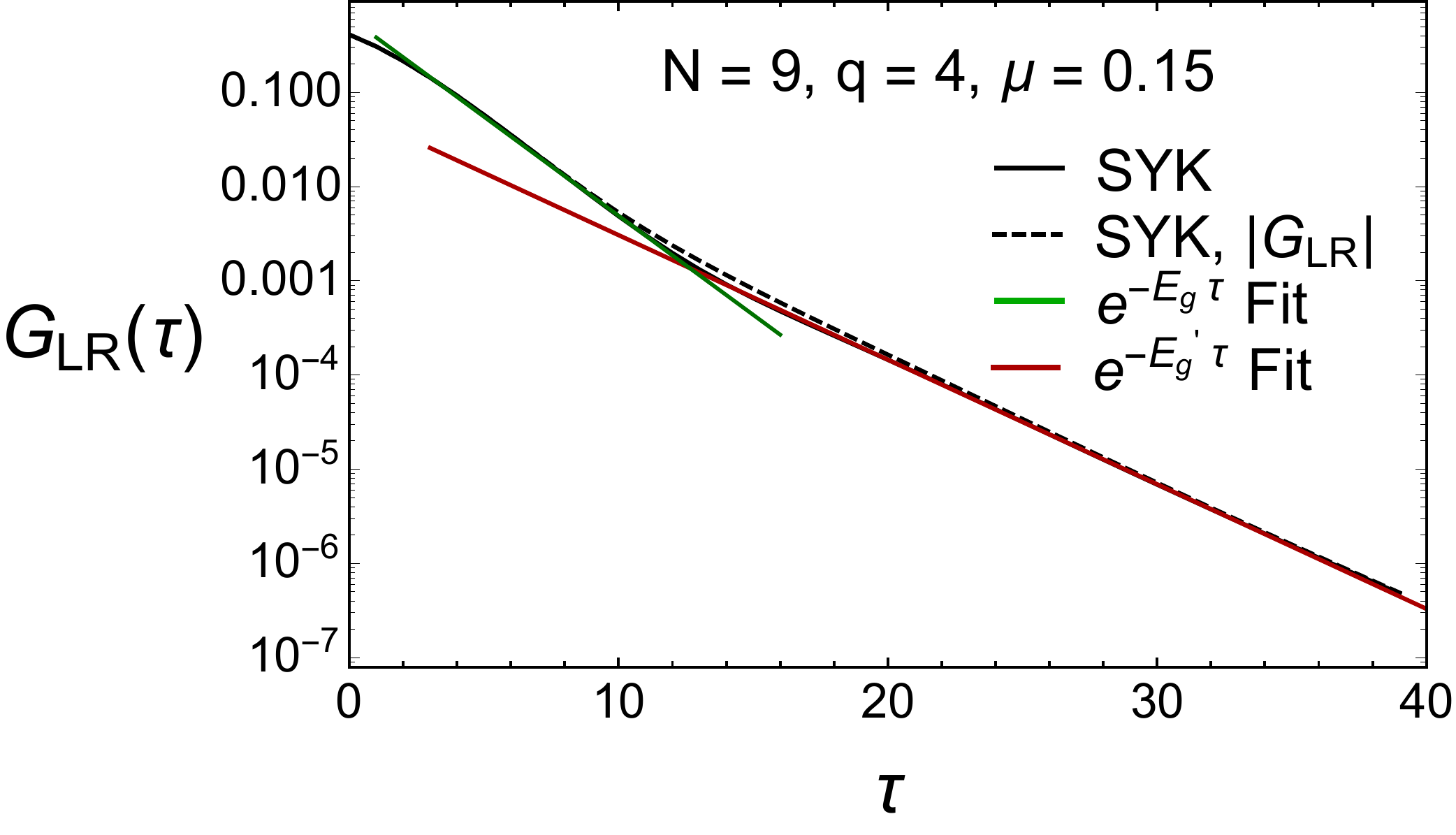}
\includegraphics[width=8cm]{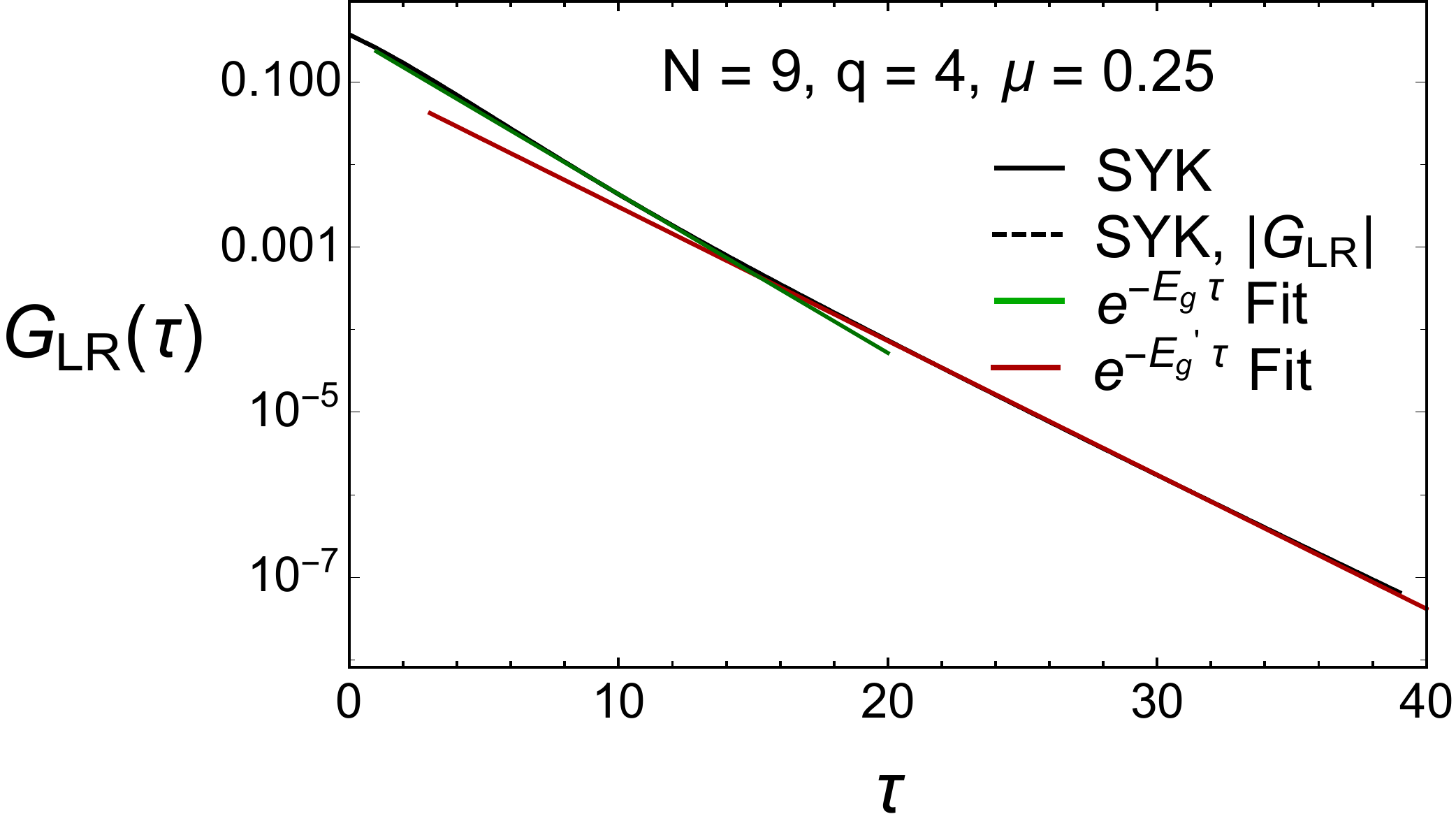}
\caption{Logarithmic plot of the time dependence of the ground state expectation value
	of the Euclidean Green's function (black curves) for $N=9$
	and $\mu= 0.01$, $\mu=0.05$, $\mu=0.15$, and $\mu=0.25$. The red line is the fit to the slope of the black curve at large times, while the green line is a fit to the initial exponential behavior. The SYK results have been obtained by averaging over 400 realizations.}
	\label{fig:green}
\end{figure}

In Fig.~\ref{fig:green}, we show results for $N=9$ and various values of $\mu$ (see legends).
For small $\mu$, we can distinguish two exponents, an early rapid decay and a much slower decay at large times. The exponent of the early rapid decay is in
qualitative agreement with the results of the Schwinger-Dyson equations, but the exponent of the long-time tail is definitely not in agreement with the SD equations for
$\mu < 0.15$.
We have performed the same analysis for $N=11$ and $N=13$, which shows a similar behavior of $G_{LR}(\tau)$.

\begin{figure}[t!]
\centerline{\includegraphics[width=8cm]{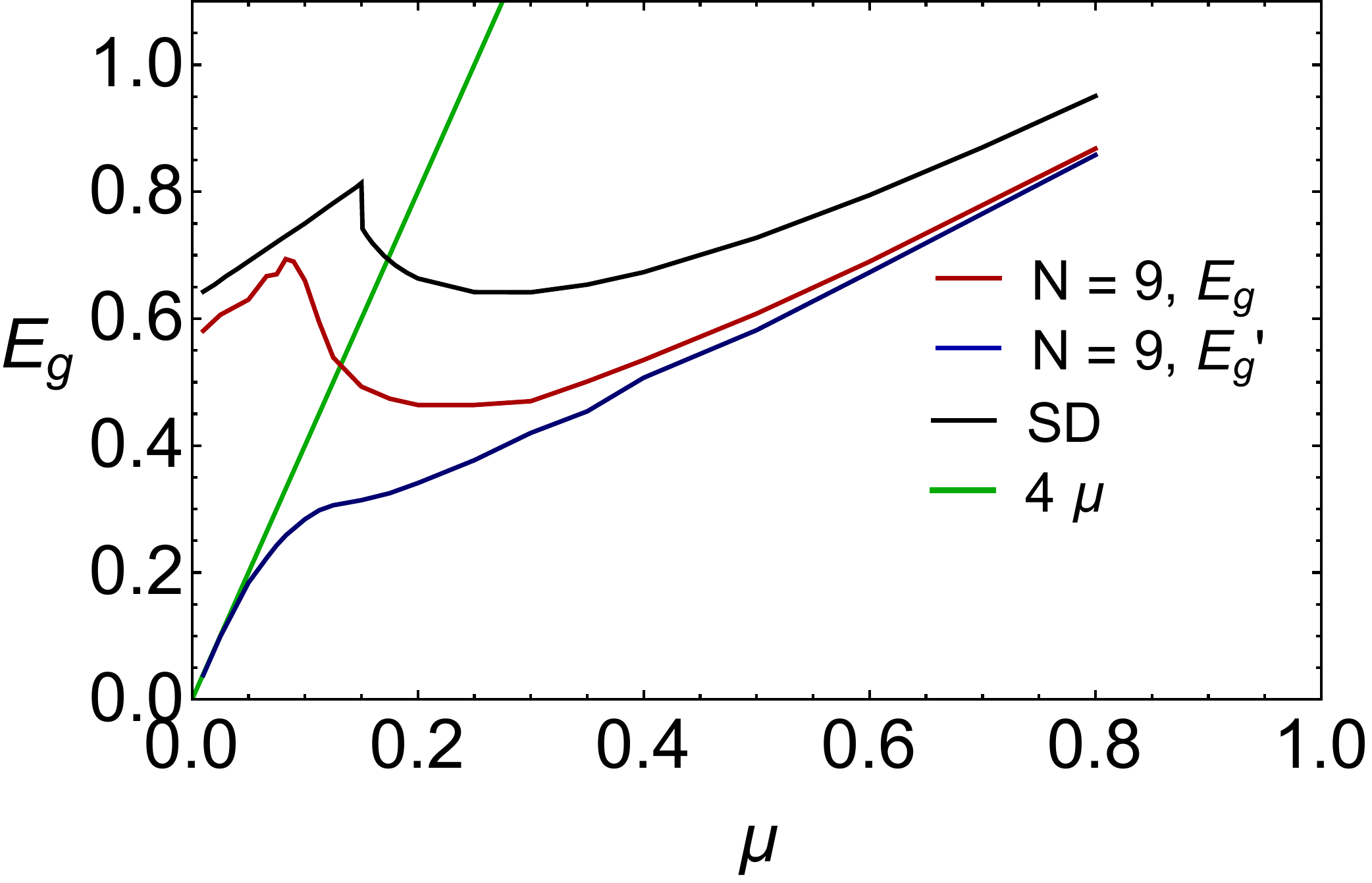}}
\caption{The gap $E_g$ as a function of $\mu$ for $N=9$ (red curve) obtained
  from the initial exponential decay of the Green's function. The green
  line shows the analytical result for the spectral gap for small $\mu$. The results obtained from the SD equations are represented by the black curve. The blue
  curve represents $E_g'$ obtained by fitting the long-time exponential decay of the Green's function. For $\mu= 0.075$ the average is over 25,000 realizations
  while for other values of $\mu$ in the range
  $0.066 \le \mu \le 0.1125$ we used 10,000
    realizations. For all other points we used 4000 realizations.
}
\label{fig:gap}
\end{figure}

To obtain a better understanding of the discrepancy between the long-time tail
of the Green's function obtained from the SD equations and from exact diagonalization, we plot in Fig. \ref{fig:green} both the ensemble average of the Green's
function (black curve) and the ensemble average of the absolute value of the
Green's function (dashed black curve). For $\mu <0.15$ we observe large cancellations in the ensemble average of the Green's function, which increases the range of the initial exponential decay. The exponent of
the long-time tail is not affected by these cancellations and remains equal
to the exponent of the tail of the ensemble average of the absolute value of
the Green's functions.
The cancellations are strongest when $\mu \approx 0.075$
and decrease both for smaller and larger values of $\mu$. They are absent
for $\mu > 0.175$.
The $\mu$ dependence of the initial exponent (red curve), denoted by $E_g$, and the long-time exponent (green curve), denoted by $E_g'$, for $N=9$ are shown
in Fig.~\ref{fig:gap}. 
The gap is obtained by fitting $\exp(-E_g \tau)$
to the initial decay (green line in Fig. \ref{fig:green}).
For the decay rate of the long-time tail, we fit $\exp(-E_g' \tau)$
(red line in Fig.~\ref{fig:green}) to the numerically
obtained long-time tail of $G_{LR}(\tau)$.
Also shown is the result obtained from the solution of the Schwinger-Dyson equations (black curve).
We observe that the result for the Schwinger-Dyson equations is in qualitative
agreement with the exponent of the initial decay rate.
For small $\mu$, the gap in the spectrum is well defined and is given by (see green line in Fig.~\ref{fig:gap})
\be
\Delta= E_1-E_0 = 4\mu.
\ee
The finite-$N$ numerical results for the long-time tail
are in good agreement with this
analytical prediction, but we did not find solutions of the SD equations
with this behavior.

Our finite $N$ results show that the cancellations that take place for
$\mu < 0.15$ increase the range of the initial exponential behavior, but it
requires a very large ensemble to realize this.
We expect that, in the large-$N$ limit, this initial exponential behavior will
become dominant. A detailed finite-size scaling analysis
required to show this will be deferred to future work.

\section{Large-$q$ limit in the Euclidean problem}
\label{app:large_q}
Following the method in Ref.~\cite{maldacena2018}, the
SD equations can be solved analytically in the large-$q$ limit at low temperature.
The main idea is to solve the equations in small-$\tau$ and large-$\tau$ region
with proper boundary conditions, then match these two solutions to fix the undetermined parameters. 

Recalling the Schwinger-Dyson equations, 
\begin{equation}\begin{aligned}
		& \partial_{\tau_1}G_{LL} -\Sigma_{LL} *G_{LL} -\Sigma_{LR}*G_{RL}=\delta(\tau_1,\tau_2), \\
		& \partial_{\tau_1}G_{LR} -\Sigma_{LL} *G_{LR} -\Sigma_{LR}*G_{RR}=0, \\
		& \Sigma_{LL}=-\frac{\mathcal{J}^2}{q}(2G_{LL})^{q-1}, \\
		& \Sigma_{LR}=(-1)^{q/2}\frac{\mathcal{J}^2}{q}(2G_{LR})^{q-1} -i\mu\delta(\tau_1-\tau_2),
		\label{eq:EoM_original}
\end{aligned}\end{equation}
where ${\mathcal{J}}^2= q J^2/2^{q-1}$, we first notice that they give rise to the free solutions in the $\tau\to 0$ limit.
Since the large-$q$ interaction provides only a weak constraint on
the momenta through the conservation law at each vertex of the Feynman diagram, $G_{ab}(\tau)$ can be expanded in powers of $1/q$ around the free solution as
\begin{equation}\begin{aligned}
		& G_{LL}=\frac{1}{2}{\rm sign}(\tau)e^{\frac{1}{q}g_{LL}}=\frac{1}{2}{\rm sign}(\tau)(1+\frac{1}{q}g_{LL}+\dots), \\
		& G_{LR}=\frac{i}{2}e^{\frac{1}{q}g_{LR}}=\frac{i}{2}(1+\frac{1}{q}g_{LR}+\dots),
\end{aligned}\end{equation}
which leads to   
\begin{equation}\begin{aligned}
		& \Sigma_{LL}=-\frac{\mathcal{J}^2}{q}e^{\frac{q-1}{q}g_{LL}}\approx -\frac{\mathcal{J}^2}{q}e^{g_{LL}}, \\
		& \Sigma_{LR}=-i\frac{\mathcal{J}^2}{q}e^{\frac{q-1}{q}g_{LR}} -i\mu\delta(\tau_1-\tau_2) \approx-i\frac{\mathcal{J}^2}{q}e^{g_{LR}} -i\mu\delta(\tau_1-\tau_2).
\end{aligned}\end{equation}
Substituting the expressions for $G_{ab}$ and $\Sigma_{ab}$ into SD equations,
and applying an additional partial derivative, the leading order terms cancel and at order
${1}/{q}$ we obtain
\begin{equation}\begin{aligned}
		& \sign(\tau)\( \partial_{\tau}^2 g_{LL} + 2\mathcal{J}^2 e^{g_{LL}(\tau)} \)= 0,  \\
		& \partial_{\tau}^2 g_{LR} + 2\mathcal{J}^2e^{g_{LR}(\tau)} +2\hat{\mu}\delta(\tau) = 0,
		\label{eq:EoM_gll_glr_app}
\end{aligned}\end{equation}
with $\hat \mu={\mu}{q}$. 

The boundary conditions for $g_{LL}$ and $g_{LR}$ at $\tau=0$ are
\begin{equation}\begin{aligned}
		g_{LL}(0^+)=0, \qquad \partial_{\tau} g_{LR}(0^+) = -\hat{\mu},
		\label{eq:bc_0}
\end{aligned}\end{equation}
which follow from the free limit of $G_{LL}$ at $\tau \to 0$, and the integral over the second saddle-point equation of Eq.~(\ref{eq:EoM_gll_glr_app}) over an infinitesimal region $\tau\in[0,\epsilon]$, respectively. In addition, when $T\to 0$, $g_{LL}$ and $g_{LR}$ satisfy the boundary condition
\begin{equation}\begin{aligned}
		g_{LL}(\tau) - g_{LR}(\tau) \to 0 ,\qquad\tau\to \infty,
		\label{eq:bc_infinity}
\end{aligned}\end{equation}
which will select the exponentially decaying solutions.
At finite but low temperatures, $g_{LL}$ and $g_{LR}$ are not equal, and we need to solve the SD equations to find the relation between $g_{LL}$ and $g_{LR}$.

With the above boundary conditions, Eqs.~(\ref{eq:EoM_gll_glr_app}) can be solved to give
\begin{equation}\begin{aligned}
		e^{g_{LL}}=\frac{\alpha^2}{\mathcal{J}^2 \cosh^2(\alpha|\tau|+\gamma)},\qquad e^{g_{LR}}=\frac{\tilde{\alpha}^2}{\mathcal{J}^2 \cosh^2(\tilde{\alpha}|\tau|+\tilde{\gamma})}.
		\label{eq:solutions_q}
\end{aligned}\end{equation}
The boundary conditions~(\ref{eq:bc_0}) and (\ref{eq:bc_infinity}) give the relations
\begin{equation}\begin{aligned}
		\frac{\alpha^2}{\mathcal{J}^2 \cosh^2 \gamma} =1,\quad 2\tilde{\alpha}\tanh\tilde{\gamma}=\hat{\mu},\quad \tilde{\alpha}=\alpha, \quad \tilde{\gamma}=\gamma +\sigma,
		\label{eq:constants_0}
\end{aligned}\end{equation}
where $\sigma$ is a parameter
to account for deviations from the zero-temperature limit.

In order to find the value of $\sigma$, the SD equations need to be solved in the long-$\tau$ region.
Note that $G_{ab}(\tau)$ decays exponentially near $\tau=0$ and $\tau=\beta$.
Therefore, $G_{ab}(\tau)$ is close to zero for large $\tau$.
Since $\Sigma_{ab}(\tau)\sim G_{ab}(\tau)^{q-1}$ we can approximate
\begin{equation}\begin{aligned}
		\Sigma_{LR}(\tau) \sim -i\nu\delta(\tau),
\end{aligned}\end{equation}
where $\nu$ is given by the integral
\begin{equation}\begin{aligned}
		\nu = i\int\Sigma_{LR}(\tau) \approx \int_{-\infty}^{\infty}\frac{\mathcal{J}^2}{q}e^{g_{LR}}d\tau =
		\frac{\hat{\mu}}{q\tanh\tilde{\gamma}}.
		\label{eq:def_nu}
\end{aligned}\end{equation}
With this approximation, and noting that $\Sigma_{LL}*G_{LL}=\int\Sigma_{LL}(\tau-\tau')G_{LL}(\tau')d\tau'$ and $\Sigma_{LL}*G_{LR}=\int\Sigma_{LL}(\tau-\tau')G_{LR}(\tau')d\tau'$ are negligible because $\Sigma_{LL}(\tau)$ is odd in $\tau$, the SD equations in Eq.~(\ref{eq:EoM_original}) are reduced to
\begin{equation}\begin{aligned}
		\partial_{\tau} G_{LL} +i\nu G_{RL}=0, \qquad \partial_{\tau} G_{LR} +i\nu G_{RR}=0.
\end{aligned}\end{equation}
Because of the symmetries $G_{LL}(\tau)=G_{LL}(\beta-\tau)$ and $G_{LR}(\tau)=-G_{RL}(\beta-\tau)$, the solutions are given by
\begin{equation}\begin{aligned}
		G_{LL}(\tau)=A\cosh[\nu(\beta/2-\tau)], \qquad G_{LR}(\tau)=iA\sinh[\nu(\beta/2-\tau)],
\end{aligned}\end{equation}
for some constant $A$.
The small-$\tau$ behavior of these solutions is matched with the small-$\tau$ behavior of Eq.~\eref{eq:solutions_q}. We find
\begin{equation}\begin{aligned}
		\left|\frac{G_{LR}}{G_{LL}}\right| =& \tanh[\nu(\beta/2-\tau)]
		\approx 1 -2 e^{-\nu\beta} \\
		=& e^{(g_{LR}-g_{LL})/q} \approx e^{2(\gamma -\tilde \gamma)/q}
		= 1- 2(\tilde \gamma -\gamma)/q +O(1/q^2),
\end{aligned}\end{equation}
implying
\begin{equation}\begin{aligned}
		\sigma=\tilde{\gamma}-\gamma=q e^{-\nu\beta}.
		\label{eq:sigma_gamma}
\end{aligned}\end{equation}

The free energy follows from the derivative
\begin{equation}\begin{aligned}
		\partial_{\beta}(\beta F)/N = E/N =& \left[\frac{1}{q}\partial_{\tau}G_{LL}+\frac{1}{q} \partial_{\tau} G_{RR} +i\mu(1-\frac{2}{q})G_{LR}\right]_{\tau\to 0^+} \\
		=& \frac{1}{q^2}\left[\partial_{\tau}\log e^{g_{LL}} -\hat{\mu}(\frac{q}{2}-1)-\frac{\hat{\mu}}{2}\log e^{g_{LR}}\right]_{\tau\to 0^+} \\
		=& \frac{\hat{\mu}}{q^2}\left[-\frac{\tanh\gamma}{\tanh\tilde{\gamma}} -(\frac{q}{2}-1)-\log\frac{\cosh\gamma}{\cosh\tilde{\gamma}}\right]
		\label{eq:E_T}
\end{aligned}\end{equation}
This result is obtained by expressing the free energy in terms of the dimensionless
variables $\beta J$ and $\beta \mu$ and using that 
$\beta\partial_{\beta} = \mathcal{J}\partial_{\mathcal{J}}+\mu\partial_{\mu}$ when acting them on $\beta F$. 
 After a series of manipulations, we finally obtain the free energy,
\begin{equation}\begin{aligned}
		\beta 
		F/N =& \frac{\tanh\tilde{\gamma}\log(q/\sigma)}{q}\left(-\frac{\tanh\gamma}{\tanh\tilde{\gamma}} -(\frac{q}{2}-1)-\log\frac{\cosh\gamma}{\cosh\tilde{\gamma}}\right) -\frac{\sigma}{q}\log\frac{q}{\sigma}-\frac{\sigma}{q}  \\
		=& \frac{\beta\hat{\mu}}{q^2}\left(-\frac{\tanh\gamma}{\tanh\tilde{\gamma}} -(\frac{q}{2}-1)-\log\frac{\cosh\gamma}{\cosh\tilde{\gamma}}\right) -\frac{\sigma}{q}(1+\log\frac{q}{\sigma}).
\end{aligned}\end{equation}
In the limit $\beta \to \infty$ we have that $\sigma=\tilde \gamma -\gamma \to 0 $ from relation (\ref{eq:sigma_gamma}), so that $\beta F/N=-\mu /2$, which is observed both in large-$N$ and finite-$N$ calculations. Furthermore, $\beta F/N=-({1}/{\beta N})\log Z \approx ({1}/{N})\log e^{-\beta E_0}=E_0/N$ in the low-temperature limit, which agrees with the value of the ground state energy discussed in Appendix~\ref{app:ED}.

\bibliography{libryu}

\end{document}